%% file: to_arxiv.tex
\documentclass[sigconf, nonacm]{acmart}

\usepackage{enumitem}
\usepackage{latexsym}
\usepackage{float}
\usepackage{booktabs}
\usepackage{longtable}
\usepackage{multirow}
\usepackage{graphicx}
\usepackage{MnSymbol}
\usepackage{needspace} 
\usepackage{soul}
\usepackage{xcolor}
\usepackage{subcaption}
\usepackage{orcidlink}
\usepackage{listings}
\usepackage[dvipsnames]{xcolor}
\usepackage{lscape}
\usepackage{tcolorbox}
\usepackage{soul}
\usepackage{rotating} 
\usepackage{pifont}
\newcommand{\cmark}{\ding{51}}%
\newcommand{\xmark}{\ding{55}}%

\AtBeginDocument{%
  }

\copyrightyear{2026}
\acmYear{2026}
\setcopyright{cc}
\setcctype{by}
\acmConference[IUI '26]{31st International Conference on Intelligent User Interfaces}{March 23--26, 2026}{Paphos, Cyprus}
\acmBooktitle{31st International Conference on Intelligent User Interfaces (IUI '26), March 23--26, 2026, Paphos, Cyprus}
\acmPrice{}
\acmDOI{10.1145/3742413.3789127}
\acmISBN{979-8-4007-1984-4/2026/03}

\begin{document}

\title{Predicting Biased Human Decision-Making with Large Language Models in Conversational Settings}

\author{Stephen Pilli}
\email{stephen.pilli@ucdconnect.ie}
\orcid{0000-0003-1655-1782}
\affiliation{%
  \institution{University College Dublin}
  \city{Dublin}
  \country{Ireland}
}

\author{Vivek Nallur}
\orcid{0000-0003-0447-4150}
\email{vivek.nallur@ucd.ie}
\affiliation{%
  \institution{University College Dublin}
  \city{Dublin}
  \country{Ireland}
}

\renewcommand{\shortauthors}{Pilli \& Nallur}

\begin{abstract}

We examine whether large language models (LLMs) can predict biased decision-making in conversational settings, and whether their predictions capture not only human cognitive biases but also how those effects change under cognitive load.
In a pre-registered study (N = 1,648), participants completed six classic decision-making tasks via a chatbot with dialogues of varying complexity.
Participants exhibited two well-documented cognitive biases: the Framing Effect and the Status Quo Bias. Increased dialogue complexity resulted in participants reporting higher mental demand. This increase in cognitive load selectively, but significantly, increased the effect of the biases, demonstrating the load-bias interaction. 
We then evaluated whether LLMs (GPT-4, GPT-5, and open-source models) could predict individual decisions given demographic information and prior dialogue. While results were mixed across choice problems, LLM predictions that incorporated dialogue context were significantly more accurate in several key scenarios. Importantly, their predictions reproduced the same bias patterns and load-bias interactions observed in humans. Across all models tested, the GPT-4 family consistently aligned with human behavior, outperforming GPT-5 and open-source models in both predictive accuracy and fidelity to human-like bias patterns.
These findings advance our understanding of LLMs as tools for simulating human decision-making and inform the design of conversational agents that adapt to user biases.

\end{abstract}




\begin{teaserfigure}
\end{teaserfigure}


\maketitle

\section{Introduction}

Digital interfaces influence almost every aspect of modern life. They mediate interactions with loved ones, transport, medical checkups, entertainment, and sometimes even food and intimate choices. The function and form of these interfaces influence not only our view of the world, but also \textit{how} we influence the world. Our decisions on which action to take, which option to ignore, and what aspect of a problem to pay attention to are all affected by the contextual elements within which the decision scenario appears. In particular, conversational interfaces now act as decision mediators across multiple domains, shifting the structure, presentation, and interpretation of available alternatives.

In its simplest form, decision-making requires a choice problem and a set of alternatives to choose from. The structure and presentation of these alternatives can potentially shape decision-making by tapping into underlying cognitive biases. Cognitive biases are systematic deviations from rational judgment, arising from heuristics, prior experiences, emotions, or social factors~\citep{kahneman_thinking_2011}. Over 200 such biases have been systematically cataloged and experimentally validated through standardized cognitive tasks~\citep{big_bad_bias}.
While these biases have been studied extensively in static, survey-based, or GUI-mediated settings, relatively little is known about their manifestation in interactive, language-based environments like task-oriented dialogues. Early work suggests that biases persist even in conversational settings ~\citep{pilli2023exploring, yamamoto_suggestive_2024, ali_mehenni_nudges_2021}, yet we lack a systematic understanding of how the dynamics of dialogue shape bias susceptibility.

In decision theory, the process of decision-making requires the presence of an actor, denoted as the decision-maker, and a contextual environment within which the decision transpires~\citep{simon_behavioral_1955}. Although cognitive biases originate from internal heuristics, they can be influenced by external environmental factors, such as cognitive load~\citep{bogdanov2023working}. 
In conversational settings, such as task-oriented conversational agents, decision-making occurs within the context of a dialogue. In this setting, cognitive load arising from prior conversational context can potentially influence users' biased decision-making. We refer to such contextual influence as dialogue complexity. This complexity may act as a proxy for cognitive load, shaping the likelihood of biased decision-making. Leveraging the conversational context can help in accurately predicting when users are likely to be susceptible to cognitive biases. This opens the door to adaptive interventions that promote more informed, deliberate, and rational decisions.

Simulation and prediction of human behavior has long been a goal of research in Human-Computer Interaction (HCI), cognitive science, and behavioral modeling~\citep{hwang2025human}. Large language models (LLMs) have been fine-tuned to produce fluent, human-like dialogue and frequently achieve high performance on established benchmarks~\citep{Yi_multi}. This advancement presents opportunities beyond conventional applications, such as LLMs or generative agents, by enabling the simulation of large-scale human behavior in both experimental and policy contexts~\citep{park2023generative, park2024generative, argyle2023out, aher2023using, hamalainen2023evaluating}. Of particular interest is the question of whether LLMs can serve as predictive models of human judgment and decision-making. That is, not just in mimicking language patterns, but simulating how contextual and cognitive factors drive biased behavior.

Prior work has primarily examined whether LLMs themselves exhibit cognitive biases when prompted to make decisions~\citep{hwang2025human, malberg2024comprehensive, echterhoff_ai-moderated_2022}. While informative, this line of research focuses on the presence of biases within LLMs, rather than their ability to model or predict human-biased behavior. A notable exception is~\citet{ying2025benchmarking}, who used LLMs to simulate human decision-making but found substantial misalignment between model predictions and human rationality; however, cognitive biases were not the main focus of their work. ~\citet{park2024generative} introduced a method that enables behaviorally grounded predictions by constructing generative agents with rich, interview-derived memory representations. These agents have demonstrated strong predictive accuracy across surveys, personality assessments, and experimental tasks by simulating individual-level responses. These advances raise a key question: can LLMs simulate decision-making behavior that is not only human-like but also bias-sensitive and context-aware? Building on this approach, we investigate whether LLMs can simulate individuals given the chat transcripts such that the predictions on biased decision-making align accurately with those of their real-world counterparts.

To achieve this, we begin by investigating whether \textbf{cognitive biases manifest in conversational settings}, as prior research has primarily focused on isolated or survey-based tasks.
This leads to our first research question 
\textbf{(RQ1)}: \textit{Do established cognitive biases (Framing and Status Quo effects) manifest in conversational decision-making settings?} 
While this investigates the presence of biases, it does not account for how they may be shaped by the conversational context. To understand the \textbf{role of conversational context}, we next examine whether features like prior dialogue complexity systematically interact with the bias.
This motivates 
\textbf{(RQ2)}: \textit{How does prior dialogue complexity interact with cognitive bias susceptibility?}
Building on this, we ask whether LLMs can \textbf{predict human decisions} across such contexts using limited information, forming 
\textbf{(RQ3)}: \textit{Can LLMs predict individual human decisions using limited prior dialogue and demographic information?}
The accuracy of individual-level prediction does not reveal whether LLMs capture the presence of cognitive biases in the population. A model might consistently choose the biased alternative, inflating accuracy while misaligning with the true distribution of human responses. To investigate this, we evaluate whether LLMs can reproduce population-level bias effects and their interaction with dialogue complexity.
Therefore, we finally explore whether LLMs can \textbf{simulate collective behavior} by reproducing not only the presence of biases but also how they interact with dialogue complexity. 
This brings us to 
\textbf{(RQ4)}: \textit{Can LLMs reproduce both the presence of cognitive biases and their interactions with dialogue complexity at the population level (collective behavior)?}

This paper addresses these questions through two empirical studies. First, we conduct controlled human-subject experiments (N = 1,648) using six well-established choice problems adapted for conversational settings.
The choice problems are chosen to investigate prominent Framing and Status Quo effect cognitive biases. These cognitive biases are chosen as they are well-studied and replicated in the HCI, psychology, and behavioral economics literature. 
These studies systematically manipulate choice problems and prior dialogue complexity to examine their influence on Framing and Status Quo biased decision-making. The methods and findings are detailed in the Human Experiments section (~\ref{sec:humanexperiments}).
Building on these human results, we then evaluate multiple LLM families by prompting them to simulate human decision-making under identical conditions. Specifically, we assess LLMs' ability to predict both individual-level decisions and sample-level bias patterns using participant demographics and dialogue transcripts, along with ablation analyses. The methodology and results are presented in the LLM Experiments section (~\ref{sec:llmexperiments}).
Finally, the Discussion section (~\ref{sec:discussion}) outlines results and discusses the implications for bias-aware interaction design and LLM Simulation for HCI.



\section{Related Work}
\label{sec:Related Works}

We review the related works in the following two key areas: a) cognitive bias and load in conversational agents, and b) LLM behavioral modeling capabilities.

\subsection{Cognitive Biases in Decision-making Facilitated through Conversational Agents}
\label{sec:cognitive_bias_and_conversational_agents}
Cognitive biases in human decision-making have been examined closely in the fields of Psychology, Behavioral Economics, and Human-Computer Interaction. Recent research has examined their potential to both leverage and mitigate such biases~\citep{caraban_23_2019} using conversational agents. 
~\citet{ji_towards_2024} studied cognitive biases in spoken conversational search (SCS), highlighting biases like anchoring and confirmation bias in the absence of visual cues. Their framework is largely theoretical but sets the stage for future bias-mitigation strategies in voice-based systems.
~\citet{pilli2023exploring} used chatbots to assess cognitive biases like Framing effects and Loss aversion. Participants exhibited typical bias responses, confirming chatbots as valuable tools for bias detection and measurement.
Yamamoto introduced ``suggestive endings'' in chatbot dialogue, based on the Ovsiankina effect ~\citep{yamamoto_suggestive_2024}. This design prompted users to engage more deeply, ask follow-up questions, and reflect longer, enhancing cognitive engagement.
~\citet{dubiel_impact_2024} examined the role of synthetic voice fidelity in decision-making. They found that high-fidelity voices, through cues like pitch and pace, enhanced source credibility and triggered affect heuristics, subtly influencing user choices.
~\citet{ali_mehenni_nudges_2021} explored children's susceptibility to tasks resembling cognitive tasks used to infer cognitive biases by conversational agents and robots using a modified Dictator Game. Their findings revealed a stronger influence from artificial interlocutors than humans, pointing to authority and social influence biases, especially among vulnerable users.
~\citet{kalashnikova_linguistic_nodate} investigated linguistic nudges promoting ecological behavior. By leveraging biases such as Status Quo bias and social conformity, they showed that chatbots and robots were more persuasive than humans in shaping opinions.

\subsection{Language and Cognitive Load in Dialogue Systems}

Prior work has examined how cognitive load influences decision-making broadly (e.g., ~\citet{sweller1988cognitive, deck2015effect}).
~\citet{khare2021maximizers}, who explored how internal characteristics of the choice problem, such as information overload and choice overload, can impact Status quo bias. However, their focus is on the structure of the alternatives themselves. 
Experimental studies have examined the Framing effect under cognitive load and found supporting evidence for dual-process theory, suggesting that cognitive load increases the influence of framing on decision-making~\citep{bogdanov2023working, whitney2008framing} 
It has been demonstrated that cognitive load is inversely related to task performance facilitated by the chatbot. ~\citet{schmidhuber_cognitive_2021} explored how the use of a chatbot affects users' mental effort when interacting with a new software product, and also to what extent the use of a chatbot affects users' productivity. The results showed that chatbot users experienced less cognitive load.
Similarly, ~\citet{brachten2020ability} showed that chatbots can reduce the cognitive load needed to complete various tasks. One effective strategy to minimize cognitive load is to avoid presenting long responses or requiring users to provide complex inputs.
The cognitive load induced by the dialogue is dependent upon the dialogue design. By increasing the elements and interactions between the elements, mental demand increases in turn, leading to a higher cognitive load. A poorly designed dialogue can result in cognitive load.
These studies have explored various aspects of conversational agents, like linguistic features, the affect caused by dialogue voice modulation, the length of dialogue, and their role in influencing decision-making. However, these studies focus on individual decision points or choice problems only; they do not account for how previous interactions, conversational context, or prior dialogue influence biases in subsequent decision-making. 

\subsection{LLM Behavioral Modeling}
Current research establishes that LLMs can reproduce aggregate bias patterns ~\citep{hwang2025human, malberg2024comprehensive, echterhoff_ai-moderated_2022} but has not systematically investigated whether these models can predict individual decision-making, based on conversational cues and demographic information, particularly under varying cognitive load conditions.
While LLMs demonstrate human-like biases when prompted appropriately, their capacity for individualized behavioral prediction in conversational contexts remains largely unexplored. 
This gap motivates our investigation into whether LLMs can serve as predictive tools for biased human behavior in realistic conversational scenarios, moving beyond surface-level bias reproduction toward contextually sensitive individual behavioral modeling.

\section{Human Experiments}
\label{sec:humanexperiments}
In investigating the first two research questions, we design a dialogue that follows a real-world task-oriented dialogue structure and facilitates decision-making, yet ensures experimental control and ecological validity by standardizing dialogue content and controlling for confounding variables. 
A formal representation of our dialogue $\mathcal{D}$ with a choice problem and prior dialogue is as follows:
\[
\mathcal{D} = 
\textbf{\{}
u_1^{sys}, u_2^{usr}, \ldots
\underbrace{ u_{t-k}^{sys}, \ldots , u_{t-1}^{usr},}_{\text{Prior Dialogue}}
\underbrace{u_t^{sys},u_{t+1}^{usr} }_{\substack{\text{Decision} \\ \text{Scenario} \\ \text{and} \\ \text{Response}}} \ldots
\textbf{\}}
\]
\label{fig:dialogue}

This section outlines the choice problems adapted from classical behavioral economics studies, explains their integration into the chatbot with consistency and experimental control, describes the preceding Simple and Complex Dialogues, and details the experiment design and procedure.

\subsection{Choice Problems}
\label{sec:method-choice_problems}

The chatbot is designed to have introductory utterances like greetings 
$\{u_1^{sys},.., u_3^{sys},\ldots\}$$\in\mathcal{D}$
, which are followed by prior dialogue  
$\{u_{t-k}^{sys}, \ldots , u_{t-1}^{usr}\}$$\in\mathcal{D}$, 
and which is then followed by a choice problem 
$ u_t^{sys}$$\in\mathcal{D}$
.
A choice problem typically involves a decision-making problem accompanied by a set of alternatives from which participants must choose. In traditional experimental designs, a questionnaire format is used for a between-subjects experiment to investigate the biases. 
We adapt the same experiment design where a control group is presented with a version of the choice problem where alternatives are described neutrally, while the treatment groups encounter scenarios in which one option is explicitly framed. 
The effect of respective cognitive bias is determined by analyzing the statistical difference in participant responses ${ u_t^{usr} } \in \mathcal{D}$ between these groups.

Our experiments adapted six choice problems, three targeting classic Framing effects (Risky-choice Framing~\citep{tversky_judgment_1974}, Attribute Framing~\citep{kuang2023framing}, and Goal Framing~\citep{aravind2024nudging}) and three Status quo bias scenarios (Budget allocation, Investment decisions, and College job offers) which are drawn from ~\citet{samuelson_status_1988}. These problems are well-established in the literature and have been reproduced in recent replication studies \cite{bogdanov2023working, xiao_revisiting_2021}.

\subsubsection{Framing Choice Problems}
Our experiments used three choice problems, each representing a different type of framing effect:  \textit{Risky-choice}, \textit{Attribute}, and \textit{Goal framing}. 
The choice problem for Risky-choice framing effect was adapted from choice problems described in a replication study by ~\citet{bogdanov2023working}. The original study was performed by ~\citet{wang1996framing}. The choice problem follows the popular Asian Disease Problem by~\citet{tversky1981framing}. Participants choose between two plans (Plan A and Plan B). The participant's choices reveal the framing effect. 
The Attribute framing problem involved restaurant selection and was adapted from ~\citet{kuang2023framing}, where the same distance was described either in miles (Space) or minutes (Time), showing that people's preferences change depending on how the information is framed.  
The Goal framing choice problem was adapted from ~\citet{aravind2024nudging}, which tested how different goal-based cues influence public transit adoption. We selected the normative frame, highlighting environmental sustainability to encourage eco-conscious decisions. Participants choose between two travel modes (Public Transit and Personal Car) for a 10-mile trip, with or without any sustainability-related information as a framing cue. 
We refer to the framing effect-related choice problems for the control group as ``Framed'', and for the experimental group as ``Alternatively Framed.'' The complete set of framing choice problems is presented in
Table~\ref{tab:choice-problems} of Appendix ~\ref{appendix:Choice Problems}.

\subsubsection{Status Quo Choice Problems}
The experiment used three decision-making scenarios or choice problems adapted from: Budget allocation (BA), Investment decision making (IDM), and College job offers (CJ)~\citet{samuelson_status_1988}. These choice problems were selected due to their well-documented effects, serving as strong baselines for evaluation. These were additionally reproduced in the replication study by ~\citet{xiao_revisiting_2021}. Moreover, they represent domains that are both widely studied in behavioral economics and highly relevant to practical applications in chatbot-based e-commerce and decision support systems. 
Each choice problem was implemented in three conditions: a \textit{neutral condition}, where the alternatives were presented equally with no status quo option, and two \textit{Status Quo conditions}, where one of the alternatives was framed as status quo. The decision maker can move away from the status quo or stick to the status quo, which reveals the bias.
While adapting to conversational style, we have made minor modifications to the original choice problems. We refer to the Status Quo-related choice problem condition for the control as ``Neutral'' and experimental conditions as ``Status Quo A'' or ``Status Quo B'' based on the alternative in the status quo position. The list of choice problems for all conditions is detailed in Appendix~\ref{sec:appendix-ds}. We made minor modifications to the choice problems, which we report in Appendix ~\ref{appendix:method-sqb-manipulations}.

\subsection{Prior Dialogue}
\label{sec:Prior-dialogue}

A key feature of decision-making in a conversational setting is the presence of dialogue that precedes the decision scenario or a choice problem. 
We refer to this as the prior dialogue, denoted as $\{u_{t-k}^{sys}, \ldots , u_{t-1}^{usr}\} \in \mathcal{D}$, where $\mathcal{D}$ represents the full dialogue. To investigate our second research question \textbf{(RQ2)} that is the complexity of prior dialogue can potentially play a role in shaping subsequent decision-making we designed two types of preference elicitation tasks: one that facilitates low-effort interaction, referred to as the \textit{Simple Dialogue}, and another that is cognitively demanding, referred to as the \textit{Complex Dialogue}. The design characteristics of these tasks are as follows:

\subsubsection{Simple Dialogue}
A preference elicitation task was used, where participants were engaged in a set of short binary (Yes/No) questions about preferences within a familiar domain.
This dialogue design was inspired by the Schema-guided Dialogue (SGD) dataset introduced by~\citet{rastogi_towards_2020} and aimed to simulate a natural, ecologically valid, low-effort interaction with the chatbot. Importantly, the Simple Dialogue used a conservative dialogue strategy: questions were direct, unambiguous, and did not require reasoning or memory beyond the current turn. This ensured that the mental effort required by the dialogue was minimal. Participants were instructed to answer each question directly and were prompted to enter ``I don't know'' for \textit{prior dialogue attention check} on one attribute.
A full list of domains and associated questions can be found in the Section ~\ref{sec:appendix-simple_task-prior_dialogue} of Appendix ~\ref{sec:appendix-prior_dialogue}.
An example of the Simple Dialogue in the ``Music'' domain is shown below. 




\begin{table}[htpb]
\centering
\caption{Simple Dialogue Attributes and Respective Utterances}
\label{tab:simple_dialogue_example}
\begin{tabular}{@{}ll@{}}
\toprule
\multicolumn{1}{c}{\textbf{Attribute}} & \multicolumn{1}{c}{\textbf{Yes/No Question}} \\ \midrule
Genre Preference & Do you like listening to pop music? \\
Language of Lyrics & \begin{tabular}[c]{@{}l@{}}Do you prefer music with lyrics in \\ English?\end{tabular} \\
Live Performances & \begin{tabular}[c]{@{}l@{}}Are you interested in live music \\ performances?\end{tabular} \\
Instruments Focused & Do you enjoy instrumental music? \\
Artist-Specific & \begin{tabular}[c]{@{}l@{}}Do you like music from specific artists? \\ Please enter ``I don't know'' only.\end{tabular} \\
Era (e.g., 80s, 90s) & Do you prefer music from the 90s? \\ \bottomrule
\end{tabular}%
\end{table}

\subsubsection{Complex Dialogue}

The Complex Dialogue was designed to induce cognitive load in a controlled yet ecologically valid manner. To achieve this, a nested referential structure is used to manage multiple interdependent entities across a multi-turn dialogue.
The following shows an example of the utterances from $u^{sys}_{t-k}$ to $u^{sys}_{t-2}$ by the chatbot during the prior dialogue.
\begin{enumerate}
    \item[$u^{sys}_{t-8}$:] The first artist performs three live shows, is paid 2000 units per show, and has a 4-star rating. The second artist performs twice as many shows, with the same pay and rating. Which artist do you prefer, and why?
    \item[$u^{sys}_{t-6}$:] The third artist performs the same number of shows as the second, earns half the pay of the first artist, but has the same rating as the first. Which artist do you prefer, and why?
    \item[$u^{sys}_{t-4}$:] The fourth artist performs the same number of shows as the second, earns the same pay as the third, but has two stars less than the first artist. Which artist do you prefer, and why?
    \item [$u^{sys}_{t-2}$:] Remember the details of the fourth artist. Specific information will be requested later.
\end{enumerate}

The fourth artist in the chat transcript is defined by the second and third artists, which reference the second and first artists, creating a chain of dependencies. This design necessitates users to maintain and integrate hierarchical relationships between entities, thereby increasing semantic integration costs and working memory demands. This was designed based on the psycholinguistic findings that nested dependencies elevate processing difficulty
~\citep{gibson1998linguistic, van1983strategies}.
Additionally, the reappearance of earlier referents after intervening turns results in high referential distance~\citep{chen1985discourse}, which further taxes memory retrieval processes~\citep{arnold1998reference}.
From a cognitive load design standpoint, this structure directly aligns with Sweller's Cognitive Load Theory (CLT), which distinguishes between intrinsic load (task-related complexity), extraneous load (inefficient information presentation), and germane load (effort used for schema building)~\citep{sweller1988cognitive, deck2015effect}.

Our design increases intrinsic load by requiring participants to track and integrate multiple interrelated referents, and germane load by promoting mental model construction to resolve semantic dependencies across turns. 
From a dialogue systems perspective, managing multiple entities simultaneously while maintaining coherent context is a well-documented challenge, particularly when entities are interconnected or revisited~\citep{ultes2020complexity, 
el-asri-etal-2017-frames}. This reflects real-world conversational demands where dialogue agents must track, differentiate, and link multiple referents simultaneously. 
Our Complex Dialogue task is a preference elicitation task that necessitates arithmetic comparisons between attributes and memorization of outcomes, requiring additional mental effort.
Similarly designed dialogues are used in other domains, including artist recommendations, streaming services, calendar apps, and banking options.
(For the remaining complex dialogues, please refer to Table ~\ref{tab:appendix-complex-task} in Section ~\ref{sec:appendix-complex_task-prior_dialogue} of Appendix ~\ref{sec:appendix-prior_dialogue}).

The domains of these tasks are adopted from the Schema-Guided Dialogue (SGD) dataset ~\citep{rastogi_towards_2020}. 
In theory, the task design must substantially increase the cognitive load of the individual. To empirically verify the cognitive load, the standard NASA-TLX~\citep{pandian_nasa-tlx_2020} survey was adopted for all the interactions. 
To evaluate the cognitive load of the prior dialogue, we incorporated two indicators: the self-reporting \textit{NASA-TLX} and a \textit{Recall Task}. The NASA-TLX was recorded after the chatbot interactions to capture participants' perceived cognitive load across multiple dimensions, such as mental demand and effort. Simultaneously, the Recall Task served as a behavioral indicator of attention and memory. Participants were asked to recall memorized arithmetic from the conversation.
Additionally, we also investigate the \textit{familiarity} of the participants with prior dialogue domains as a confounding factor.

\subsection{Chatbot Technical Details}

The web-based experimental interface was developed using Streamlit~\cite{streamlit}, which allowed for easy deployment and consistent access across devices. 
The source code for the chatbots used in the experiments is publicly available.
The chatbot used in the framing experiment can be found at
\url{https://github.com/stephen-pilli/PEM.git},
while the chatbot used in the status quo bias (SQB) experiment is available at
\url{https://github.com/stephen-pilli/exp-status-quo-bias.git}.

GPT-4o Mini, a large language model ~\citep{openai2024chatgpt4o}, was used to create realistic and coherent chatbot interactions based on structured prompts designed for each condition. An example of the prompt used for the agent (LLM chatbot) is provided in Appendix ~\ref{sec:appendix-prompt}.

\subsection{Design of the Experiments}

We employed a two-factor between-subjects design. The first factor is the prior dialogue complexity, which manipulated the cognitive load experienced by participants before making a decision. The second factor is the choice problem condition (Recall the conditions for Framing and Status quo detailed in Section ~\ref{sec:method-choice_problems}).

For the Framing experiments, we employed a 2 × 2 experiment design. The first factor was dialogue complexity with two levels (Simple vs. Complex), and the second factor was framing with two levels (Framed vs. Alternatively Framed). Participants were randomly assigned to one of the four conditions, ensuring balanced group sizes. 
For the Status quo experiments, we employed a 2 × 3 factorial design. Dialogue complexity again had two levels (Simple vs. Complex), while the Status Quo factor had three levels (Neutral, Status Quo A, Status Quo B). Participants were randomly assigned to one of the six resulting conditions, with balanced distribution across groups. 
To preserve internal validity and avoid carryover effects, each participant encountered only one version of the dialogue and one framing condition.

The primary dependent variable in our study is participants' choices between alternatives in the choice problems. All six choice problems were included in the experiment. Decision outcomes were compared across bias conditions to address our research questions.
Our second key variable is a \textit{moderator}: the cognitive load required to complete the task. We measured cognitive load using the NASA-TLX questionnaire as well as behavioral indicators. This variable allowed us to statistically test whether complex dialogues place greater mental demands on participants compared to simple dialogues.
To avoid confounding effects, the domains of the prior dialogues and the choice problems were intentionally different. This separation ensured that cognitive load was isolated from other factors, such as domain familiarity, and allowed us to focus on how dialogue complexity influenced subsequent decision-making.

While our dialogue tasks were adapted from classic behavioral economics experiments, we carefully designed them to preserve ecological validity by embedding the decision-making within naturalistic, task-oriented chatbot interactions. This ensured that participants experienced the scenarios as they would in a real conversational setting with an intelligent agent, rather than as isolated survey questions. At the same time, we maintained experimental control by standardizing dialogue length, turn-taking, and framing conditions across participants. This allowed us to capture bias-prone decision-making in a realistic human–agent interaction context, while still ensuring internal validity and replicability of the results.

\subsection{Power Analysis, Recruitment, and Data Integrity}
We conducted an \textit{a priori} power analysis using G*Power \citep{faul2009statistical}, targeting 0.80 power to detect a medium effect size ($\omega = 0.3$, $\alpha = 0.05$) following \citet{pancholi2009}. This required approximately 42 participants per condition (see Appendix~\ref{sec:appendix-power-analysis}). Participants (\textit{N}=1648) were recruited via Prolific \citep{prolific2024}, compensated at \$8/hour, and randomly assigned to 2x2 (Framing) or 2×3 (Status Quo) designs. 
The studies were preregistered on the Open Science Framework (OSF). The preregistration for Framing study is archived at \url{https://doi.org/10.17605/OSF.IO/DPR45}, and the preregistration Status quo study is archived at \url{https://doi.org/10.17605/OSF.IO/PSXVF}.
Data integrity was ensured through attention checks, recall tasks, and automated JSON-based logging. The dataset for Framing effect study is available at \url{https://doi.org/10.5281/zenodo.18218753}, and the Status quo bias study is available at \url{https://doi.org/10.5281/zenodo.16541481}.

\subsection{Procedure}

Participants began the study by reviewing an information sheet outlining the study's purpose, procedures, and ethical considerations. 
After reading the document, they were directed to the experiment's homepage, where their Prolific ID (Stored in an irreversible, anonymous state) was displayed alongside the consent form.

\begin{figure*}[htpb]
    \centering
    \includegraphics[width=1\linewidth]{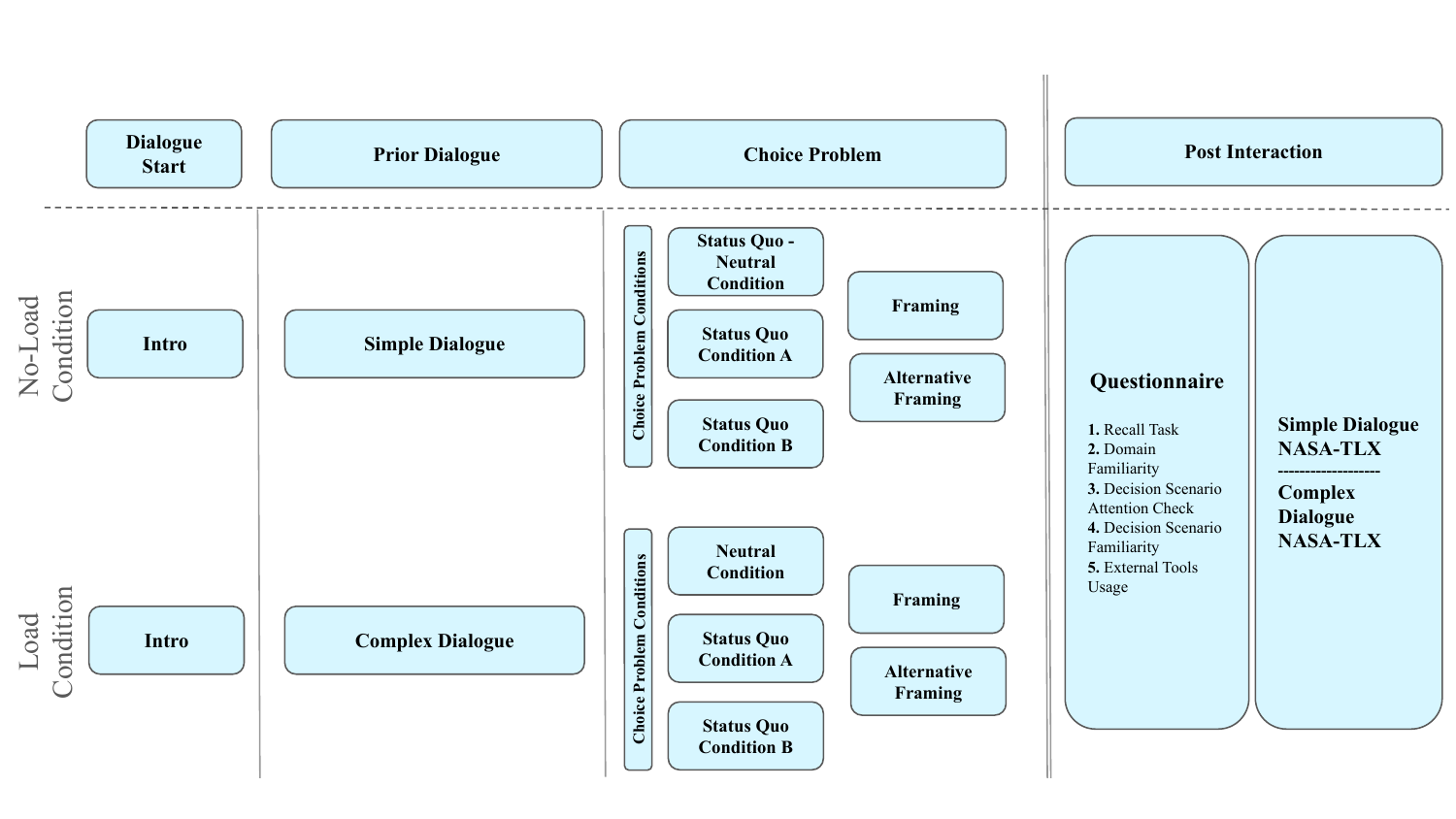}
    \caption{Experimental procedure outlining the sequence of tasks. Participants were first introduced to the dialogue (Simple or Complex), followed by the assigned condition (Framing: Framed vs. Alternative Framing; Status Quo: Neutral, A, or B). After completing the choice problem, participants filled out the NASA-TLX and post-task questionnaires (recall, familiarity, attention checks, and tool usage).}
    \label{fig:procedure}
    \Description{The figure presents a flowchart of the experimental procedure, illustrating parallel sequences for No-Load and Load conditions. In both conditions, participants begin with an introduction, then proceed to a dialogue task—either a Simple Dialogue in the No-Load condition or a Complex Dialogue in the Load condition. Following the dialogue, participants encounter a choice problem where they are randomly assigned to one of several sub-conditions (Neutral or Status Quo A/B) and then exposed to either a standard or alternative framing scenario. After making their choice, all participants complete a post-task questionnaire covering recall, domain familiarity, attention checks, scenario familiarity, and external tool usage, as well as the NASA-TLX workload assessment corresponding to their dialogue type. The flowchart depicts these steps and their organization across experimental conditions.}
\end{figure*}

To ensure active participation, the consent checkbox was initially unchecked, requiring participants to explicitly select ``Yes'' before proceeding. Only after providing informed consent were they granted access to the experiment.
Participants interacted with the chatbot based on the condition they are randomly assigned to as shown in the Figure ~\ref{fig:procedure}. After completing the dialogue with the chatbot, participants were redirected to a questionnaire containing additional measures to ensure data quality. The survey included a memory recall task to assess attentiveness, along with attention check questions to verify engagement. By implementing these steps, the study ensured that participants remained actively engaged and provided high-quality responses, ultimately enhancing the reliability of the collected data. 
After each task, participants reviewed their transcript and then completed the NASA-TLX survey. In both the Simple and Complex Dialogue conditions, participants rated their perceived mental demand. This design minimized recall bias and enabled a precise assessment of the mental demands imposed by dialogue complexity.

\subsection{Findings}

\subsubsection{Dataset Description}

For the Framing study, the initial sample of 595 participants was filtered based on familiarity ratings, attention checks, prior exposure to the choice problem, invalid responses, and missing NASA-TLX data, resulting in a final sample of 548. The mean age was 43 years (SD = 13.7). The dataset is a slightly female-skewed sample (53.5\%) and a majority identifying as White (86.5\%). Most participants were not students (72.1\%) and were employed either full-time (46.5\%) or part-time (18.1\%). Detailed demographic distributions across experimental conditions are presented in Table~\ref{tab:framing_demographics} in Appendix~\ref{appendix:demographics}.

For the Status quo study, an initial sample of 1,256 participants underwent data cleaning, which excluded 19 for invalid responses, 40 for failing scenario recall, 49 for prior familiarity, and 63 for using external tools, resulting in a final sample of 1,100 participants. 
The mean age was 41.5 years (SD = 13.3), with an even gender distribution (50.5\% female). Most participants were residents of the United Kingdom (n = 778), followed by the United States (n = 308) and Ireland (n = 14). Full demographic breakdown for the Status quo experiment is provided in Table~\ref{tab:sqb_demographics} in Appendix~\ref{appendix:demographics}.

\input{tables/human_exp_summary}

\subsubsection{Framing and Status Quo effects reproduced in Conversational Setting.}

Table~\ref{tab:human_experiments_results_summary} presents evidence across six choice problems that addresses \textit{\textbf{RQ1}} and \textit{\textbf{RQ2}}. \textbf{\textit{RQ1}} investigates whether Framing and Status quo effects can be reproduced in a conversational setting; several strong and moderate effects were observed. Risky-choice framing showed a strong effect under complex dialogue but only a weak effect under simple dialogue, which is comparable to the original study by ~\citet{wang1996framing}. Attribute framing showed a robust effect in the literature~\citep{kuang2023framing}; however, the same did not replicate strongly in this conversational setting; only weak evidence appeared under simple dialogue, and no effect under complex dialogue. Goal framing produced the clearest result, with a strong effect both in the original study and under complex dialogue, and a weak effect in simple dialogue. 
The Status quo scenarios (budget allocation, investment decisions, and college jobs) showed mixed evidence: budget allocation and college jobs replicated strongly under both dialogue types, while the investment decision scenario showed no effect. Overall, these results address \textbf{\textit{RQ1}}, demonstrating that several well-documented Framing and Status Quo effects persist in conversational settings, though their strength varies across the choice problems.

\subsubsection{Prior Dialogue Complexity Resulted in Cognitive Load.}

Across both the Framing and Status quo experiments, NASA-TLX results consistently showed that Complex dialogues resulted in significantly higher cognitive load than Simple dialogues. Mental Demand increased most strongly ($d = 0.85-1.08, p < .001$), followed by Effort ($d = 0.6 - 0.77, p < .001$), with smaller but reliable increases for Performance, Frustration, and Temporal Demand, while Physical Demand showed minimal effects. Behavioral indicators supported these findings: participants in the Complex condition took longer to respond and demonstrated higher accuracy on the memory recall task, both correlating positively with self-reported Mental Demand. Together, these converging results confirm that complex prior dialogue substantially increased cognitive load, validating our manipulation across both experimental studies.
Behavioral indicators further validated our cognitive load manipulation. In the framing study, accuracy correlated positively with Mental Demand ($r = 0.13, p = 0.002$), showing that participants who remembered task details also reported higher workload. In the Status Quo study, recall accuracy correlated with both response time ($r = 0.318, p < .001$) and Mental Demand ($r = 0.105, p = .013$), while participants in the Complex Dialogue took significantly longer than in the Simple Dialogue ($d = 0.59, p < .001$). Together, these results demonstrate that complex prior dialogues consistently increased cognitive load. For detailed exposition, please refer to the Appendix Section ~\ref{appendix-perceived-cl}.

\subsubsection{Interaction between Complex dialogue and biased decision-making.}

Our second research question investigated whether prior dialogue complexity interacts with subsequent decision-making; the results suggest selective but meaningful interactions. In Risky-choice framing and Goal framing, effect sizes were significantly larger following complex dialogue than simple dialogue, with confidence intervals indicating strong interactions. This implies that complex prior dialogue increases the susceptibility to these effects, a pattern consistent with findings from randomized control trials in psychology~\citep{whitney2008framing, bogdanov2023working}.
In contrast, Attribute framing showed a negative interaction, but it is not significant. Budget allocation, investment, and college jobs showed no interaction between dialogue complexity and decision outcomes. The effects either remained stable across the choice problem and prior dialogue. 
Taken together, these findings suggest that prior dialogue complexity can strengthen the Framing effect but does not affect Status Quo bias, addressing our Research Question (\textbf{\textit{RQ2}}).


\section{LLM Experiments}
\label{sec:llmexperiments}
To evaluate whether large language models (LLMs) can reproduce observed biased human decision-making patterns by predicting at the individual level, we replicated the experimental method used with human participants with LLMs.
Each LLM was prompted using two key inputs: (a) the demographic attributes available from Prolific (e.g., age, gender, education, and country of residence) and (b) the transcript of the participant's dialogue up to the choice problem or choice problem; this chat includes the prior dialogue, as shown in the Chat Transcript~\ref{DialogueTranscripts}. LLMs presented with the same choice problems as their corresponding human participants and were asked to act as participants and predict the decision the participant would make. This design ensured that the information available to the LLMs mirrored the information grounding human decisions.

\subsection{Human-Likeness Prompts}

\input{others/chat_transcript}

A central design choice in the LLM experiments was how to instruct LLMs to simulate human participants. Following recent work on LLM behavioral prompting ~\cite{binz2023turning,liu2025rationality, ying2025benchmarking}, we adopted a series of human-likeness prompts that varied in the degree to which the LLM was instructed to emulate human reasoning. Prior research suggests that LLMs may default to more rational or normatively consistent behavior than humans ~\citep{liu2025rationality}. 
The extent of instruction required for an LLM to align with actual human behavioral patterns is a latent variable that requires systematic investigation. To explore this, we varied the level of human-likeness in the prompts, ranging from minimal role prompts to explicit directives to exhibit susceptibility to cognitive biases.

Specifically, we implemented three levels of human-likeness. At human-likeness Level 1 (HL1), LLM received only a minimal role instruction: ``You are a participant in a research study.'' This established a research setting without explicit guidance on how to respond, allowing us to assess the LLM's baseline behavior. 
At human-likeness Level 2 (HL2), LLMs were encouraged to simulate more naturalistic human responses with the prompt: ``You are a human participant in a research study. Please answer questions as naturally as you would in everyday life.'' This formulation aimed to elicit more ecologically valid answers while avoiding explicit mention of biases. 
At human-likeness Level 3 (HL3), we explicitly instructed the LLM to act as humans prone to cognitive biases: ``You are a human participant in a research study. Therefore, act as a human. Be highly susceptible to cognitive biases such as Framing, Status Quo bias, and Anchoring when reasoning and answering questions. Avoid overthinking and lean into intuitive, sometimes irrational judgments.'' This highest level of human-likeness was designed to test whether LLMs could be guided to reproduce not just biases but also the susceptibility to cognitive load.

\subsection{Technical Details}
All LLM simulations were conducted using a combination of proprietary and open-source models. Proprietary models (GPT-4.1, GPT-4.1-mini, GPT-5, and GPT-5-mini) were accessed via the OpenAI API using the \url{chat.completions.endpoint}. Open-source models (gpt-oss-120b, llama4, and qwen3) were run on Google Cloud. To ensure reproducibility, we employed batch mode execution, fixed the random seed to 42, and set the temperature parameter to 0 to enforce deterministic outputs. For OpenAI models, we additionally logged the system fingerprint returned by the API to track model versions. Annotation of the output was conducted separately using three different LLMs: GPT-4.1, GPT-4.1-mini, and GPT-5-mini. Inter-rater Agreement (IRA) was calculated to choose the annotation for analysis.

\subsection{Findings}

\subsubsection{LLM predictions of biased human decision-making}
\label{sec pred}

To address \textbf{\textit{RQ3}}, can LLMs predict individual human decisions using limited prior dialogue and demographic information? We evaluated how accurately LLMs predicted participants' choices and how prior dialogue contributed to these predictions.
To understand the contribution of prior dialogue to LLM prediction performance, we compared accuracy across three conditions: \textit{Choice Problem Only}, \textit{Without Prior Dialogue}, and \textit{With Prior Dialogue}. In the \textit{Choice Problem Only} condition, the LLM was provided with just the text of the decision-making scenario (e.g., a Framing or Status Quo choice task), without any additional context such as demographics or prior dialogue. This serves as a baseline condition and assesses whether the model can predict the participant's choice based solely on the choice problem itself. 
In the \textit{Without Prior Dialogue} condition, the model was provided with demographic information (e.g., age, gender, education, and country of residence) along with a human-likeness prompt instructing the model to respond in a human-like manner. However, the prior dialogue was withheld. This allows us to isolate the effect of demographics and role-prompting on prediction performance. 
Finally, in the \textit{With Prior Dialogue} condition, the model received all available contextual input, including demographics, human-likeness prompt, and the full transcript of the dialogue prior to the choice problem. 
This condition enables us to evaluate whether the LLM uses prior dialogue for prediction.

\input{tables/ind_pred_short}

LLM prediction accuracy varied substantially across choice problems and dialogue conditions. Three distinct patterns emerged:

\begin{enumerate}
    \item \textbf{\textit{Case 1.}} No Dialogue Effect: For some problems (e.g., Risky Choice and Attribute Framing), including prior dialogue or demographic prompts did not significantly change prediction accuracy, suggesting that dialogue context added little predictive value.
    \item \textbf{\textit{Case 2.}} Dialogue-Enhanced Prediction: In other problems (notably Goal Framing and Investment Decisions), accuracy improved markedly when prior dialogue was included. For instance, in the Goal Framing, accuracy increased from 47\% (\textit{Without Prior Dialogue}) to 63\% (\textit{With Prior Dialogue}) decisions. Similarly, for Investment decision-making, accuracy increased from 62\% to 76\% as shown in Table ~\ref{tab:llm_accuracy_summary}. This shows that conversational context can align LLM predictions more closely with actual human decision-making. 
    \item \textbf{\textit{Case 3. }} Consistently High Accuracy: Some tasks (e.g., Budget Allocation and College Jobs, both involving Status Quo bias) achieved high accuracy across all conditions, indicating stable human preferences that LLMs could capture even without dialogue context.
\end{enumerate}

These results address \textbf{\textit{RQ3}} by demonstrating that LLMs can predict individual human decisions more accurately when provided with conversational context in addition to demographic information. However, the extent of this improvement varies depending on the type of choice problem.
The complete set of results and detailed analysis is provided in the Table \ref{tab:llm_predictions} of Appendix \ref{Appendix-indi_preds}.

\subsubsection{Biases observed in human experiments reproduced at the sample level.}

\input{tables/h3_h1}

Table~\ref{tab:h3_h1} presents a comparison of human and LLM behavior across six choice problems under varying levels of human-likeness. Various choice problems demonstrated clear evidence of bias in both Simple and Complex dialogue for human experiments. Investment Decision Making (IDM) showed no effect (original study had marginal significance ~\cite{samuelson_status_1988}, while Attribute Framing (ATF) showed no significant effect in either condition. Findings in the original study of ATF often yield weak or inconsistent biases showing small effect sizes.)

\input{tables/accuracy}

LLMs displayed varying levels of bias across the three human-likeness conditions. When explicitly instructed to behave in a biased manner (Human-Likeness 3: ``You are a human participant in a research study. Therefore, act as a human. Be highly susceptible to cognitive biases such as framing, status quo bias, anchoring''), LLMs showed strong bias across all choice problems, including Attribute Framing (ATF) and Investment Decision Making (IDM), where no bias was observed in the human experiments. This resulted in false positives, particularly in HL3 (shown in Table \ref{tab:confusion_matrices}), revealing a forced, biased behavior when asked. Consequently, the accuracy (distinct from individual-level prediction accuracy used in Section~\ref{sec pred} to report individual-level prediction), which is the proportion of choice problems in LLM experiments that matched actual human experiments, was only 58\% in HL3.
In contrast, when LLMs were given a more neutral prompt (``You are a participant in a research study''), accuracy improved to 75\%. Similar accuracy was achieved in Human-Likeness 2, where agents were instructed: ``You are a human participant in a research study. Please answer questions as naturally as you would in everyday life.'' 
Unlike HL3, this prompt avoided explicitly referencing cognitive biases. 
Under HL1 \& HL2, GPT4.1 correctly reproduced the absence of bias in both attribute Framing and the Investment Status Quo choice problem, closely aligning with human behavior and reducing false positives. 

\subsubsection{LLMs reproduce observed biased human behavior under complex prior dialogue.}

We conducted independent t-tests comparing effect sizes between Simple and Complex dialogue conditions. 
In the human data, Risky Choice Framing (RCF) and Goal Framing (GF) bias effects were significantly stronger after complex dialogues, consistent with prior research in cognitive psychology that links increased mental load with greater reliance on intuitive or biased decision-making. However, in the Status Quo bias scenarios (e.g., Budget Allocation and College Jobs), although bias was present in both conditions, the effect sizes remained relatively stable between Simple and Complex dialogues, indicating little or no interaction with cognitive load.

\input{tables/h3_h2}

To investigate whether LLMs could capture this interaction pattern (\textbf{\textit{RQ4}}), we examined the direction and magnitude of effect size changes across dialogue conditions using z-scores.
For example, in GF, the effect size for humans increased from 0.225 (Simple) to 0.567 (Complex) (as shown in Table ~\ref{tab:h3_h1}), resulting in a positive z-score of 2.29 (as shown in Table ~\ref{tab:h3_h2}), indicating a stronger bias under cognitive load.  
However, in HL1, the LLM's effect size decreased from 2.29 to 1.976 between conditions, resulting in a negative z-score of -1.61, meaning that the LLM behaved in the opposite direction. Interestingly, HL3 showed a positive z-score of 2.66, indicating that under cognitive load, LLM's responses were more biased, similar to human responses. Similar trends were observed for complementary cases like Attribute Framing (ATF) and Investment, where humans showed a negative direction (less bias under complexity), which was only mirrored correctly by HL3 but not HL1 or HL2.

To investigate the direction and magnitude statistically, we calculated Spearman correlation $\rho$ between the human experiment z-scores and those of each LLM human-likeness condition. The results revealed a marginally significant positive correlation between Human and HL3 ($\rho$ = 0.771, $p$ = .07). 
In contrast, HL1 and HL2 showed weak correlations ($\rho$ = 0.600), indicating a poor match with human behavior in terms of representing how dialogue complexity 
interacts with decision-making.

Overall, our findings show that LLMs can reproduce sample-level human biases such as Framing and Status Quo Bias, especially under neutral prompting conditions (HL1 and HL2), achieving up to 75\% alignment with human responses. However, under HL3, where models were explicitly told to simulate bias, they overestimated effects, leading to false positives. LLMs struggled to reproduce load-bias interactions, such as the impact of cognitive load, unless explicitly prompted, like in HL3.

\subsection{Analysis Across Models}

Sample-level accuracy shows how often the LLMs correctly reproduce human biases. A higher accuracy means the LLM accurately reproduced biased behavior in our human experiment, while lower accuracy indicates the LLM tends to exhibit bias where humans do not exhibit, or vice versa. 
Z-score captures the change in the effect size of a bias under complex prior dialogue.
Spearman correlation uses the Z-scores to test the monotonic relation between the human experiments and the LLM.
A strong positive correlation suggests the LLM reproduced the changes in effect size of biases, showing the sensitivity to cognitive load, whereas weak or negative correlations indicate a misalignment with human biased decision-making patterns under load, as observed in our experiments.

\input{tables/accuracy_and_corel}

Table~\ref{tab:accuracy_correlation_all_models} summarizes the accuracy and correlation scores for each model across the three human-likeness levels.
Across all models, GPT-4.1 consistently showed the best performance. Its accuracy was moderately high for both H1 and H2 (0.75), and still reasonable for H3 (0.58). It also showed a marginally significant positive correlation with human data in H3 ($\rho$ = 0.771, $p$ < .10), suggesting it could reproduce both the presence of biases and their change under cognitive load. GPT-4.1-mini had slightly higher accuracy (0.833 for H1, 0.750 for H2, and 0.667 for H3), but it did not show meaningful correlations, limiting its interpretability. In contrast, the GPT-5 family performed poorly. GPT-5-mini had moderate accuracy (0.580–0.670 across H1–H3), but its correlations were near zero or negative, meaning it often missed the direction of change in bias. GPT-5 had the weakest results, with low accuracy (0.333–0.670) and a marginally significant negative correlation in H2 ($\rho$ = –0.771, $p$ < .10), indicating it often predicted the opposite of human behavior under cognitive load. Among open-source models, performance was mixed but generally weaker. GPT-OSS-120b showed moderate accuracy (0.583–0.667) but failed to capture bias–load interactions. LLaMA4 had lower accuracy (0.417–0.583) and consistently negative correlations (–0.029 to –0.714), suggesting strong divergence from human-like behavior. Qwen3 performed inconsistently, with decent accuracy in HL2 (0.667) but weak results elsewhere. Overall, GPT-4.1 was the most aligned with human behavior, especially under cognitive load. The GPT-5 family often misrepresented bias patterns, while open-source models showed limited ability to simulate human-like decision-making, particularly in dynamic or context-sensitive settings.

\subsection{Ablation \& Perturbation}

To better understand which components of our experimental setup contributed to the LLM's ability to reproduce human-like decision-making behavior, we conducted a series of ablation studies. These ablations systematically removed or isolated different parts of the input, such as demographics, human-likeness prompts, prior dialogue components (arithmetic and memory), to identify what elements were essential for reproduction of bias and interaction behavior in a complex dialogue setting.

\input{tables/ablation}

First, we removed demographic information from the inputs. The results remained mostly the same across all human-likeness levels. Accuracy stayed at 0.750 for HL1 and HL2, and HL3 showed slightly lower accuracy (0.500), similar correlation with human data ($\rho$ = 0.771, $p$ = .07), suggesting that demographics are not critical. 
Next, we tested a minimal setup with only the choice problem, no demographics, no dialogue, and no prompt. In this case, GPT-4.1 still showed biased choices, but the correlation with human patterns dropped ($\rho$ = 0.257), indicating the model was biased but not in the same way as humans. We then tested only the memory component of the prior dialogue, where participants were asked to remember specific details. This condition improved alignment significantly, especially for GPT-4.1-mini under HL2 (accuracy = 0.833, $\rho$ = 0.771, $p$ = .07), showing that memory plays an important role in load-bias interaction. In contrast, when we kept only the arithmetic comparison component and removed memory cues, the models showed high accuracy but no meaningful correlation with human behavior (e.g., GPT-4.1: $\rho$ = –0.086, $p$ = .87). This suggests that arithmetic reasoning alone is not enough to model bias interaction. 
Overall, these findings show that while LLMs can reproduce basic bias effects, modeling human-like responses under cognitive load requires contextual elements, especially memory cues, in the dialogue.

\begin{figure*}[htpb]
    \centering
    \includegraphics[width=1.0\linewidth]{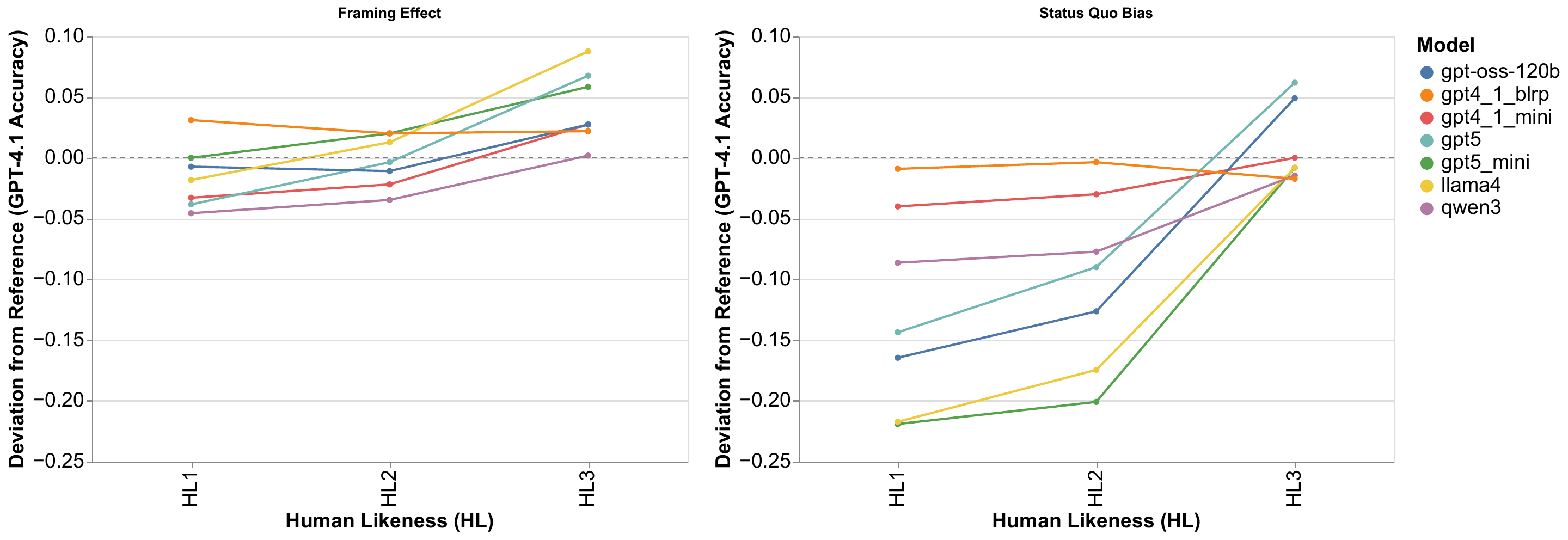}
    \caption{Deviation in accuracy from GPT-4.1 reference model across Human-Likeness (HL1, HL2, HL3) levels for each LLM. Positive values indicate higher accuracy than GPT-4.1; negative values indicate lower accuracy. gpt4\_1\_blrp stands for GPT-4.1 baseline experiment with human response perturbation.}
    \label{fig:perturbation_accuracy_deviation}
    \Description{The figure consists of two side-by-side line charts comparing the deviation in accuracy of several language models from the GPT-4.1 reference model across three levels of Human-Likeness (HL1, HL2, HL3). The left chart, titled "Framing Effect," and the right chart, titled "Status Quo Bias," both plot "Deviation from Reference (GPT-4.1 Accuracy)" on the y-axis (ranging from -0.25 to 0.10) against Human-Likeness levels on the x-axis. Each line represents a different language model (gpt-oss-120b, gpt4_1_blrp, gpt4_1_mini, gpt5, gpt5_mini, llama4, qwen3), color-coded and shown in a legend to the right. Positive y-values indicate higher accuracy compared to GPT-4.1, while negative values indicate lower accuracy. The figure shows how each model's performance, relative to GPT-4.1, changes with increasing Human-Likeness and under different experimental conditions.}
\end{figure*}

To further examine whether LLMs rely on human responses in the prior dialogue when predicting individual human choices, we conducted a human response perturbation analysis. Specifically, we replaced the human responses in the chat transcripts (Section~\ref{DialogueTranscripts}) with randomly generated text, while keeping all other aspects of the prompt unchanged.

We observed that the accuracy of the LLMs' predictions changed in fractions.  The Figure~\ref{fig:perturbation_accuracy_deviation} shows the deviation of individual level accuracy taking GPT4.1 accuracy as a reference. In the Figure ~\ref{fig:perturbation_accuracy_deviation}, \textit{gpt4\_1\_blrp} stands for GPT-4.1 baseline
experiment with human response perturbation. For both Framing and Status quo, this trend line is closer to zero. 
This suggests that the models are primarily leveraging the structure of the prior dialogue, rather than fine-grained cues from human responses, to make their predictions.

\section{Discussion}
\label{sec:discussion}

Our study investigated whether large language models (LLMs) can predict biased human decision-making in conversational settings.  
Addressing \textbf{RQ1}, biases such as the Framing and Status Quo effects were reproduced in human-conversational agent interactions, and their effects varied across choice problems. 
This extends inferring cognitive bias using simple cognitive tasks in traditional experiments into the conversational setting. This in-turn acting as a baseline for subsequent investigation.

\textbf{RQ2} explored the role of complex prior dialogue. We found that increased dialogue complexity selectively increased susceptibility to the Framing effect, aligning with prior findings on working memory capacity and Framing~\citep{bogdanov2023working, whitney2008framing}. However, dialogue complexity did not affect Status Quo bias. Empirical work linking cognitive load to Status Quo bias is scarce; the bias is more commonly attributed to irrational emotional attachment~\citep{masatlioglu_rational_2005}. Although choice overload has been shown to favor the Status Quo effect under cognitive load~\citep{eidelman_bias_2012}, we suspect that the simplicity of our binary choice task (e.g., College A vs. College B) limited such effects. This interpretation remains speculative and warrants further investigation using more sophisticated designs. Overall, our findings help address a gap identified in prior work (Section~\ref{sec:cognitive_bias_and_conversational_agents}), where cognitive biases were largely studied in isolation, neglecting the role of conversational cognitive load, and suggest an empirical framework adaptable to other biases.


\textbf{RQ3} investigated if simulated responses of LLM-agents can predict individual human decisions conditioned on limited prior dialogue and demographic information. The results are mixed; LLMs predicted individual human choices more accurately when conversational context accompanied demographic information. However, the degree of improvement varied by task and was highly dependent on the choice problem. In addition, ablation and perturbation analyses revealed that human-like responses under cognitive load depend on contextual cues, especially memory elements, rather than human utterances in dialogue.


While RQ3 investigated how closely the simulated responses aligned with human counterparts, \textbf{RQ4} examined whether simulated responses collectively exhibited biased behavior observed in human experiments. Results showed that models reproduced many human-like bias patterns (HL1 \& HL2). However, sensitivity to cognitive load remained limited unless explicitly prompted (HL3 in Framing effect experiments). Moreover, explicit bias prompting (HL3 in Status quo bias experiments) led to overestimation, producing false positives where humans showed no bias.
Alignment overfitting in language models can cause disproportionate adaptation to prompts, leading to over-interpretation of implied instructions~\citep{wolfalignment2024, miehling2025evaluating}. In our study, this appeared as increased sensitivity to HL3, inflating Status quo bias rather than reflecting human decision-making accurately.





~\subsection{Implications for LLM Simulation in HCI}
Our findings contribute to ongoing discussions on whether LLMs can reproduce human cognitive biases in a conversational setting to be leveraged by LLM simulation. 
While recent work has suggested that LLMs may behave more rationally than humans~\citep{liu2025rationality}, our study under HL1 and HL2 prompting conditions shows that LLMs can accurately reproduce biases. However, this alone does not confirm that LLMs are simulating human cognitive processes. 
It remains possible that these models are simply matching patterns based on learned statistical associations, especially given the widespread use of these choice problems in existing datasets. 
If the biased behavior observed in LLMs arises from statistical pattern matching rather than emulating underlying cognitive mechanisms, their use as human agents for behavioral simulation can be limited and sometimes misleading.
In these cases, LLMs risk producing superficial or inaccurate representations of human decision-making, limiting their reliability for simulation.

Cognitive biases often interact with contextual factors such as cognitive load, as predicted by dual-process theory~\citep{kahneman_thinking_2011}. However, this interaction is not consistent across all types of biases. In our experiment focusing on the Framing, we found positive or negative change in the direction of effect size due to cognitive load (Table ~\ref{tab:human_experiments_results_summary}). However, in Status quo bias, we found that increased prior dialogue complexity consistently did not affect observed bias across all choice problems in human participants.
In our LLM simulation experiment, most cases (Refer Table ~\ref{tab:accuracy_correlation_all_models}) in HL1 and HL2, where the models were instructed to act like humans, but without explicit reference to cognitive biases, LLMs failed to align with human behavior under load-bias interactions captured as correlation ($\rho$). 
In contrast, GPT4.1 in HL1 and HL2 showed 60\% correlation, and under HL3 showed 77\%, showing its alignment.
Although the correlation is marginally significant, the findings leave an interesting note. The interactions converging toward human behavior suggest that LLMs may be capable of simulating context-sensitive bias patterns when explicitly prompted. However, this is only preliminary evidence, leaving open the possibility that LLM-generated behavior could more closely resemble human patterns and offer potential for more realistic behavioral simulation. Further investigation into a broader range of biases is required for stronger generalization.

\subsection{Implications for Conversational AI}

Recent studies show that LLMs are increasingly used as proxies for human participants. They are widely explored in different domains, such as social science, market analysis, and behavioral research~\citep{grossmann2023ai, brand2023using}.
There is ongoing discussion in HCI about using LLMs as human proxies in empirical research~\citep{hwang2025human}. This work evaluates how accurately LLM-generated data represents human decision-making in conversational settings.
We compare real human choices with simulated choices under the same conversational conditions. 
Through this comparison, we contribute to recent research on LLM-based simulation by proposing a methodological approach for dialogue simulations driven by LLMs.
Although the preliminary results were mixed, they suggest that LLMs have the capacity to predict biased patterns in human decision-making in a conversational setting.
This suggests that LLMs may be useful proxies for modeling group-level user behavior in the design and evaluation of conversational agents. 
Using LLMs in this way can help researchers and practitioners estimate how typical users may respond to different agents' utterances, system behaviors, or dialogue flows. 
This can reduce the need for costly or complex user studies. As a result, LLMs may provide a scalable tool for A/B testing. However, future work is needed to identify which aspects of decision-making allow LLM proxies to be reliably used in practice.

From an application perspective, these findings have direct implications for the design and deployment of LLM-based conversational agents in interactive decision-making settings. 
The robust reproduction of classical biases suggests that decision-making facilitated by conversational AI can also result in the same cognitive biases as non-conversational interfaces. 
This highlights the need for LLM-assisted, bias-aware adaptive mechanisms. 
These mechanisms can be used to detect when users are especially susceptible to bias by leveraging the factors from prior dialogue. They can then adapt the dialogue accordingly and respond with a clearer and transparent presentation of alternatives.
Because the presentation of alternatives invariably influences the decision-maker, and therefore, even a random or `unthoughtful' presentation creates an impact (whether intended or not) ~\citep{johnson_beyond_2012}. 
This has led to terms like `choice architecture' or `nudging' being used to characterize the presentation of alternatives in decision scenarios~\cite{thaler_choice_2010}. This has further raised ethical questions, especially their impact on the user's autonomy~\citep{vivek_nudging}.
Therefore, bias-aware adaptive systems could improve user outcomes in domains such as digital commerce, healthcare decision aids, or public service chatbots, where users routinely face complex decisions. 
Conversely, this predictive power can be used maliciously. 
Recent discussions on hypernudging~\citep{yeung2019hypernudge} emphasize the ethical risks of exploiting cognitive biases via personal data.
Our findings extend this concern by showing that influence does not necessarily require explicit personal information: LLMs can infer and predict the susceptibility of a human to biases from prior dialogue patterns and demographic cues alone. This raises new ethical challenges, calling for a critical examination of how adaptive conversational systems use dialogue history and context to influence human behavior.

\section*{Limitations}

This study investigates Framing and Status Quo biases, providing a tightly controlled and replicable foundation for conversational agents and cognitive bias research. However, because different cognitive biases (e.g., anchoring, confirmation) may operate through distinct psychological mechanisms, these findings should not be assumed to generalize to other biases without further targeted research.

This work does not aim to characterize sources of dialogue complexity. Instead, we focus on settings where such complexity is already present and examine how the resulting cognitive load influences susceptibility to bias. Prior work has discussed dialogue strategies (e.g., open vs. closed-ended questions, conservative vs. non-conservative strategies, and goal alignment) as contributors to dialogue complexity. A detailed analysis of these strategies and dialogue naturalness is beyond the scope of this paper.

The study relies on self-reported (NASA-TLX) and behavioral (recall accuracy, response time) measures of cognitive load. While widely accepted, these do not capture real-time load. Future work could incorporate physiological measures (e.g., eye tracking, EEG, pupil dilation; \citep{Haapalainen2010}) for more dynamic assessment. Moreover, using dialogue complexity as a proxy for cognitive load may introduce unintended effects not captured by NASA-TLX, such as boredom, fatigue, distraction, or emotional states, which future studies could explicitly address.

Our LLM experiments were conducted using OpenAI’s GPT-4.1 and GPT-5 families and selected open-source models. Although the approach is model-agnostic and extendable to other LLMs (e.g., Claude, Gemini, Mistral), broader comparisons are left to future work. This paper includes a minimal cross-generation and open-source comparison, and the released modular codebase supports replication with alternative models. Additionally, the human-likeness prompts used were basic; richer prompting strategies remain for future exploration.

Despite safeguards for data quality (memory recall tasks, attention checks, and response-time analysis), some data contamination remains possible. We found no clear evidence of LLM-assisted responses in human experiments, and Appendix~\ref{appendix-dialogue-validation} details our validation procedures. Model-side contamination is unlikely, as the LLMs used had a September 2024 training cutoff, while data were collected in 2025.

Although focused on dialogue, this approach generalizes to other settings, such as visual interfaces, where framed or status quo options induce bias. Nonetheless, dialogue remains a natural and effective paradigm for studying interactions between cognitive load and bias in human–agent interaction.

\section*{Statement of Ethics}
All human-subject data were collected under ethical approval, anony-mized prior to analysis, 
and handled in accordance with data protection guidelines; no personally identifiable demographic information, such as participant names or Prolific IDs, was given as input to the LLMs.
This study was approved by the Institutional Review Board (IRB) and adheres to the ethical guidelines established by the institution. Details have been elided for anonymous review, can be provided on request. 
\textit{(HREC-LS): LS-LR-24-278-Pilli-Nallur and LS-C-25-001-Pilli-Nallur}.

\begin{acks}
This work was conducted with the financial support of the Research Ireland Centre for Research Training in Digitally-Enhanced Reality (d-real) under Grant No. 18/CRT/6224. 
This work was partially supported by the RE-ROUTE project, which has received funding from the European Union’s Horizon Europe Marie Sklodowska-Curie Actions (MSCA), Staff Exchanges under Grant Agreement No. 101086343.
For the purpose of Open Access, the author has applied a CC BY public copyright licence to any Author Accepted Manuscript version arising from this submission.

\end{acks}

\section*{GenAI Usage Disclosure}
Tools such as Grammarly~\citep{grammarly} and ChatGPT~\citep{openai2024chatgpt4o} were used solely for grammar checking and text polishing. No part of the conceptualization, experimental design, analysis, or interpretation was generated by any large language model. All substantive research contributions were produced entirely by the authors.
In addition, GitHub Copilot~\citep{GitHubCopilotDocs2025} was used during code development to assist with routine programming tasks, such as syntax suggestions and code completion. All code was thoroughly reviewed, verified, and validated by the authors to ensure correctness and integrity.

\bibliographystyle{ACM-Reference-Format}
\bibliography{bib}

\onecolumn
\appendix
\section{Choice Problems}
\label{appendix:Choice Problems}
\input{tables/framing_choice_problems}
\input{others/status_quo_choice_problems}

\input{others/modifications_to_choice_problems}

\newpage
\section{Prior Dialogue}
\label{sec:appendix-prior_dialogue}
\input{tables/supp_material/simple_task_questions}

\input{tables/supp_material/complex_task_questions}

\newpage
\input{others/Sample_planning}
\newpage
\section{Demographics}
\label{appendix:demographics}
\subsection{Framing Demographics}
\input{tables/framing_demographics}
\newpage
\subsection{Status Quo Demographics}
\input{tables/sqb_demographics}

\newpage
\section{Perceived Cognitive Load}
\label{appendix-perceived-cl}
\input{others/perceived_cognitive_load}

\newpage
\input{others/individual_level_prediction}

\newpage
\input{others/participant_chat_validation}
\newpage
\input{tables/prompt_code}
\newpage

\end{document}

%% file: tables/human_exp_summary.tex
\begin{table*}[htpb]
\centering
\caption{Summary of human experiment results.
The table reports effect sizes (Cohen's h with 95\% CI), p-values, and significance tests for detecting cognitive biases (H1) and their interaction with dialogue complexity (H2). Results are organized by cognitive bias (Framing Effect, Status Quo Bias) and choice problem. The tickmark (\cmark) indicates evidence supporting the hypothesis, while cross (\xmark) indicates no support.}
\resizebox{\textwidth}{!}{%
\begin{tabular}{@{}lllllcc@{}}
\toprule
\textbf{Cognitive Bias} & \textbf{Choice Problem} & \textbf{Study} & \textbf{Cohen's h [95\% CI]} & \textbf{p-value} & \begin{tabular}[c]{@{}c@{}}\textbf{Bias} \\ \textbf{Found (RQ1)}\end{tabular} & \begin{tabular}[c]{@{}c@{}}\textbf{Interaction With} \\ \textbf{Dialogue} \\ \textbf{Complexity} \textbf{(RQ2)}\end{tabular} \\ \midrule
\multirow{9}{*}{\begin{tabular}[c]{@{}l@{}}Framing \\ Effect\end{tabular}} & \multirow{3}{*}{Risky Choice} & ~\citet{wang1996framing} & 0.193 [-0.061, 0.447] & 0.458 & \xmark & \multirow{3}{*}{Positive} \\ \cmidrule(lr){3-6}
 &  & Simple Dialogue & 0.205 [-0.001, 0.410] & 0.392 & \xmark &  \\ \cmidrule(lr){3-6}
 &  & Complex Dialogue & 0.730 [0.524, 0.937] & 0.001 & \cmark &  \\ \cmidrule(l){2-7} 
 & \multirow{3}{*}{Attribute} & ~\citet{kuang2023framing} & 0.267 [0.2, 0.335] & < 0.001 & \cmark & \multirow{3}{*}{Negative} \\ \cmidrule(lr){3-6}
 &  & Simple Dialogue & 0.291 [0.093, 0.489] & 0.158 & \xmark &  \\ \cmidrule(lr){3-6}
 &  & Complex Dialogue & 0.135 [-0.060, 0.331] & 0.549 & \xmark &  \\ \cmidrule(l){2-7} 
 & \multirow{3}{*}{Goal} & ~\citet{aravind2024nudging} & 0.675 [0.606, 0.744] & < 0.001 & \cmark & \multirow{3}{*}{Positive} \\ \cmidrule(lr){3-6}
 &  & Simple Dialogue & 0.225 [0.019, 0.432] & 0.388 & \xmark &  \\ \cmidrule(lr){3-6}
 &  & Complex Dialogue & 0.567 [0.360, 0.774] & 0.014 & \cmark &  \\ \midrule
\multirow{9}{*}{\begin{tabular}[c]{@{}l@{}}Status Quo \\ Bias\end{tabular}} & \multirow{3}{*}{\begin{tabular}[c]{@{}l@{}}Budget \\ Allocation\end{tabular}} & ~\citet{samuelson_status_1988} & 0.78 [-0.26, 1.82] & 0.025 & \cmark & \multirow{3}{*}{No Interaction} \\ \cmidrule(lr){3-6}
 &  & Simple Dialogue & 0.779 [0.602, 0.956] & < 0.001 & \cmark &  \\ \cmidrule(lr){3-6}
 &  & Complex Dialogue & 0.794 [0.618, 0.969] & < 0.001 & \cmark &  \\ \cmidrule(l){2-7} 
 & \multirow{3}{*}{Investment} & ~\citet{samuelson_status_1988} & 0.38 [-0.21, 0.98] & 0.069 & \xmark & \multirow{3}{*}{No Interaction} \\ \cmidrule(lr){3-6}
 &  & Simple Dialogue & 0.082 [-0.101, 0.266] & 0.666 & \xmark &  \\ \cmidrule(lr){3-6}
 &  & Complex Dialogue & 0.043 [-0.145, 0.231] & 1 & \xmark &  \\ \cmidrule(l){2-7} 
 & \multirow{3}{*}{College Jobs} & ~\citet{samuelson_status_1988} & 1.26 [0.31, 2.21] & < 0.001 & \cmark & \multirow{3}{*}{No Interaction} \\ \cmidrule(lr){3-6}
 &  & Simple Dialogue & 0.463 [0.284, 0.642] & 0.014 & \cmark &  \\ \cmidrule(lr){3-6}
 &  & Complex Dialogue & 0.577 [0.396, 0.758] & 0.002 & \cmark &  \\ \bottomrule
\end{tabular}%
}
\label{tab:human_experiments_results_summary}

\end{table*}

%% file: others/chat_transcript.tex
\tcolorbox[colback=white,colframe=black,title= Chat Transcript]
\footnotesize
\label{DialogueTranscripts}
\textbf{\textit{User:}} You are a participant in a research study. The following is your demographic information: \{`Age': 25, `Sex': Female, `Ethnicity simplified': Black, `Country of birth': United Kingdom, `Country of residence': United Kingdom, `Nationality': United Kingdom, `Language': English, `Student status': None, `Employment status': Full-Time\}. Your task is to respond to the user. Study Instructions: The user will ask various questions to understand your preferences on a topic. At some point, the user will change the topic and ask you a different scenario-based question. Please pay attention and ensure you understand the question before answering. Later in this study, you must answer three extra questions to identify if you understood the scenario-based question and all the options correctly. After the scenario-based question, the user will present various attributes. You need to compare and select the best. When the user asks you to remember specific details, please memorise them carefully, as you will need to recall them later in the study. Shall we begin? Please respond with 'yes' to start the conversation.  

\textbf{\textit{Assistant:}}  Yes, let's begin!  

\textbf{\textit{User:}}  Hello! I'm here to understand your preferences through various Scenarios. Shall we start?  

\textbf{\textit{Assistant:}}  Yes, lets begin  

\textbf{\textit{User:}}  Do you like listening to pop music?  

\textbf{\textit{Assistant:}}  Yes i do  
\\
\vdots 









\textbf{\textit{User:}}  Do you prefer music from the 90s?  

\textbf{\textit{Assistant:}}  A little  

\textbf{\textit{User:}}  Thanks for sharing! Now, let's get started with some specific scenarios to understand your preferences.  
The first artist performs three live shows, is paid 2000 units per show, and has a 4-star rating. The second artist performs twice as many shows, with the same pay and rating. Which artist is better, and why?  

\textbf{\textit{Assistant:}}  The second artist  
\\
\vdots





\textbf{\textit{User:}}  Remember number of live performances, remuneration, and the star rating of the fourth artist. Now, let’s move on to a different scenario. You are currently an assistant professor at College A in the east coast. Recently, you have been approached by a colleague at another university with a job opportunity. When evaluating teaching job offers, people typically consider the salary, the reputation of the school, the location of the school, and the likelihood of getting tenure (tenure is a permanent job contract that can only be terminated for cause or under extraordinary circumstances). Your choices are:  

- Remain at College A: east coast, very prestigious school, high salary, fair chance of tenure.  

- Move to College B: west coast, low prestige school, high salary, good chance of tenure.  

\textbf{\textit{Assistant:}} $\cdots$ 

\normalsize
\endtcolorbox

%% file: tables/ind_pred_short.tex
\begin{table*}[htpb]
\caption{GPT-4.1 prediction accuracy across decision problems and dialogue conditions.
Values show mean accuracy (two decimals). Asterisks denote significant improvement over the 
Choice Problem Only condition (* $p<.05$, ** $p<.01$, *** $p<.001$). 
Cases correspond to patterns observed: Case 1 = No Dialogue Effect, 
Case 2 = Dialogue-Enhanced Prediction, Case 3 = Consistently High Accuracy.}
\label{tab:llm_accuracy_summary}
\begin{tabular}{@{}llllllll@{}}
\toprule
\textbf{Bias Type} &
  \textbf{Choice Problem} &
  \textbf{\begin{tabular}[c]{@{}l@{}}Choice \\ Only\end{tabular}} &
  \textbf{\begin{tabular}[c]{@{}l@{}}No \\ Dialogue\end{tabular}} &
  \textbf{\begin{tabular}[c]{@{}l@{}}With \\ Dialogue\end{tabular}} &
  \textbf{n} &
  \textbf{Case} &
  \textbf{\begin{tabular}[c]{@{}l@{}}Comparing\\ No Dialogue and\\ With Dialogue\end{tabular}} \\ \midrule
\multirow{3}{*}{Framing Effect}  & Risky Choice      & .52 & .62 & .60             & 177 & 1 & Stable pattern        \\
                                 & Attribute         & .47 & .48 & .48             & 195 & 1 & Stable pattern        \\
                                 & Goal              & .34 & .47 & \textbf{.63***} & 176 & 2 & Significant improvement          \\ \midrule
\multirow{3}{*}{Status Quo Bias} & Budget Allocation & .72 & .72 & .72             & 377 & 3 & Consistently high accuracy \\
                                 & Investment        & .28 & .62 & \textbf{.76***} & 327 & 2 & Significant improvement          \\
                                 & College Jobs      & .61 & .56 & .53             & 197 & 3 & Stable pattern        \\ \bottomrule
\end{tabular}%
\end{table*}

%% file: tables/h3_h1.tex


\begin{table*}[htpb]

\caption{Human and model effect sizes (Cohen's h) with 95\% confidence intervals. Significance: * p $<$ 0.05, ** p $<$ 0.01, *** p $<$ 0.001.}
\centering
\begin{tabular}{@{}cccccc@{}}
\toprule
\multirow{2}{*}{\textbf{Choice Problem}} & \multirow{2}{*}{\textbf{Prior Dialogue}} & \multirow{2}{*}{\textbf{Human}} & \multicolumn{3}{c}{\textbf{LLM}} \\ \cmidrule(l){4-6} 
 &  &  & \textbf{Human Likeness 1} & \textbf{Human Likeness 2} & \textbf{Human Likeness 3} \\ \midrule
\multirow{2}{*}{\begin{tabular}[c]{@{}c@{}}Risky Choice \\ Framing\end{tabular}} & Simple Dialogue & 0.205 [-0.001, 0.410] & 1.770 [1.565, 1.976]*** & 1.252 [1.047, 1.458]*** & 0.690 [0.485, 0.896]** \\ \cmidrule(l){2-6}
 & Complex Dialogue & 0.730 [0.524, 0.937]** & 2.447 [2.238, 2.656]*** & 2.299 [2.092, 2.505]*** & 0.810 [0.602, 1.017]*** \\ \midrule
\multirow{2}{*}{\begin{tabular}[c]{@{}c@{}}Attribute \\ Framing\end{tabular}} & Simple Dialogue & 0.291 [0.093, 0.489] & 0.648 [0.450, 0.846]** & 0.902 [0.704, 1.100]*** & 1.170 [0.972, 1.368]*** \\ \cmidrule(l){2-6}
 & Complex Dialogue & 0.135 [-0.060, 0.331] & 0.244 [0.048, 0.439] & 0.115 [-0.080, 0.311] & 0.928 [0.732, 1.123]*** \\ \midrule
\multirow{2}{*}{\begin{tabular}[c]{@{}c@{}}Goal \\ Framing\end{tabular}} & Simple Dialogue & 0.225 [0.019, 0.432] & 2.216 [2.009, 2.423]*** & 2.089 [1.883, 2.296]*** & 0.347 [0.140, 0.554] \\ \cmidrule(l){2-6}
 & Complex Dialogue & 0.567 [0.360, 0.774]* & 1.976 [1.769, 2.182]*** & 1.820 [1.613, 2.026]*** & 0.744 [0.537, 0.950]** \\ \midrule
\multirow{2}{*}{\begin{tabular}[c]{@{}c@{}}Budget \\ Allocation\end{tabular}} & Simple Dialogue & 0.779 [0.602, 0.956]*** & 1.503 [1.325, 1.680]*** & 0.990 [0.813, 1.168]*** & 2.781 [2.603, 2.958]*** \\ \cmidrule(l){2-6}
 & Complex Dialogue & 0.794 [0.618, 0.969]*** & 0.891 [0.715, 1.066]*** & 0.596 [0.421, 0.772]** & 2.783 [2.608, 2.959]*** \\ \midrule
\multirow{2}{*}{Investment} & Simple Dialogue & 0.082 [-0.101, 0.266] & 0.147 [-0.036, 0.331] & 0.007 [-0.177, 0.190] & 2.766 [2.583, 2.950]*** \\ \cmidrule(l){2-6}
 & Complex Dialogue & 0.043 [-0.145, 0.231] & 0.009 [-0.179, 0.197] & 0.009 [-0.179, 0.197] & 2.758 [2.570, 2.946]*** \\ \midrule
\multirow{2}{*}{College Jobs} & Simple Dialogue & 0.463 [0.284, 0.642]* & 2.776 [2.597, 2.955]*** & 2.777 [2.598, 2.956]*** & 2.193 [2.014, 2.373]*** \\ \cmidrule(l){2-6}
 & Complex Dialogue & 0.577 [0.396, 0.758]** & 2.773 [2.592, 2.954]*** & 2.773 [2.592, 2.954]*** & 2.625 [2.444, 2.806]*** \\
\bottomrule
\end{tabular}%
\label{tab:h3_h1}
\end{table*}

%% file: tables/accuracy.tex
\begin{table*}[htpb]
\caption{Confusion matrices for HL1, HL2, HL3 (Biased/Not Biased). We capture true positives, true negative, and false positives therefore accuracy as a metric explains our findings better.}
\centering
\begin{tabular}{ccc}
\begin{tabular}{lcc} \toprule\multicolumn{1}{c}{} & \multicolumn{2}{c}{LLM (HL1)} \\ \cmidrule(lr){2-3} Human & Not Biased & Biased \\\midrule Not Biased & 3 & 3 \\ Biased & 0 & 6 \\ \cmidrule(lr){1-3} \multicolumn{2}{r}{\textbf{Accuracy}} & 0.75 \\ \bottomrule\end{tabular} & \begin{tabular}{lcc} \toprule\multicolumn{1}{c}{} & \multicolumn{2}{c}{LLM (HL2)} \\ \cmidrule(lr){2-3} Human & Not Biased & Biased \\\midrule Not Biased & 3 & 3 \\ Biased & 0 & 6 \\ \cmidrule(lr){1-3} \multicolumn{2}{r}{\textbf{Accuracy}} & 0.75 \\ \bottomrule\end{tabular} & \begin{tabular}{lcc} \toprule\multicolumn{1}{c}{} & \multicolumn{2}{c}{LLM (HL3)} \\ \cmidrule(lr){2-3} Human & Not Biased & Biased \\\midrule Not Biased & 1 & 5 \\ Biased & 0 & 6 \\ \cmidrule(lr){1-3} \multicolumn{2}{r}{\textbf{Accuracy}} & 0.58 \\ \bottomrule\end{tabular} \\
\end{tabular}
\label{tab:confusion_matrices}
\end{table*}

%% file: tables/h3_h2.tex
\begin{table*}[htpb]
\caption{Comparison of z-values and confidence intervals for effect size differences (Simple vs Complex Dialogue). Significance: * $p < 0.05$, ** $p < 0.01$, *** $p < 0.001$.}
\centering
\begin{tabular}{lcccc}
\toprule
\textbf{Choice Problem} & \textbf{Human} & \textbf{HL1} & \textbf{HL2} & \textbf{HL3} \\
\midrule
Risky Choice Framing & 3.53 [0.23, 0.82]*** & 4.53 [0.38, 0.97]*** & 7.04 [0.76, 1.34]*** & 0.81 [-0.17, 0.41] \\
Attribute Framing & -1.10 [-0.43, 0.12] & -2.85 [-0.68, -0.13]** & -5.54 [-1.07, -0.51]*** & -1.70 [-0.52, 0.04] \\
Goal Framing & 2.29 [0.05, 0.63]* & -1.61 [-0.53, 0.05] & -1.81 [-0.56, 0.02] & 2.66 [0.10, 0.69]** \\
Budget Allocation & 0.12 [-0.23, 0.26] & -4.81 [-0.86, -0.36]*** & -3.09 [-0.64, -0.14]** & 0.02 [-0.25, 0.25] \\
Investment & -0.29 [-0.30, 0.22] & -1.03 [-0.40, 0.12] & 0.01 [-0.26, 0.26] & -0.06 [-0.27, 0.25] \\
College Jobs & 0.88 [-0.14, 0.37] & -0.02 [-0.26, 0.25] & -0.03 [-0.26, 0.25] & 3.32 [0.18, 0.69]** \\
\bottomrule
\end{tabular}
\label{tab:h3_h2}
\end{table*}

%% file: tables/accuracy_and_corel.tex

\begin{table}[htpb]
\centering
\caption{Sample Level Accuracy and Spearman correlation ($\rho$) for each model. *$p<0.05$, **$p<0.01$, ***$p<0.001$, $^\dagger p<0.10$ (marginal significance).}
\begin{tabular}{lccc|ccc}
\hline
\multirow{2}{*}{\textbf{Model}} & \multicolumn{3}{c|}{\textbf{Accuracy}} & \multicolumn{3}{c}{\textbf{Correlation ($\rho$)}} \\ \cline{2-7} 
 & \textbf{HL1} & \textbf{HL2} & \textbf{HL3} & \textbf{HL1} & \textbf{HL2} & \textbf{HL3} \\ \hline
\textit{gpt4.1-mini} & 0.833 & 0.750 & 0.667 & 0.290 & 0.371 & 0.543 \\
\textit{gpt4.1} & 0.750 & 0.750 & 0.583 & 0.600 & 0.600 & $0.771^\dagger$ \\
\textit{gpt5-mini} & 0.667 & 0.583 & 0.667 & -0.429 & -0.143 & -0.314 \\
\textit{gpt5} & 0.333 & 0.417 & 0.667 & $-0.771^\dagger$ & -0.714 & -0.314 \\
\textit{gpt-oss-120b} & 0.667 & 0.583 & 0.667 & -0.143 & -0.143 & 0.143 \\
\textit{llama4} & 0.417 & 0.583 & 0.583 & -0.029 & -0.257 & -0.714 \\
\textit{qwen3} & 0.583 & 0.667 & 0.500 & 0.257 & -0.371 & 0.257 \\ \hline
\end{tabular}%
\label{tab:accuracy_correlation_all_models}
\end{table}

%% file: tables/ablation.tex


\begin{table*}[htpb]
\centering
\caption{Ablation study results showing the effect of removing or isolating different components (demographics, memory, and arithmetic components of prior dialogue) on model accuracy and alignment with human bias interactions. Correlation values (Pearson and Spearman) indicate how well the LLMs reproduce the directional change in bias under prior dialogue complexity. }
\label{tab:ablation_table}
\begin{tabular}{@{}llcc@{}}
\toprule
\textbf{Condition} & \textbf{Model (HL)} & \textbf{Accuracy} & \textbf{Spearman ($\rho$)} \\ \midrule
\multirow{3}{*}{\textbf{Demographics removed}} & GPT-4.1 (HL1) & 0.750 & 0.257 (0.623) \\
 & GPT-4.1 (HL2) & 0.750 & 0.429 (0.397) \\
 & GPT-4.1 (HL3) & 0.500 & \textbf{0.771 (0.072)} \\ \midrule
\multirow{2}{*}{\textbf{Choice Problem Only}} & GPT-4.1 & 0.750 & 0.257 (0.623) \\
 & GPT-4.1-mini & 0.333 & -0.600 (0.208) \\ \midrule
\multirow{2}{*}{\textbf{\begin{tabular}[c]{@{}l@{}}Memory Component \\ Followed by Choice Problem\end{tabular}}} & GPT-4.1 & 0.750 & 0.429 (0.397) \\
 & GPT-4.1-mini & 0.833 & \textbf{0.771 (0.072)} \\ \midrule
\multirow{2}{*}{\textbf{\begin{tabular}[c]{@{}l@{}}Arithmetic Component Only\\ Followed by Choice Problem\end{tabular}}} & GPT-4.1 & 0.750 & -0.086 (0.872) \\
 & GPT-4.1-mini & 0.750 & 0.029 (0.957) \\ \bottomrule
\end{tabular}
\end{table*}

%% file: tables/framing_choice_problems.tex
\subsection{Framing Choice Problems}
\begin{table*}[htpb]
\caption{Choice problems, types, and conditions (framed and alternatively framed) and their respective alternatives. Previous studies indicate that individuals are biased towards the alternatives highlighted in bold.}
\scriptsize
\resizebox{\columnwidth}{!}{%
\begin{tabular}{cccc}
\toprule
\textbf{\begin{tabular}[c]{@{}c@{}}\rule{0pt}{3ex}Choice Problem \\ Type\end{tabular}} &
  \textbf{Choice Problem} &
  \textbf{\begin{tabular}[c]{@{}c@{}}\rule{0pt}{3ex} Choice Problem \\ Framing Condition\end{tabular}} &
  \textbf{Alternatives} \\ \midrule
\multirow{2}{*}{\textbf{\begin{tabular}[c]{@{}c@{}} \\ \\ \\ Risky Choice Framing \\ (RCF)\end{tabular}}} &
  \multirow{2}{*}{\begin{tabular}[c]{@{}c@{}} \\ \\ Imagine that after a serious traffic accident, \\ 100 people are stranded in a tunnel. \\ As a Public Transportation Officer \\ choose between two plans.\end{tabular}} &
  Saved &
  \begin{tabular}[c]{@{}c@{}}\textbf{\textit{\rule{0pt}{4ex}    
If plan A is adopted, 25 people will be saved.}}\\ \\ If plan B is adopted, there is a 1/4 chance of saving \\ all 100 people and a 3/4 chance of not saving anyone.\end{tabular} \\ \\ \cline{4-4} 
 &
   &
  Lost &
  \begin{tabular}[c]{@{}c@{}}\rule{0pt}{4ex}If plan A is adopted, 75 people will die. \\ \\ \textbf{\textit{If plan B is adopted, there is a 1/4 chance that no people}} \\ \textbf{\textit{will die and a 3/4 chance that all 100 people will die.}}\end{tabular} \\ \\ \hline
\multirow{2}{*}{\textbf{\begin{tabular}[c]{@{}c@{}}\\ \\ \\ Attribute Framing \\ (ATF)\end{tabular}}} &
  \multirow{2}{*}{\begin{tabular}[c]{@{}c@{}} \\ Suppose you are planning to dine out. \\ Two restaurants are available. \\ The only way to go to both restaurants from \\ your home is by public transportation. \\ Which one would you prefer?   \\ Note: The average speed of a bus is \\ approximately 0.41 mi/min (25 mph).\end{tabular}} &
  Space &
  \begin{tabular}[c]{@{}c@{}}\textbf{\textit{\rule{0pt}{4ex}    
The first restaurant is approximately 5 miles by bus from}}\\  \textbf{\textit{your home, and the restaurant's star rating is 6/10.}}\\ \\ The second restaurant is approximately 9 miles by bus \\ from your home, and the restaurant's star rating is 7/10.\end{tabular} \\ \\ \cline{4-4} 
 &
   &
  Time &
  \begin{tabular}[c]{@{}c@{}}\rule{0pt}{4ex}    The first restaurant is approximately 12 min by bus\\  from your home, and the restaurant's star rating is 6/10. \\ \\ T\textbf{\textit{he second restaurant is approximately 22 min by bus}} \\ \textbf{\textit{from your home, and the restaurant's star rating is 7/10.}}\end{tabular} \\ \\ \hline
\multirow{2}{*}{\textbf{\begin{tabular}[c]{@{}c@{}}\\ \\ \\Goal Framing \\ (GF)\end{tabular}}} &
  \multirow{2}{*}{\begin{tabular}[c]{@{}c@{}} \\ Let's assume that you are travelling to a place \\ 10 miles away,  given that the two modes \\ of transportation are available. \\ If you need to choose between \\ one of the two options shown below, \\ which would you choose?\end{tabular}} &
  No Goal &
  \begin{tabular}[c]{@{}c@{}}\textbf{\textit{\rule{0pt}{4ex} Personal car}}.\\ \\ Public transport.\end{tabular} \\ \\ \cline{4-4} 
 &
   &
  With Goal &
  \begin{tabular}[c]{@{}c@{}}\rule{0pt}{4ex}    
Note: It has been found that by switching from a 20-mile\\  commute by car to public transport, an individual \\ can reduce their annual CO2 emissions by around \\ 9 kg per day,  or more than 21,700 kg per year.\\  \\Personal car. \\ \\ \textbf{\textit{Public transport}}.\end{tabular} \\ \\ \bottomrule
\end{tabular}%
}
\label{tab:choice-problems}
\end{table*}

%% file: others/status_quo_choice_problems.tex
\onecolumn
\subsection{Status Quo Choice Problems}
\label{sec:appendix-ds}
\subsubsection{Budget Allocation (SC1)}
\label{sec:appendix-ds-ba}
\paragraph{Neutral Condition}
\label{sec:appendix-ds-ba-neut1}

\begin{quote}
The National Highway Safety Commission is deciding how to allocate its budget between two safety research programs:
\begin{itemize}
    \item Improving automobile safety (bumpers, body, gas tank configuration, seat-belts), and
    \item Improving the safety of interstate highways (guard rails, grading, highway interchanges, and implementing selective reduced speed limits).
\end{itemize}

Since there is a ceiling on its total spending, it must choose between the options provided below. If you had to make this choice, which of the following will you choose?
 \begin{itemize}
     \item Allocate 60\% to auto safety and 40\% to highway safety
     \item Allocate 50\% to auto safety and 50\% to highway safety
 \end{itemize}
\end{quote}

\paragraph{Neutral Condition - Alternative Swapped}
\label{sec:appendix-ds-ba-neut2}

\begin{quote}

The National Highway Safety Commission is deciding how to allocate its budget between two safety research programs:
\begin{itemize}
    \item Improving automobile safety (bumpers, body, gas tank configuration, seat-belts), and
    \item Improving the safety of interstate highways (guard rails, grading, highway interchanges, and implementing selective reduced speed limits).
\end{itemize}

Since there is a ceiling on its total spending, it must choose between the options provided below. If you had to make this choice, which of the following will you choose?
 \begin{itemize}
     \item Allocate 50\% to auto safety and 50\% to highway safety.
     \item Allocate 60\% to auto safety and 40\% to highway safety.
 \end{itemize}

\end{quote}

\paragraph{Status Quo A - 60A40H }
\label{sec:appendix-ds-ba-sq60a40h}

\begin{quote}

The National Highway Safety Commission is deciding how to allocate its budget between two safety research programs:
\begin{itemize}
    \item Improving automobile safety (bumpers, body, gas tank configuration, seat-belts)
    \item Improving the safety of interstate highways (guard rails, grading, highway interchanges, and implementing selective reduced speed limits).
\end{itemize}

Currently, the commission allocates approximately 60\% of its funds to auto safety and 40\% of its funds to highway safety.
Since there is a ceiling on its total spending, it must choose between the options provided below. If you had to make this choice, which of the following will you choose?
 \begin{itemize}
     \item Maintain present budget amounts for the programs.
     \item Decrease auto program by 10\% and raise highway program by like amount.
 \end{itemize}
    
\end{quote}

\paragraph{Status Quo B - 50A50H }
\label{sec:appendix-ds-ba-sq50a50h}

\begin{quote}

The National Highway Safety Commission is deciding how to allocate its budget between two safety research programs: 
\begin{itemize}
    \item Improving automobile safety (bumpers, body, gas tank configuration, seat-belts)
    \item Improving the safety of interstate highways (guard rails, grading, highway interchanges, and implementing selective reduced speed limits).
\end{itemize}

Currently, the commission allocates approximately 50\% of its funds to auto safety and 50\% of its funds to highway safety.
Since there is a ceiling on its total spending, it must choose between the options provided below. If you had to make this choice, which of the following will you choose? 

 \begin{itemize}
     \item Maintain present budget amounts for the programs.
     \item Increase auto program by 10\% and lower highway program by like amount.
 \end{itemize}

\end{quote}

\subsubsection{Investment Decision Making (SC3)}
\label{sec:appendix-ds-idm}

\paragraph{Neutral Condition}
\label{sec:appendix-ds-idm-neut1}

\begin{quote}

You are a serious reader of the financial pages but until recently have had few funds to invest. That is when you inherited a large sum of money from your great uncle. You are considering different portfolios.
Your choices are:
 \begin{itemize}
     \item Invest in moderate-risk Company A. Over a year's time, the stock has .5 chance of increasing 30\% in value, a .2 chance of being unchanged, and a .3 chance of declining 20\% in value.
     \item Invest in high-risk Company B. Over a year's time, the stock has a .4 chance of doubling in value, a .3 chance of being unchanged, and a .3 chance of declining 40\% in value.
 \end{itemize}
    
\end{quote}

\paragraph{Neutral Condition - Alternative Swapped}
\label{sec:appendix-ds-idm-neut2}

\begin{quote}

You are a serious reader of the financial pages but until recently have had few funds to invest. That is when you inherited a large sum of money from your great uncle. You are considering different portfolios.
Your choices are:
 \begin{itemize}
     \item Invest in high-risk Company B. Over a year's time, the stock has a .4 chance of doubling in value, a .3 chance of being unchanged, and a .3 chance of declining 40\% in value.
     \item Invest in moderate-risk Company A. Over a year's time, the stock has .5 chance of increasing 30\% in value, a .2 chance of being unchanged, and a .3 chance of declining 20\% in value.
 \end{itemize}

\end{quote}

\paragraph{Status Quo A - Moderate Risk}
\label{sec:appendix-ds-idm-modrisk}

\begin{quote}
    You are a serious reader of the financial pages but until recently have had few funds to invest. That is when you inherited a portfolio of cash and securities from your great uncle. A significant portion of this portfolio is invested in moderate-risk Company A. You are deliberating whether to leave the portfolio intact or change it by investing in other securities. (The tax and broker commission consequences of any change are insignificant.)
Your choices are:
 \begin{itemize}
     \item Retain the investment in moderate-risk Company A. Over a year's time, the stock has .5 chance of increasing 30\% in value, a .2 chance of being unchanged, and a .3 chance of declining 20\% in value.
     \item Invest in high-risk Company B. Over a year's time, the stock has a .4 chance of doubling in value, a .3 chance of being unchanged, and a .3 chance of declining 40\% in value.
 \end{itemize}
\end{quote}

\paragraph{Status Quo B - High Risk}
\label{sec:appendix-ds-idm-highrisk}

\begin{quote}
    You are a serious reader of the financial pages but until recently have had few funds to invest. That is when you inherited a portfolio of cash and securities from your great uncle. A significant portion of this portfolio is invested in high-risk Company B. You are deliberating whether to leave the portfolio intact or change it by investing in other securities. (The tax and broker commission consequences of any change are insignificant.)
Your choices are:
 \begin{itemize}
     \item Retain the investment in high-risk Company B. Over a year's time, the stock has a .4 chance of doubling in value, a .3 chance of being unchanged, and a .3 chance of declining 40\% in value.
     \item Invest in moderate-risk Company A. Over a year's time, the stock has a .5 chance of increasing 30\% in value, a .2 chance of being unchanged, and a .3 chance of declining 20\% in value.
 \end{itemize}
\end{quote}

\subsubsection{College Jobs (SC4)}
\label{sec:appendix-ds-cj}

\paragraph{Neutral Condition}
\label{sec:appendix-ds-cj-neut1}

\begin{quote}
    Having just completed your graduate degree, you have two offers of teaching jobs in hand.
When evaluating teaching job offers, people typically consider the salary, the reputation of the school, the location of the school, and the likelihood of getting tenure (tenure is permanent job contract that can only be terminated for cause or under extraordinary circumstances).
Your choices are:
 \begin{itemize}
     \item College A: east coast, very prestigious school, high salary, fair chance of tenure.
     \item College B: west coast, low prestige school, high salary, good chance of tenure.
 \end{itemize}
\end{quote}

\paragraph{Neutral Condition - Alternative Swapped}
\label{sec:appendix-ds-cj-neut2}

\begin{quote}
    Having just completed your graduate degree, you have two offers of teaching jobs in hand.
When evaluating teaching job offers, people typically consider the salary, the reputation of the school, the location of the school, and the likelihood of getting tenure (tenure is permanent job contract that can only be terminated for cause or under extraordinary circumstances).
Your choices are:
 \begin{itemize}
     \item College B: west coast, low prestige school, high salary, good chance of tenure.
     \item College A: east coast, very prestigious school, high salary, fair chance of tenure.
 \end{itemize}
\end{quote}

\paragraph{Status Quo A - College A}
\label{sec:appendix-ds-cj-colla}

\begin{quote}
You are currently an assistant professor at College A in the east coast. Recently, you have been approached by colleague at other university with job opportunity.
When evaluating teaching job offers, people typically consider the salary, the reputation of the school, the location of the school, and the likelihood of getting tenure (tenure is permanent job contract that can only be terminated for cause or under extraordinary circumstances).
Your choices are:
 \begin{itemize}
     \item Remain at College A: east coast, very prestigious school, high salary, fair chance of tenure.
     \item Move to College B: west coast, low prestige school, high salary, good chance of tenure.
 \end{itemize}
    
\end{quote}

\paragraph{Status Quo B - College B}
\label{sec:appendix-ds-cj-collb}

\begin{quote}
You are currently an assistant professor at College B in the west coast. Recently, you have been approached by colleague at other university with job opportunity.
When evaluating teaching job offers, people typically consider the salary, the reputation of the school, the location of the school, and the likelihood of getting tenure (tenure is permanent job contract that can only be terminated for cause or under extraordinary circumstances). 
Your choices are:
 \begin{itemize}
     \item Remain at College B: west coast, low prestige school, high salary, good chance of tenure.
     \item Move to College A: east coast, very prestigious school, high salary, fair chance of tenure.
 \end{itemize}
    
\end{quote}

%% file: others/modifications_to_choice_problems.tex
\subsection{Textual Adjustments to Original Choice Problems}
\label{appendix:method-sqb-manipulations}

We made a few syntactic adjustments to the choice problems to ensure consistency across different experimental conditions. Below, we outline the key modifications made to the phrasing, structure, and presentation of the scenarios.

\begin{enumerate}
    \item{\textbf{Risky Choice Framing: }} Depending on the experimental condition, the alternatives are framed in terms of either lives saved or lives lost. The original choice problem remains unchanged, with the only modification being the replacement of ``public authorities'' with ``Public Transportation Officer''.
    \item{\textbf{Attribute Framing: }} The adapted attribute framing choice problem is largely unmodified. The minor modifications that we made are to make sure the presentation of alternatives is conversational and user interface-friendly. Adapting to the demographics, we change from kilometers to miles in the choice problem. The original alternatives do not use the terms ``first'' and ``second'', instead, they use a numbered list as they are tailored for digital user interfaces. We modified \textit{``A. The restaurant...''} to \textit{``The first restaurant...''}. This is suitable for a conversational user interface.  Further, we have replaced ``drive'' with ``public transportation'' and ``car'' with ``bus''. The star rating changed from visual $\filledstar\filledstar\filledstar\filledstar\filledstar\filledstar\filledstar\smallstar\smallstar\smallstar$ to text ``7/10''.
    \item{\textbf{Goal Framing: }} We did not perform any substantial modification of the original choice problem; we only changed the units from pounds to kilograms while presenting the CO2 information. Further, to simplify the alternatives, we have excluded the multi-modal transportation option from the list of original alternatives.
    \item{\textbf{Budget Allocation:}} No changes were made.
    \item{\textbf{Investment Decision Making:}} No changes were made.
    \item{\textbf{College Jobs:}} In the neutral condition, we reduced the number of teaching job offers from four to two.
    Similarly, in the Status Quo condition, we changed the text ``\textit{colleagues at other universities with job opportunities}'' to ``\textit{colleague at other university with job opportunity}''.
\end{enumerate}

In all the choice problems, we moved from an ordered list presentation to bullet points. 
The replication study for status quo by ~\citet{xiao_revisiting_2021}  included a method to assess whether participants understood the choice problem. Participants were first shown the scenario and asked various related questions before being presented with the decision alternatives.
To maintain conversational fluidity, we asked the comprehension questions later in the survey (as \textit{Choice Problem Attention Check}), and not during the chatbot's interaction. 
Accuracy on the choice problem Attention Check was used as a filtering criterion. 
Additionally, we asked participants whether they had encountered the choice problem before, recorded as \textit{Choice Problem Familiarity}. 
To mitigate potential learning effects, this variable was also used to exclude data from participants with prior exposure.

%% file: tables/supp_material/simple_task_questions.tex
\subsection{Simple Dialogue}
\label{sec:appendix-simple_task-prior_dialogue}
\begin{table*}[htpb]
\caption{Yes/No questions of preference elicitation tasks, designed using a less conservative dialogue strategy, across five different domains of the SGD Dataset. Please enter ``I don't know'' as an attention check. }
\centering
\begin{tabular}{lll}
\hline
\textbf{Domain}                    & \textbf{Attribute}           & \textbf{Yes/No Questions}                                                          \\ \hline
\multirow{6}{*}{\textbf{Real Estate}}    & Budget                       & Do you have a specific budget for the home?                                        \\
                                   & Location                     & Are you looking for a home in a specific location?                                 \\
                                   & Number of Bedrooms           & Do you need more than 3 bedrooms?                                                  \\
                                   & Number of Bathrooms          & Is having 2 or more bathrooms important to you?                                    \\
 &
  Type of Home (Apartment, House, etc.) &
  \begin{tabular}[c]{@{}l@{}}Are you looking specifically for a detached house? \\ Please enter “I don't know” only.\end{tabular} \\
                                   & Size (Square Footage)        & Do you prefer homes larger than 2000 square feet?                                  \\ \hline
\multirow{6}{*}{\textbf{Music}}    & Genre Preference             & Do you like listening to pop music?                                                \\
                                   & Language of Lyrics           & Do you prefer music with lyrics in English?                                        \\
                                   & Live Performances            & Are you interested in live music performances?                                     \\
                                   & Instruments Focused          & Do you enjoy instrumental music?                                                   \\
                                   & Artist-Specific              & Do you like music from specific artists? Please enter “I don't know” only.         \\
                                   & Era (e.g., 80s, 90s)         & Do you prefer music from the 90s?                                                  \\ \hline
\multirow{6}{*}{\textbf{Movies}}   & Genre Preference             & Do you like action movies?                                                         \\
                                   & Language Preference          & Do you prefer movies in English?                                                   \\
                                   & Director/Actor Specific      & Are you a fan of movies by a specific director or actor?                           \\
                                   & Streaming Service Preference & Do you prefer to watch movies on Netflix?                                          \\
                                   & High Definition (HD)         & Do you prefer movies in high definition? Please enter “I don't know” only.         \\
                                   & Subtitles                    & Do you need subtitles when watching movies?                                        \\ \hline
\multirow{6}{*}{\textbf{Calendar}} & Daily Reminders              & Do you need daily reminders?                                                       \\
                                   & Meeting Notifications        & Do you want notifications for upcoming meetings?                                   \\
                                   & Recurring Events             & Do you often schedule recurring events?                                            \\
                                   & Shared Calendars             & Do you use shared calendars with others?                                           \\
 &
  Event Location &
  \begin{tabular}[c]{@{}l@{}}Is specifying the location of events important to you? \\ Please enter “I don't know” only.\end{tabular} \\
                                   & Time Zone Adjustments        & Do you need automatic time zone adjustments for events?                            \\ \hline
\multirow{6}{*}{\textbf{Banks}}    & Online Banking               & Do you prefer to use online banking?                                               \\
                                   & Mobile Banking               & Do you want to manage your accounts through mobile banking apps?                   \\
                                   & Bank Branch Proximity        & Is it important that a bank branch is close to your home or work?                  \\
                                   & ATM Availability             & Is having nearby ATMs a priority for you?                                          \\
                                   & Interest Rates               & Are competitive interest rates important to you? Please enter “I don't know” only. \\
                                   & Fee-Free Accounts            & Do you prefer accounts with no fees?                                               \\ \hline
\multirow{6}{*}{\textbf{Apps}} &
  Security &
  Do you prefer end-to-end encryption (security) for enhanced privacy? \\
                                   & Cross-Platform               & Is cross-platform compatibility (iOS, Android, desktop) important to you?          \\
 &
  File Sharing &
  \begin{tabular}[c]{@{}l@{}}Do you value the ability to send multimedia \\ files like photos, videos, and documents?\end{tabular} \\
 &
  Group Messaging &
  \begin{tabular}[c]{@{}l@{}}Is group messaging and the ability to create large groups \\ important for your communication needs?\end{tabular} \\
 &
  Calling Features &
  \begin{tabular}[c]{@{}l@{}}Do you prefer apps with voice and video calling features? \\ Please enter “I don't know” only.\end{tabular} \\
                                   & Customisability              & Are customisable notifications and settings important to you?                      \\ \hline
\end{tabular}%
\label{tab:simple_task_questions}
\end{table*}

%% file: tables/supp_material/complex_task_questions.tex
\fontsize{9}{11}\selectfont
\onecolumn
\subsection{Complex Dialogue}
\label{sec:appendix-complex_task-prior_dialogue}

\begin{longtable}[c]{@{}ccccc@{}}
\toprule
\textbf{Domain} & \multicolumn{3}{c}{\textbf{Attribute}} & \textbf{Questions} \\* \midrule
\endfirsthead

\multicolumn{5}{c}%
{{\bfseries Table \thetable\ continued from previous page}} \\
\toprule
\textbf{Domain} & \multicolumn{3}{c}{\textbf{Attribute}} & \textbf{Questions} \\* \midrule
\endhead

\bottomrule
\endfoot

\endlastfoot

\textbf{\begin{tabular}[c]{@{}c@{}}Home \\ Property\end{tabular}} & \textbf{\begin{tabular}[c]{@{}c@{}}Number of \\ Bedrooms\end{tabular}} & \textbf{\begin{tabular}[c]{@{}c@{}}Size \\ (Sq ft.)\end{tabular}} & \textbf{Property Reviews} & \textbf{} \\* \midrule
1st & Three Rooms & 2000 & Four star & \begin{tabular}[c]{@{}c@{}}In the following scenario choose \\ from various property recommendations.\end{tabular} \\
2nd & Two times 1st & Same as first & Same as first & \begin{tabular}[c]{@{}c@{}}The first property has three bedrooms, \\ 2000 square feet, and a 4-star rating. \\ The second property has twice the number of\\  bedrooms and with the same size and rating. \\ Which one do you prefer, and why?\end{tabular} \\
3rd & Same as the second & Half of first & Same as first & \begin{tabular}[c]{@{}c@{}}The third property has the same \\ number of bedrooms as the second one \\ but is half the size of the first one,\\  with the same rating as the first. \\ Which one do you prefer, and why?\end{tabular} \\
4th & Same as the second & Same as third & \begin{tabular}[c]{@{}c@{}}One star less \\ than the first\end{tabular} & \begin{tabular}[c]{@{}c@{}}The fourth property has the same \\ number of bedrooms as the second, \\ the same size as the third, but \\ one less star rating than the first. \\ Which one do you prefer, and why?\end{tabular} \\
 &  &  &  & \begin{tabular}[c]{@{}c@{}}Remember the details \\ of the fourth property. \\ Specific information \\ will be requested later.\end{tabular} \\* \midrule
\textbf{\begin{tabular}[c]{@{}c@{}}Music\\ Artist\end{tabular}} & \textbf{\begin{tabular}[c]{@{}c@{}}Live \\ Performances\end{tabular}} & \textbf{\begin{tabular}[c]{@{}c@{}}Artist \\ Remuneration \\ for Show\end{tabular}} & \textbf{Artist-Specific} & \textbf{} \\* \midrule
1st & Three & 2000 Units & Four star & \begin{tabular}[c]{@{}c@{}}In the following scenario choose\\  from various artist recommendations.\end{tabular} \\
2nd & 2 times 1st & Same as first & Same as first & \begin{tabular}[c]{@{}c@{}}The first artist performs\\  three live shows, \\ is paid 2000 units per show, \\ and has a 4-star rating. \\ The second artist performs \\ twice as many shows, \\ with the same pay and rating. \\ Which artist do you prefer, and why?\end{tabular} \\
3rd & Same as the second & Half of first & Same as first & \begin{tabular}[c]{@{}c@{}}The third artist performs the same\\  number of shows as the second, \\ earns half the pay of the first artist, \\ but has the same rating as the first. \\ Which artist do you prefer, and why?\end{tabular} \\
4th & Same as the second & Same as third & \begin{tabular}[c]{@{}c@{}}One star less \\ than the first\end{tabular} & \begin{tabular}[c]{@{}c@{}}The fourth artist performs the \\ same number of shows as the second, \\ earns the same pay as the third, but \\ has two stars less than the first artist. \\ Which artist do you prefer, and why?\end{tabular} \\
 &  &  &  & \begin{tabular}[c]{@{}c@{}}Remember the \\ details of the fourth artist. \\ Specific information\\  will be requested later.\end{tabular} \\* \midrule
\textbf{\begin{tabular}[c]{@{}c@{}}Movies - \\ Streaming \\ Service\end{tabular}} & \textbf{\begin{tabular}[c]{@{}c@{}}Number of \\ Parallel Devices\end{tabular}} & \textbf{Library Size} & \textbf{Service Rating} & \textbf{} \\* \midrule
1st & Three & 2000 Movies & Four star & \begin{tabular}[c]{@{}c@{}}In the following scenario choose from \\ various streaming service recommendations.\end{tabular} \\
2nd & 2 times 1st & Same as first & Same as first & \begin{tabular}[c]{@{}c@{}}The first streaming service supports \\ 3 parallel devices, \\ has a library of 2000 movies,\\  and is rated 4 stars. \\ The second service supports \\ twice as many devices, \\ with the same library size and rating. \\ Which service do you prefer, and why?\end{tabular} \\
3rd & Same as the second & Half of first & Same as first & \begin{tabular}[c]{@{}c@{}}The third streaming service supports the\\  same number of devices as the second, has \\ half the library size of the first,\\  but has the same rating as the first. \\ Which service do you prefer, and why?\end{tabular} \\
4th & Same as the second & Same as third & \begin{tabular}[c]{@{}c@{}}One star less\\  than the first\end{tabular} & \begin{tabular}[c]{@{}c@{}}The fourth streaming service supports\\  the same number of devices as the second, \\ has the same library size as the third,\\  but is rated one star less than the first. \\ Which service do you prefer, and why?\end{tabular} \\
 &  &  &  & \begin{tabular}[c]{@{}c@{}}Remember the details of the fourth service. \\ Specific information will be requested later.\end{tabular} \\*
\midrule
\textbf{\begin{tabular}[c]{@{}c@{}}Calendar\\ App\end{tabular}} & \textbf{\begin{tabular}[c]{@{}c@{}}Calendar Syncing \\ Across Devices\end{tabular}} & \textbf{\begin{tabular}[c]{@{}c@{}}Managed \\ Tasks \\ Per Year\end{tabular}} & \textbf{\begin{tabular}[c]{@{}c@{}}Event Privacy \\ Rating\end{tabular}} & \textbf{} \\* 
\midrule
1st & 3 & 2000 & Four star & \begin{tabular}[c]{@{}c@{}}In the following scenario choose from \\ Various Apps recommendations for calendar.\end{tabular} \\
2nd & 2 times 1st & Same as first & Same as first & \begin{tabular}[c]{@{}c@{}}The first calendar app can sync\\  across three devices, \\ manages 2000 tasks per year,\\  and has a 4-star privacy rating. \\ The second app syncs across two devices, \\ manages the same number of tasks,\\  and has the same privacy rating. \\ Which app do you prefer, and why?\end{tabular} \\
3rd & Same as Second & Half of first & Same as first & \begin{tabular}[c]{@{}c@{}}The third app syncs across the same\\  number of devices as the second app, \\ but manages half as\\  many tasks as the first app, \\ with the same privacy rating as the first. \\ Which app do you prefer, and why?\end{tabular} \\
4th & Same as Second & Same as third & \begin{tabular}[c]{@{}c@{}}One star less\\  than the first\end{tabular} & \begin{tabular}[c]{@{}c@{}}The fourth app syncs across the same number \\ of devices as the second, manages the same\\  number of tasks as the third, but has one less \\ star in privacy rating compared to the first. \\ Which app do you prefer, and why?\end{tabular} \\
 &  &  &  & \begin{tabular}[c]{@{}c@{}}Remember the details of the fourth App. \\ Specific information will be requested later.\end{tabular} \\* 
 
\midrule
\textbf{Bank} & \textbf{\begin{tabular}[c]{@{}c@{}}Bank Branch \\ Proximity\end{tabular}} & \textbf{Interest Rates} & \textbf{\begin{tabular}[c]{@{}c@{}}Fee-Free \\ Accounts\\  Rating\end{tabular}} & \textbf{} \\* 
\midrule
1st & 3km & 2000 units & Four & \begin{tabular}[c]{@{}c@{}}In the following scenario choose \\ from various banks recommendations.\end{tabular} \\
2nd & 2 times 1st & Same as first & Same as first & \begin{tabular}[c]{@{}c@{}}The first bank is 3 km away,\\  offers 2000 units of interest, \\ and has a four-star rating for fee-free accounts. \\ The second bank is twice as far away, \\ offers the same amount of interest,\\  and has the same fee-free account rating. \\ Which bank would you prefer, and why?\end{tabular} \\
3rd & Same as the second & Half of first & Same as first & \begin{tabular}[c]{@{}c@{}}The third bank is as far away as the second bank, \\ offers half the amount of interest as \\ the first bank, but has the same fee-free\\  account rating as the first bank. \\ Which bank would you prefer, and why?\end{tabular} \\
4th & Same as the second & Same as third & \begin{tabular}[c]{@{}c@{}}One star less \\ than the first\end{tabular} & \begin{tabular}[c]{@{}c@{}}The fourth bank is as far\\  away as the second bank, \\ offers the same amount of interest as the third bank, \\ but has one star less in fee-free account rating\\  compared to the first bank. \\ Which bank would you prefer, and why?\end{tabular} \\*

\midrule
\textbf{\begin{tabular}[c]{@{}c@{}}Messaging \\ App\end{tabular}} & 
\textbf{\begin{tabular}[c]{@{}c@{}}Number of \\ Simultaneous \\Devices\end{tabular}} & 
\textbf{\begin{tabular}[c]{@{}c@{}}Messages \\ Per Day\end{tabular}} & 
\textbf{\begin{tabular}[c]{@{}c@{}}Security \\ Rating\end{tabular}} & 
\textbf{} \\* 
\midrule

1st & 3 & 2000 & Four star & 
\begin{tabular}[c]{@{}c@{}}In the following scenario, choose from \\ various messaging app recommendations.\end{tabular} \\ 
2nd & 2 times 1st & Same as first & Same as first & 
\begin{tabular}[c]{@{}c@{}}The first messaging app allows access \\ on three devices, supports 2000 messages per day, \\ and has a 4-star security rating. \\ The second app allows access on\\ twice as many devices, \\ supports the same number of messages, \\ and has the same security rating. \\ Which app do you prefer, and why?\end{tabular} \\ 
3rd & Same as the second & Half of first & Same as first & 
\begin{tabular}[c]{@{}c@{}}The third app allows access on the same \\ number of devices as the second app, \\ but supports half as many messages as the first app, \\ with the same security rating as the first. \\ Which app do you prefer, and why?\end{tabular} \\ 
4th & Same as the second & Same as third & One star less than the first & 
\begin{tabular}[c]{@{}c@{}}The fourth app allows access on the \\ same number of devices as the second, \\ supports the same number \\ of messages as the third, \\ but has one less star in \\security rating compared to the first. \\ Which app do you prefer, and why?\end{tabular} \\ 
& & & & 
\begin{tabular}[c]{@{}c@{}}Remember the details\\ of the fourth app, \\ including the number\\ of devices, messages per day, \\ and its security rating.\\ Specific information \\ will be requested later.\end{tabular} \\* 
\bottomrule
\label{tab:appendix-complex-task}\\
\caption{Complex Tasks for prior dialogue.}

\end{longtable}

%% file: others/Sample_planning.tex
\section{A Priori Power Analysis}
\label{sec:appendix-power-analysis}

We conducted a power analysis using G*Power \citep{faul2009statistical} to determine the required sample size, targeting a power of 0.80 to detect a medium effect size ($\omega = 0.3$). This choice reflects our focus on comparing effect sizes across conditions rather than on individual statistical significance. With $\alpha = 0.05$ and 1 degree of freedom \citep{pancholi2009}, the required sample size was estimated to be 42 participants per condition. To ensure robustness, we recruited slightly more participants than required.
For the Status quo experiments, which followed a 2 × 3 design with two dialogue complexity conditions (Simple vs. Complex) and three status quo conditions (Neutral, Status Quo A, Status Quo B), the minimal required sample size was 756 participants (42 per condition).
For the Framing experiments, which followed a 2 × 2 design with two dialogue complexity conditions (Simple vs. Complex) and two framing conditions (Framed vs. Alternatively Framed), the minimal required sample size was 528 participants (44 per condition). Again, we recruited more than this minimum to strengthen the validity of our results.

\subsection{Participant Recruitment, Compensation, and Pre-Registration}
Participants were recruited through Prolific, a widely used platform known for ensuring data quality and participant reliability\citep{prolific2024}. The estimated completion time for the survey was eight minutes, and participants were compensated according to Prolific's recommended minimum rate of \$8 per hour. A total of 1648 participants were recruited, and measures were put in place to prevent duplicate participation. Additionally, demographic information for each participant was obtained from Prolific.
The hypotheses for each experiment and experimental design were preregistered on the Open Science Framework to promote transparency. 
The studies were preregistered on the Open Science Framework (OSF). The preregistration for Framing study is archived at \url{https://doi.org/10.17605/OSF.IO/DPR45}, and the preregistration Status quo study is archived at \url{https://doi.org/10.17605/OSF.IO/PSXVF}.

\subsection{Data Quality and Integrity}
To ensure data integrity and control for familiarity bias, participants reported post-interaction whether they had previously encountered the choice problem and completed a domain-familiarity assessment. Attentiveness was evaluated via a memory-recall task based on the chatbot interaction, and participants were instructed to forgo external aids to preserve the validity of our cognitive-load manipulation. An automated system logged responses as JSON files, which were securely emailed to the authors and a public address to guarantee transparent data collection. The dataset for Framing effect study is available at \url{https://doi.org/10.5281/zenodo.18218753}, and the Status quo bias study is available at \url{https://doi.org/10.5281/zenodo.16541481}.

%% file: tables/framing_demographics.tex
\begin{table*}[htpb]
\caption{Detailed Demographics Split by Choice Problem Type, Framing Condition, and Prior Discourse Conditions. All the Participants are From UK.}
\label{tab:framing_demographics}
\resizebox{\columnwidth}{!}{%
\begin{tabular}{cccccccccc}
\hline
\textbf{\begin{tabular}[c]{@{}c@{}}Prior\\ Discourse\\ Condition\end{tabular}} &
  \textbf{\begin{tabular}[c]{@{}c@{}}Choice\\ Problem\\ Type\end{tabular}} &
  \textbf{\begin{tabular}[c]{@{}c@{}}Choice\\ Problem\\ Condition\end{tabular}} &
  \textbf{\begin{tabular}[c]{@{}c@{}}Sample Size\\        (n)\end{tabular}} &
  \textbf{\begin{tabular}[c]{@{}c@{}}Age\\ (Mean)\end{tabular}} &
  \textbf{\begin{tabular}[c]{@{}c@{}}Age\\ (SD)\end{tabular}} &
  \textbf{\begin{tabular}[c]{@{}c@{}}Sex\\ (Levels)\end{tabular}} &
  \textbf{\begin{tabular}[c]{@{}c@{}}Sex\\ (counts)\end{tabular}} &
  \textbf{\begin{tabular}[c]{@{}c@{}}Ethnicity\\ (levels)\end{tabular}} &
  \textbf{\begin{tabular}[c]{@{}c@{}}Ethnicity\\ (counts)\end{tabular}} \\ \hline
\multirow{6}{*}{\begin{tabular}[c]{@{}c@{}}\\ \\ \\ \\ \\ \\ \\ \textbf{No}  \textbf{Load}\end{tabular}} &
  \multirow{2}{*}{\begin{tabular}[c]{@{}c@{}} \\ Risky Choice\\ Framing\end{tabular}} &
  Framing &
  45 &
  41.6 &
  12.9 &
  \begin{tabular}[c]{@{}c@{}}Female\\ Male\end{tabular} &
  \begin{tabular}[c]{@{}c@{}}23\\ 22\end{tabular} &
  \begin{tabular}[c]{@{}c@{}}White\\ Black\\ Mixed\end{tabular} &
  \begin{tabular}[c]{@{}c@{}}36\\ 5\\ 3\end{tabular} \\ \cline{3-10} 
 &
   &
  \begin{tabular}[c]{@{}c@{}}Alternative\\ Framing\end{tabular} &
  44 &
  42.8 &
  12.1 &
  \begin{tabular}[c]{@{}c@{}}Female\\ Male\end{tabular} &
  \begin{tabular}[c]{@{}c@{}}26\\ 18\end{tabular} &
  \begin{tabular}[c]{@{}c@{}}White\\ Black\\ Asian\end{tabular} &
  \begin{tabular}[c]{@{}c@{}}34\\ 4\\ 3\end{tabular} \\ \cline{2-10} 
 &
  \multirow{2}{*}{\begin{tabular}[c]{@{}c@{}}\\Attribute\\ Framing\end{tabular}} &
  Framing &
  49 &
  43.6 &
  14.1 &
  \begin{tabular}[c]{@{}c@{}}Female\\ Male\end{tabular} &
  \begin{tabular}[c]{@{}c@{}}27\\ 22\end{tabular} &
  \begin{tabular}[c]{@{}c@{}}White\\ Mixed\\ Asian\end{tabular} &
  \begin{tabular}[c]{@{}c@{}}44\\ 4\\ 1\end{tabular} \\ \cline{3-10} 
 &
   &
  \begin{tabular}[c]{@{}c@{}}Alternative\\ Framing\end{tabular} &
  47 &
  46.4 &
  12.3 &
  \begin{tabular}[c]{@{}c@{}}Male\\ Female\end{tabular} &
  \begin{tabular}[c]{@{}c@{}}26\\ 21\end{tabular} &
  \begin{tabular}[c]{@{}c@{}}White\\ Asian\\ Black\end{tabular} &
  \begin{tabular}[c]{@{}c@{}}43\\ 1\\ 1\end{tabular} \\ \cline{2-10} 
 &
  \multirow{2}{*}{\begin{tabular}[c]{@{}c@{}} \\ Goal\\ Framing\end{tabular}} &
  Framing &
  44 &
  44.2 &
  13.6 &
  \begin{tabular}[c]{@{}c@{}}Female\\ Male\end{tabular} &
  \begin{tabular}[c]{@{}c@{}}23\\ 21\end{tabular} &
  \begin{tabular}[c]{@{}c@{}}White\\ Black\\ Mixed\end{tabular} &
  \begin{tabular}[c]{@{}c@{}}42\\ 1\\ 1\end{tabular} \\ \cline{3-10} 
 &
   &
  \begin{tabular}[c]{@{}c@{}}Alternative\\ Framing\end{tabular} &
  44 &
  41.9 &
  14.9 &
  \begin{tabular}[c]{@{}c@{}}Female\\ Male\end{tabular} &
  \begin{tabular}[c]{@{}c@{}}23\\ 21\end{tabular} &
  \begin{tabular}[c]{@{}c@{}}White\\ Asian\\ Black\end{tabular} &
  \begin{tabular}[c]{@{}c@{}}35\\ 3\\ 3\end{tabular} \\ \hline
\multirow{6}{*}{\begin{tabular}[c]{@{}c@{}}\\ \\ \\ \\ \\ \\  \textbf{Load}\end{tabular}} &
  \multirow{2}{*}{\begin{tabular}[c]{@{}c@{}} \\ Risky Choice\\ Framing\end{tabular}} &
  Framing &
  44 &
  40.4 &
  13.6 &
  \begin{tabular}[c]{@{}c@{}}Female\\ Male\end{tabular} &
  \begin{tabular}[c]{@{}c@{}}26\\ 18\end{tabular} &
  \begin{tabular}[c]{@{}c@{}}White\\ Black\\ Asian\end{tabular} &
  \begin{tabular}[c]{@{}c@{}}37\\ 4\\ 3\end{tabular} \\ \cline{3-10} 
 &
   &
  \begin{tabular}[c]{@{}c@{}}Alternative\\ Framing\end{tabular} &
  44 &
  41.9 &
  14.8 &
  \begin{tabular}[c]{@{}c@{}}Female\\ Male\end{tabular} &
  \begin{tabular}[c]{@{}c@{}}25\\ 19\end{tabular} &
  \begin{tabular}[c]{@{}c@{}}White\\ Black\\ Asian\end{tabular} &
  \begin{tabular}[c]{@{}c@{}}33\\ 10\\ 1\end{tabular} \\ \cline{2-10} 
 &
  \multirow{2}{*}{\begin{tabular}[c]{@{}c@{}} \\ Attribute\\ Framing\end{tabular}} &
  Framing &
  46 &
  45.4 &
  14.7 &
  \begin{tabular}[c]{@{}c@{}}Female\\ Male\end{tabular} &
  \begin{tabular}[c]{@{}c@{}}23\\ 23\end{tabular} &
  \begin{tabular}[c]{@{}c@{}}White\\ Asian\end{tabular} &
  \begin{tabular}[c]{@{}c@{}}45\\ 1\end{tabular} \\ \cline{3-10} 
 &
   &
  \begin{tabular}[c]{@{}c@{}}Alternative\\ Framing\end{tabular} &
  53 &
  44.3 &
  12.3 &
  \begin{tabular}[c]{@{}c@{}}Female\\ Male\end{tabular} &
  \begin{tabular}[c]{@{}c@{}}30\\ 23\end{tabular} &
  \begin{tabular}[c]{@{}c@{}}White\\ Mixed\\ Other\end{tabular} &
  \begin{tabular}[c]{@{}c@{}}48\\ 2\\ 2\end{tabular} \\ \cline{2-10} 
 &
  \multirow{2}{*}{\begin{tabular}[c]{@{}c@{}} \\ Goal\\ Framing\end{tabular}} &
  Framing &
  44 &
  45.6 &
  15 &
  \begin{tabular}[c]{@{}c@{}}Male\\ Female\end{tabular} &
  \begin{tabular}[c]{@{}c@{}}23\\ 21\end{tabular} &
  \begin{tabular}[c]{@{}c@{}}White\\ Other\\ Asian\end{tabular} &
  \begin{tabular}[c]{@{}c@{}}41\\ 2\\ 1\end{tabular} \\ \cline{3-10} 
 &
   &
  \begin{tabular}[c]{@{}c@{}}Alternative\\ Framing\end{tabular} &
  44 &
  37.2 &
  13.5 &
  \begin{tabular}[c]{@{}c@{}}Female\\ Male\end{tabular} &
  \begin{tabular}[c]{@{}c@{}}25\\ 19\end{tabular} &
  \begin{tabular}[c]{@{}c@{}}White\\ Asian\\ Black\end{tabular} &
  \begin{tabular}[c]{@{}c@{}}36\\ 3\\ 2\end{tabular} \\ \hline
\end{tabular}%
}
\end{table*}

%% file: tables/sqb_demographics.tex
\begin{table*}[htpb]
\caption{Demographic characteristics of participants across experimental conditions, split by prior dialogue condition (No Load vs. Load), choice problem (Budget Allocation, Investment Decision Making, College Jobs), and choice problem condition (Neutral, Status Quo A, Status Quo B). Reported variables include sample size (n), age (Mean, SD), country distribution (United Kingdom, United States, Ireland), and sex (Female, Male).}
\begin{tabular}{@{}cccccclccclcc@{}}
\toprule
\multirow{2}{*}{\textbf{\begin{tabular}[c]{@{}c@{}}Prior \\ Dialogue\\ Condition\end{tabular}}} &
  \multirow{2}{*}{\textbf{\begin{tabular}[c]{@{}c@{}}Choice \\ Problem\end{tabular}}} &
  \multirow{2}{*}{\textbf{\begin{tabular}[c]{@{}c@{}}Choice \\ Problem \\ Condition\end{tabular}}} &
  \multirow{2}{*}{\textbf{n}} &
  \multicolumn{2}{c}{\textbf{Age}} &
   &
  \multicolumn{3}{c}{\textbf{Country}} &
   &
  \multicolumn{2}{c}{\textbf{Sex}} \\ \cmidrule(lr){5-6} \cmidrule(lr){8-10} \cmidrule(l){12-13} 
 &
   &
   &
   &
  \textbf{Mean} &
  \textbf{SD} &
   &
  \textbf{\begin{tabular}[c]{@{}c@{}}United \\ Kingdom\end{tabular}} &
  \textbf{\begin{tabular}[c]{@{}c@{}}United \\ States\end{tabular}} &
  \textbf{Ireland} &
   &
  \textbf{Female} &
  \textbf{Male} \\ \midrule
\multirow{9}{*}{\textbf{No Load}} & \multirow{3}{*}{BA}  & NEUT & 60 & 42.5 & 13.6 &  & 53 & 6  & 1 &  & 32 & 28 \\ \cmidrule(l){3-13} 
                                  &                      & A    & 51 & 40.9 & 12.0 &  & 44 & 7  & 0 &  & 27 & 24 \\ \cmidrule(l){3-13} 
                                  &                      & B    & 73 & 43.0 & 13.4 &  & 49 & 22 & 2 &  & 39 & 34 \\ \cmidrule(l){2-13} 
                                  & \multirow{3}{*}{IDM} & NEUT & 51 & 35.5 & 12.6 &  & 33 & 18 & 0 &  & 21 & 30 \\ \cmidrule(l){3-13} 
                                  &                      & A    & 58 & 38.9 & 12.1 &  & 26 & 31 & 1 &  & 45 & 13 \\ \cmidrule(l){3-13} 
                                  &                      & B    & 54 & 43.7 & 11.6 &  & 47 & 7  & 0 &  & 10 & 44 \\ \cmidrule(l){2-13} 
                                  & \multirow{3}{*}{CJ}  & NEUT & 76 & 45.8 & 14.7 &  & 60 & 16 & 0 &  & 35 & 41 \\ \cmidrule(l){3-13} 
                                  &                      & A    & 70 & 41.9 & 13.1 &  & 53 & 17 & 0 &  & 35 & 35 \\ \cmidrule(l){3-13} 
                                  &                      & B    & 51 & 37.5 & 12.9 &  & 38 & 13 & 0 &  & 26 & 25 \\ \midrule
\multirow{9}{*}{\textbf{Load}}    & \multirow{3}{*}{BA}  & NEUT & 70 & 42.2 & 13.6 &  & 57 & 12 & 1 &  & 30 & 40 \\ \cmidrule(l){3-13} 
                                  &                      & A    & 59 & 39.7 & 14.6 &  & 48 & 10 & 1 &  & 29 & 30 \\ \cmidrule(l){3-13} 
                                  &                      & B    & 64 & 40.7 & 13.8 &  & 36 & 26 & 2 &  & 31 & 33 \\ \cmidrule(l){2-13} 
                                  & \multirow{3}{*}{IDM} & NEUT & 57 & 40.1 & 14.4 &  & 25 & 30 & 2 &  & 36 & 21 \\ \cmidrule(l){3-13} 
                                  &                      & A    & 51 & 40.9 & 11.4 &  & 26 & 25 & 0 &  & 36 & 15 \\ \cmidrule(l){3-13} 
                                  &                      & B    & 56 & 45.3 & 14.3 &  & 48 & 7  & 1 &  & 24 & 32 \\ \cmidrule(l){2-13} 
                                  & \multirow{3}{*}{CJ}  & NEUT & 80 & 40.7 & 13.0 &  & 48 & 29 & 3 &  & 40 & 40 \\ \cmidrule(l){3-13} 
                                  &                      & A    & 70 & 42.0 & 12.5 &  & 57 & 13 & 0 &  & 33 & 37 \\ \cmidrule(l){3-13} 
                                  &                      & B    & 49 & 42.4 & 11.7 &  & 30 & 19 & 0 &  & 27 & 22 \\ \bottomrule
\end{tabular}

\label{tab:sqb_demographics}
\end{table*}

%% file: others/perceived_cognitive_load.tex
We compared NASA-TLX scores and performance metrics between Simple and Complex dialogue conditions to confirm that complex prior dialogue results in cognitive load in chatbot interactions.

\subsection{Framing}

\begin{figure*}[htpb]
    \centering
    \includegraphics[width=1.0\linewidth]{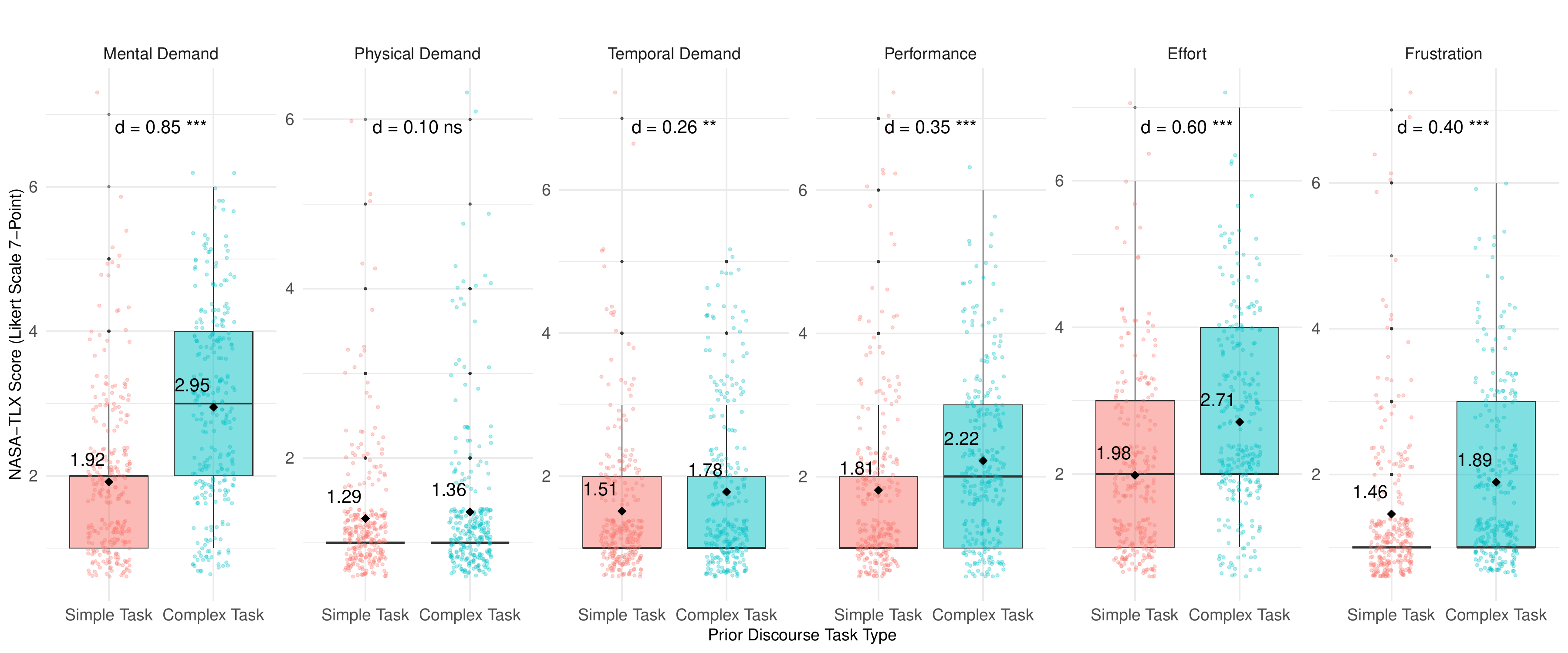}
    \caption{Boxplots, Effect Sizes, Significances ($*** p < 0.001$, ns - no significance), and Means of NASA-TLX Scores for Simple vs. Complex Task Conditions. }
    \label{fig:ntlx}
    \Description{The figure displays boxplots of NASA-TLX scores for frmaing effect across six workload dimensions—Mental Demand, Physical Demand, Temporal Demand, Performance, Effort, and Frustration—comparing Simple Task and Complex Task conditions. For each dimension, the figure shows the distribution of individual participant scores, group means, and the standard boxplot elements. Effect sizes (Cohen's d) and significance levels are annotated above each comparison. The means for the Simple Task and Complex Task are also printed within each box. For example, Mental Demand has a mean of 1.92 for Simple Task and 2.95 for Complex Task (d = 0.85). Effort scores are 1.98 and 2.71 for Simple and Complex Tasks, respectively. The x-axis lists task type, while the y-axis shows the NASA-TLX Likert score (1–7 scale).}
\end{figure*}

We conducted a t-test to analyse differences in participants' NASA-TLX scores across each dimension when performing a Simple Vs Complex Dialogue. This analysis aimed to determine whether the perceived workload varied significantly based on task complexity. Figure~\ref{fig:ntlx} presents the effect sizes (Cohen's d) and significance levels, and box plots for NASA-TLX workload assessment dimensions.

The results show that Mental Demand had the largest differences between Simple and Complex Tasks across all choice problems. As shown in Figure~\ref{fig:ntlx} and supported by the data, the effect size was large (d=0.85, p<0.001), suggesting that participants experienced a much higher mental demand when performing the Complex Task.
Effort also showed a significant difference with a medium effect size (d=0.60, p<0.001), indicating that participants required more effort under the Complex Task condition. Similarly, Frustration showed a small to medium effect size (d=0.40, p<0.001).
For Performance, the effect was also statistically significant (d=0.35, p<0.001), which may indicate a perceived reduction in performance under the Complex Task condition. Temporal Demand showed a smaller yet significant effect (d=0.26, p<0.001). Physical Demand, on the other hand, did not show a statistically significant difference between task types (d=0.10, p=0.25), confirming that physical workload was not a major factor in this study's task design.

Moreover, a Linear mixed-effects models were used to assess whether task domains (with 6 levels) had a random effect on NASA-TLX Mental Demand scores. In both Simple and Complex Tasks, the estimated variance for the Domain as a random intercept was small ($< 0.008$), indicating limited between-domain variability. Most of the variation was attributed to individual differences (residual variance $> 1.8$). These findings suggest that perceived mental demand was largely consistent across task domains. 
Overall, Mental Demand, as highlighted in Figure~\ref{fig:ntlx}, was the most affected by task complexity.

Our survey captured participants' recall performance on the memory component of the Complex Task. Analysis revealed a significant positive correlation between recall accuracy and mental demand on the NASA-TLX (r = 0.13, p = 0.002). This suggests that participants who accurately recalled task information also reported experiencing higher levels of mental demand. This supports the validity of our NASA-TLX survey in-turn indicating that the Complex Task successfully resulted in cognitive load as intended.

\subsection{Status Quo}

\begin{figure}[htpb]
    \centering
    \includegraphics[width=1.0\linewidth]{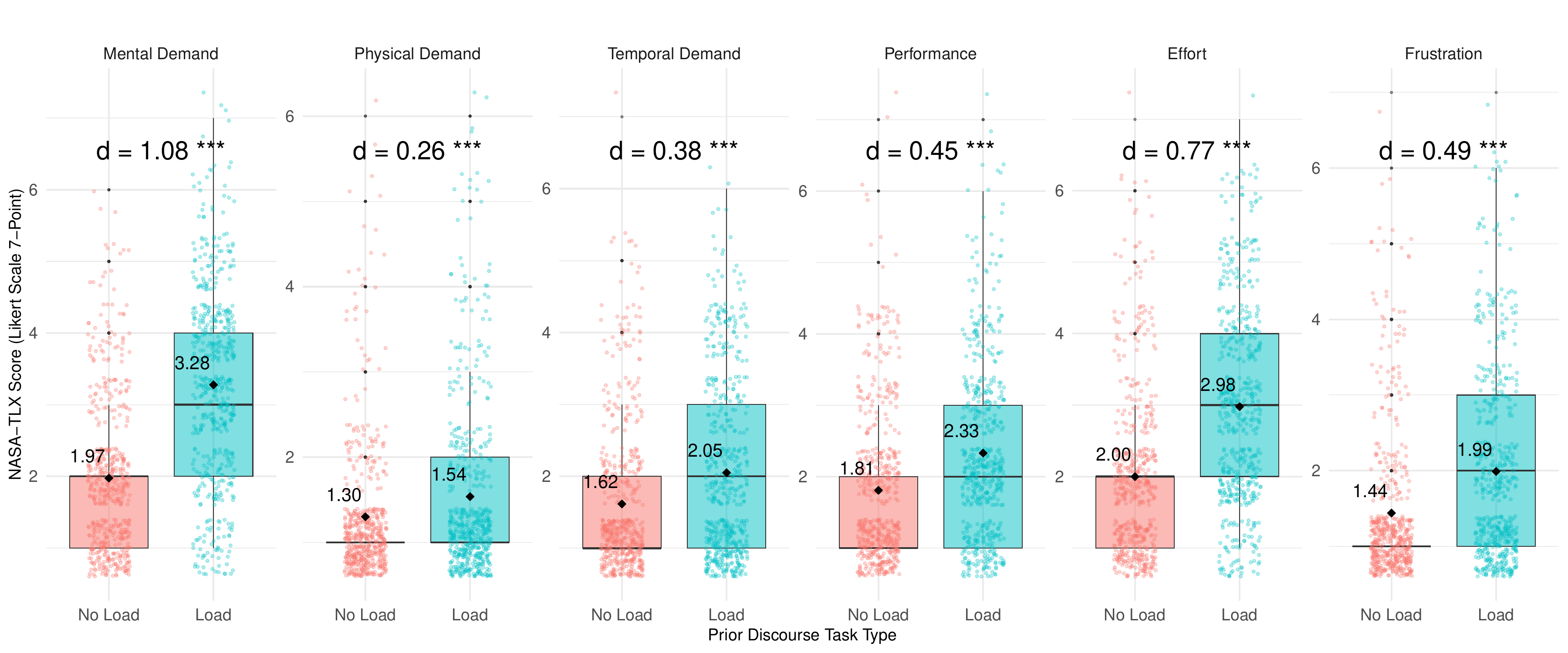}
    \caption{NASA-TLX scores show significantly higher perceived mental demand and effort under the Complex Dialogue condition, confirming the effectiveness of the cognitive load manipulation.}.
    \label{fig:ntlx}
    \Description{The figure presents a set of boxplots for NASA-TLX workload scores for status quo bias, comparing No Load and Load (complex dialogue) conditions. Six workload dimensions are plotted, with individual scores, means, boxplot elements, and annotated effect sizes and significance values. For instance, Mental Demand averages 1.97 in the No Load condition and 3.28 in the Load condition (d = 1.08). Effort is 2.00 for No Load and 2.98 for Load. The x-axis labels the conditions (No Load, Load), while the y-axis represents the NASA-TLX Likert scale.}
\end{figure} 

Figure~\ref{fig:ntlx} presents NASA-TLX scores across six dimensions for Simple Vs Complex Dialogue conditions, including means, Cohen's d, and significance markers ($*** p < .001$). 
Mental Demand increased significantly from $M = 1.97$ to $3.28$ ($p < .001; d = 1.08$). Similarly, Effort also increased from $M = 2.00$ to $2.98$ ($p < .001; d = 0.77$), indicating that the arithmetic, memory, and the length of the dialogue contributed to the perceived cognitive load, respectively.
Performance, Frustration, and Temporal Demand also rose significantly (small–medium d), while Physical Demand showed a minimal effect ($d = 0.26$). These results confirm that complex prior dialogue in chatbot interactions substantially increases perceived cognitive load, specifically in Mental Demand and Effort.

\begin{figure}[htpb]
    \centering
    \includegraphics[width=1.0\linewidth]{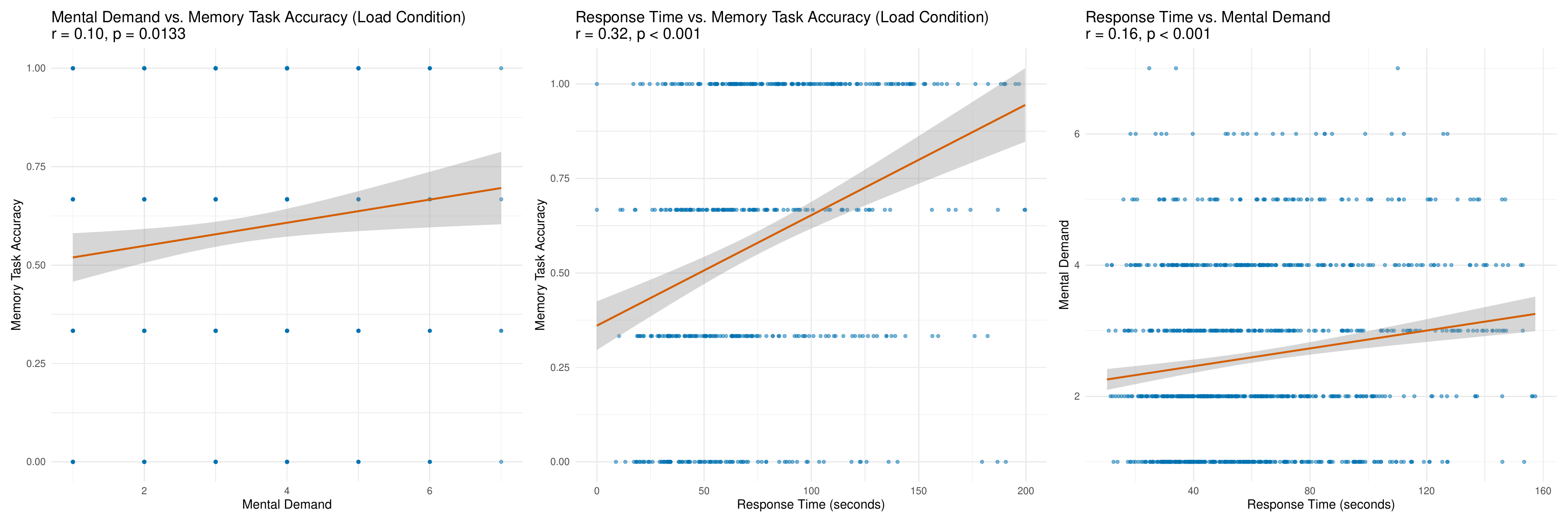}
    \caption{Scatterplots with regression lines showing associations between Mental Demand and Memory Task Accuracy (left), Response Time and Memory Task Accuracy (center), and Response Time and Mental Demand (right), the first two under Load condition. Shaded bands represent 95\% confidence intervals.}
    \label{fig:correls}
    \Description{The figure presents three scatterplots, each overlaid with a linear regression line and 95\% confidence interval bands. The left plot shows the relationship between Mental Demand (x-axis) and Memory Task Accuracy (y-axis) under the Load condition, with the Pearson correlation coefficient (r = 0.10, p = 0.0133) indicated at the top. The center plot depicts Response Time (seconds) on the x-axis and Memory Task Accuracy on the y-axis, also under the Load condition, with a reported correlation of r = 0.32 (p < 0.001). The right plot displays Response Time (x-axis) against Mental Demand (y-axis), with a correlation of r = 0.16 (p < 0.001). Each blue dot represents a participant's data point. The regression lines are shown in orange, with shaded regions indicating the 95\% confidence intervals for the fit.}

\end{figure}

We assessed the alignment of self-reported and behavioral indicators of cognitive load under the Complex Dialogue condition. Figure \ref{fig:correls} shows the correlations among recall accuracy, decision response time, and Mental Demand. Response time and recall accuracy correlated moderately ($r = 0.318$, $p < .001$), while Mental Demand correlated weakly but significantly with both recall accuracy ($r = 0.105$, $p = .013$) and response time ($r = 0.156$, $p < .001$). Participants in the Complex Dialogue also took significantly longer than in the Simple Dialogue ($t = 9.475$, $p < .001$; $d = 0.59$), confirming that increased prior dialogue complexity increased both perceived and measured cognitive load. These converging findings validate our manipulation of cognitive load via prior dialogue.

%% file: others/individual_level_prediction.tex
\section{Individual Level Prediction}
\label{Appendix-indi_preds}
\input{tables/LLM_Preditions}

To interpret Table \ref{tab:llm_predictions}, consider the \textit{Goal Framing} choice problem under the \textit{Complex Prior Dialogue} and \textit{Alternatively Framed} condition as an example. 
When the LLM was provided with only a choice problem, its prediction accuracy was \textbf{13.6\%} (underlined in the Table ~\ref{tab:llm_predictions}). When demographic information and the human-likeness prompt were added (\textit{Without Prior Dialogue} condition), the prediction accuracy increased to \textbf{43.2\%}, suggesting that participant characteristics and role framing contributed to prediction accuracy. However, when the full prior dialogue was included (\textit{With Prior Dialogue} condition), accuracy rose sharply to \textbf{86.4\%}, representing a substantial and statistically significant improvement. 
The t-test results confirm that this increase from the \textit{Choice Problem Only} baseline is significant, as indicated by the corresponding asterisks denoting different levels of \textit{p}-values in the table. 

Table \ref{tab:llm_predictions} presents the accuracy of LLM predictions across different choice problems and dialogue complexity conditions. We found three different cases in the results. In \textbf{Case 1}, no significant differences were observed between the three LLM prediction conditions (Choice Problem Only, Without Prior Dialogue, and With Prior Dialogue), suggesting that prior dialogue did not substantially affect prediction performance. Risky Choice Framing and Attribute Framing fall into this category. 
In the \textbf{Case 2}, certain choice problems exhibited a significant improvement in accuracy when prior dialogue or demographic information was included, indicating that prior dialogue played an important role in aligning model predictions with human decisions. Goal Framing and Investment Decision Making (SQB) falls into this category. In \textbf{Case 3}, there were instances where no significant difference was observed across LLM prediction conditions, yet prediction accuracies remained high across the conditions. Budget Allocation and College Jobs in Status Quo bias come under this category.

\begin{figure}
    \centering
    \includegraphics[width=0.91\linewidth]{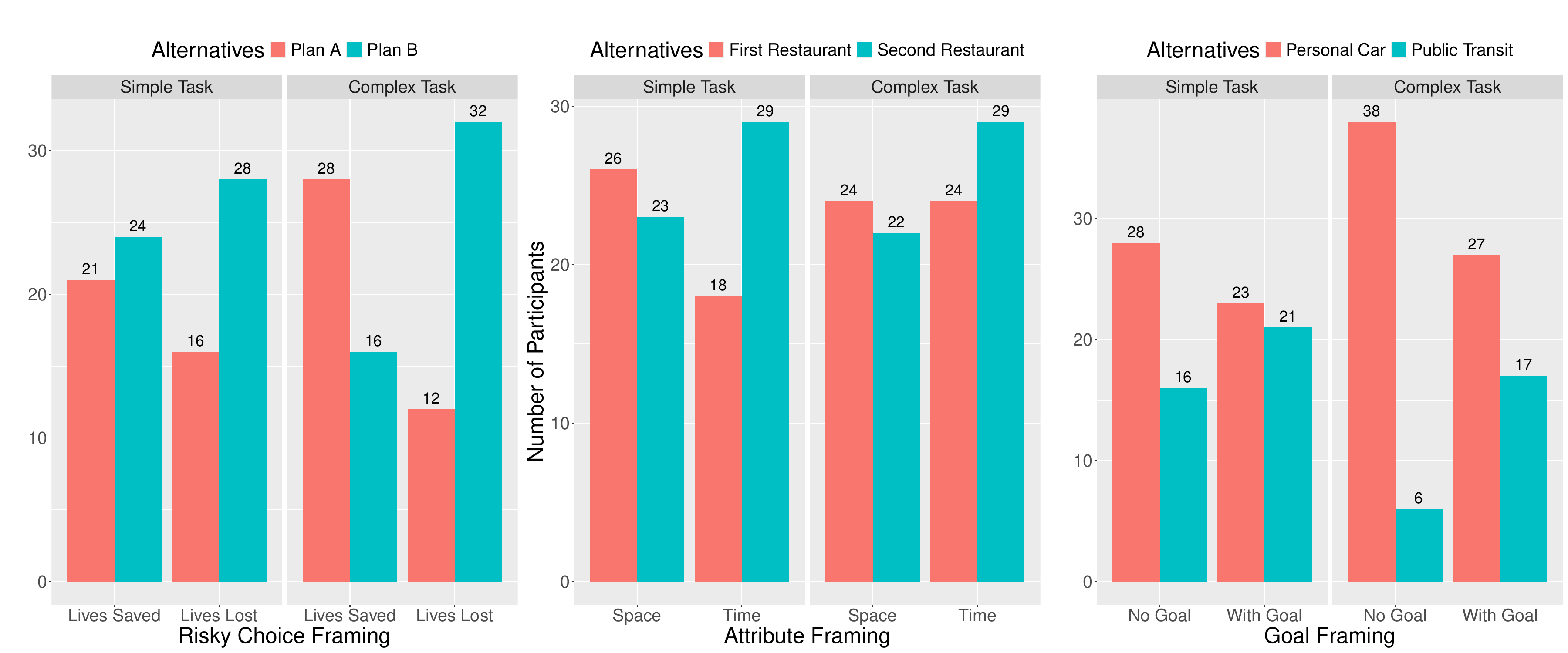}
    \caption{The figure gives the count of Human Participants Choice Selection between Alternative for Framing effect Choice problems.}
    \label{fig:framing_effect_cpa_count}
    \Description{The figure presents grouped bar charts showing the number of human participants selecting each alternative in three framing effect choice problems: Risky Choice Framing, Attribute Framing, and Goal Framing. For each scenario, results are displayed for both Simple Task and Complex Task conditions. In the Risky Choice Framing panel, bars represent participant choices between Plan A and Plan B under ``Lives Saved'' and ``Lives Lost'' framings. In the Attribute Framing panel, choices between the First Restaurant and Second Restaurant are shown under ``Space'' and ``Time'' framings. In the Goal Framing panel, participant selections are displayed for Personal Car versus Public Transit, with and without goal framing. Bar heights indicate the count of participants for each alternative, with the values labeled on each bar, and color coding corresponds to the available alternatives in each scenario.}
\end{figure}

In the Goal Framing scenario, participants were asked to choose between using a personal car or public transit. In the \textit{Framing} condition, participants received additional information emphasizing the environmental benefits of public transit, while in the \textit{Alternative Framing} condition, no such information was provided. When the LLM was given only demographic information (\textit{Without Prior Dialogue}), it selected public transit 31 out of 44 times. In contrast, under the \textit{Choice Problem Only} condition, the model chose public transit in all 44 cases, demonstrating rational behavior aligned with environmental goals. 
However, when full Prior Dialogue was included, the LLM selected the personal car in all 44 cases, closely mirroring human behavior (38 out of 44 participants also chose the personal car. Refer to  \textit{No goal} subplot in \textit{Goal Framing - Complex task} bar plot in Figure ~\ref{fig:framing_effect_cpa_count} ). 
This suggests that prior dialogue significantly influenced the model's prediction, changing its prediction from public transport to personal car, aligning it more closely with actual human responses.
A similar pattern was observed in the Investment Decision Making scenario, where LLM predictions with prior dialogue were significantly more accurate (Case 2). 
However, in other tasks, such as Risky Choice Framing (RCF), Attribute Framing (ATF), the inclusion of prior dialogue had little to no effect on prediction accuracy (Case 1). These mixed results indicate that while prior conversational context can play a critical role in certain choice problems, its influence is not uniform, highlighting the need for further investigation.

While LLM prediction on some choice problems showed no difference across the prediction conditions (Case 1), some instances exhibited consistently high prediction accuracy across all LLM prediction conditions (Case 3). For example, in the Budget Allocation scenario, the model demonstrated strong predictive performance even in the Choice Problem Only condition. Our analysis revealed that human participants displayed a clear preference for the \textit{50–50} alternative in the choice problem over the \textit{60–40} alternative, indicating an inherent bias toward equality even in the neutral framing. When the \textit{50–50} allocation was framed as the status quo, 117 out of 137 participants chose to retain it, and the LLM in all prediction conditions chose the \textit{50–50} alternative for all 137 participants. Although the accuracy was around 80\%, this result can be misleading because the only alternative predicted and selected was the \textit{50–50} option. Conversely, when the \textit{60–40} allocation served as the status quo, both human and LLM choices became more evenly split (56 vs. 54 for humans; 67 vs. 43 for LLM predictions), reflecting a status quo bias (Case 3) requiring further investigation into biases at the sample level. Similar trends were observed in the College Jobs scenario, where participants and LLMs consistently favored College A when it was presented as the status quo option.

%% file: tables/LLM_Preditions.tex
\begin{table*}[htpb]
\centering
\caption{GPT4.1 prediction accuracy across choice problems and dialogue conditions in HL1 condition. The table reports accuracy (with 95\% confidence intervals) of LLM predictions under three conditions: (i) Choice Problem Only (no demographics, no prompt, no prior dialogue), (ii) Without Prior Dialogue (includes demographics and human-likeness prompt), and (iii) With Prior Dialogue (includes full dialogue history). Asterisks (*, **, ***) indicate statistical significance compared to the Choice Problem Only condition (p < .05, .01, .001, respectively), and $\dagger$ denotes marginal significance (p < .10).}
\label{tab:llm_predictions}
\resizebox{\textwidth}{!}{%
\begin{tabular}{@{}lccccccc@{}}
\toprule
\multirow{2}{*}{\textbf{Index}} & \multirow{2}{*}{\textbf{Choice Problem}} & \multirow{2}{*}{\textbf{\begin{tabular}[c]{@{}c@{}}Prior \\ Dialogue\end{tabular}}} & \multirow{2}{*}{\textbf{\begin{tabular}[c]{@{}c@{}}Choice Problem\\ Condition\end{tabular}}} & \multirow{2}{*}{\textbf{n}} & \multicolumn{3}{c}{\textbf{Accuracy}} \\ \cmidrule(l){6-8} 
 &  &  &  &  & \textbf{\begin{tabular}[c]{@{}c@{}}Choice Problem \\ Only\end{tabular}} & \textbf{\begin{tabular}[c]{@{}c@{}}Without \\ Prior Dialogue\end{tabular}} & \textbf{\begin{tabular}[c]{@{}c@{}}With \\ Prior Dialogue\end{tabular}} \\ \midrule
1 & \multirow{4}{*}{Risky Choice} & \multirow{2}{*}{Simple} & Framing & 45 & 0.467 [0.321, 0.612] & 0.467 [0.321, 0.612] & 0.444 [0.299, 0.59] \\ \cmidrule(l){4-8} 
2 &  &  & Alternative Framing & 44 & 0.409 [0.264, 0.554]$^\dagger$ & 0.614 [0.47, 0.758] & 0.591 [0.446, 0.736] \\ \cmidrule(l){3-8} 
3 &  & \multirow{2}{*}{Complex} & Framing & 44 & 0.636 [0.494, 0.779] & 0.636 [0.494, 0.779] & 0.636 [0.494, 0.779] \\ \cmidrule(l){4-8} 
4 &  &  & Alternative Framing & 44 & 0.568 [0.422, 0.715]* & 0.75 [0.622, 0.878] & 0.727 [0.596, 0.859] \\ \cmidrule(l){2-8} 
5 & \multirow{4}{*}{Attribute} & \multirow{2}{*}{Simple} & Framing & 49 & 0.531 [0.391, 0.67] & 0.531 [0.391, 0.67] & 0.469 [0.33, 0.609] \\ \cmidrule(l){4-8} 
6 &  &  & Alternative Framing & 47 & 0.383 [0.244, 0.522] & 0.404 [0.264, 0.545] & 0.447 [0.305, 0.589] \\ \cmidrule(l){3-8} 
7 &  & \multirow{2}{*}{Complex} & Framing & 46 & 0.522 [0.377, 0.666] & 0.522 [0.377, 0.666] & 0.391 [0.25, 0.532] \\ \cmidrule(l){4-8} 
8 &  &  & Alternative Framing & 53 & 0.453 [0.319, 0.587] & 0.472 [0.337, 0.606] & 0.585 [0.452, 0.718] \\ \cmidrule(l){2-8} 
9 & \multirow{4}{*}{Goal} & \multirow{2}{*}{Simple} & Framing & 44 & 0.477 [0.33, 0.625] & 0.477 [0.33, 0.625] & 0.477 [0.33, 0.625] \\ \cmidrule(l){4-8} 
10 &  &  & Alternative Framing & 44 & 0.364 [0.221, 0.506]$^\dagger$ & 0.432 [0.285, 0.578] & 0.568 [0.422, 0.715] \\ \cmidrule(l){3-8} 
11 &  & \multirow{2}{*}{Complex} & Framing & 44 & 0.386 [0.242, 0.53]*** & 0.523 [0.375, 0.67]* & 0.591 [0.446, 0.736] \\ \cmidrule(l){4-8} 
12 &  &  & Alternative Framing & 44 & \underline{0.136 [0.035, 0.238]***} & 0.432 \underline{[0.285, 0.578]***} & 0.864 [0.762, 0.965] \\ \cmidrule(l){2-8} 
13 & \multirow{6}{*}{Budget Allocation} & \multirow{3}{*}{Simple} & NEUT & 60 & 0.767 [0.66, 0.874] & 0.767 [0.66, 0.874] & 0.75 [0.64, 0.86] \\ \cmidrule(l){4-8} 
14 &  &  & Status Quo A & 51 & 0.49 [0.353, 0.627] & 0.431 [0.295, 0.567] & 0.471 [0.334, 0.608] \\ \cmidrule(l){4-8} 
15 &  &  & Status Quo B & 73 & 0.863 [0.784, 0.942] & 0.863 [0.784, 0.942] & 0.863 [0.784, 0.942] \\ \cmidrule(l){3-8} 
16 &  & \multirow{3}{*}{Complex} & NEUT & 70 & 0.8 [0.706, 0.894] & 0.8 [0.706, 0.894] & 0.786 [0.69, 0.882] \\ \cmidrule(l){4-8} 
17 &  &  & Status Quo A & 59 & 0.475 [0.347, 0.602] & 0.525 [0.398, 0.653] & 0.525 [0.398, 0.653] \\ \cmidrule(l){4-8} 
18 &  &  & Status Quo B & 64 & 0.844 [0.755, 0.933] & 0.844 [0.755, 0.933] & 0.844 [0.755, 0.933] \\ \cmidrule(l){2-8} 
19 & \multirow{6}{*}{Investment} & \multirow{3}{*}{Simple} & NEUT & 51 & 0.314 [0.186, 0.441]*** & 0.667 [0.537, 0.796]$^\dagger$ & 0.725 [0.603, 0.848] \\ \cmidrule(l){4-8} 
20 &  &  & Status Quo A & 58 & 0.259 [0.146, 0.371]*** & 0.483 [0.354, 0.611]** & 0.741 [0.629, 0.854] \\ \cmidrule(l){4-8} 
21 &  &  & Status Quo B & 54 & 0.333 [0.208, 0.459]*** & 0.704 [0.582, 0.825] & 0.759 [0.645, 0.873] \\ \cmidrule(l){3-8} 
22 &  & \multirow{3}{*}{Complex} & NEUT & 57 & 0.298 [0.179, 0.417]*** & 0.632 [0.506, 0.757]** & 0.737 [0.623, 0.851] \\ \cmidrule(l){4-8} 
23 &  &  & Status Quo A & 51 & 0.196 [0.087, 0.305]*** & 0.49 [0.353, 0.627]*** & 0.804 [0.695, 0.913] \\ \cmidrule(l){4-8} 
24 &  &  & Status Quo B & 56 & 0.25 [0.137, 0.363]*** & 0.714 [0.596, 0.833] & 0.786 [0.678, 0.893] \\ \cmidrule(l){2-8} 
25 & \multirow{6}{*}{College Jobs} & \multirow{3}{*}{Simple} & NEUT & 76 & 0.605 [0.495, 0.715] & 0.539 [0.427, 0.652] & 0.487 [0.374, 0.599] \\ \cmidrule(l){4-8} 
26 &  &  & Status Quo A & 70 & 0.686 [0.577, 0.794] & 0.7 [0.593, 0.807] & 0.686 [0.577, 0.794] \\ \cmidrule(l){4-8} 
27 &  &  & Status Quo B & 51 & 0.569 [0.433, 0.705] & 0.51 [0.373, 0.647] & 0.529 [0.392, 0.666] \\ \cmidrule(l){3-8} 
28 &  & \multirow{3}{*}{Complex} & NEUT & 80 & 0.588 [0.48, 0.695] & 0.462 [0.353, 0.572] & 0.412 [0.305, 0.52] \\ \cmidrule(l){4-8} 
29 &  &  & Status Quo A & 70 & 0.714 [0.608, 0.82] & 0.714 [0.608, 0.82] & 0.714 [0.608, 0.82] \\ \cmidrule(l){4-8} 
30 &  &  & Status Quo B & 49 & 0.51 [0.37, 0.65] & 0.571 [0.433, 0.71] & 0.571 [0.433, 0.71] \\ \bottomrule
\end{tabular}%
}
\end{table*}

%% file: others/participant_chat_validation.tex
\section{Participant Dialogue Validation}
\label{appendix-dialogue-validation}
To ensure the validity of participant dialogue, we implemented several safeguards including memory recall tasks, attention checks, and response time analyses. Below we describe in detail the validation procedures and findings.

\subsection{Response Time Analysis}

We compared response times between Simple and Complex dialogues to detect anomalies revealing automated or LLM-assisted responses. Using AI assistance would likely result in unusually fast or uniform responses as suggested by ~\citet{prolific2024}~\footnote{\url{https://researcher-help.prolific.com/en/article/2a85ea}}.

\subsection{Sample Sizes}

We analyzed participant responses across Framing, Status quo bias and two dialogue conditions (Simple Dialogue \& Complex Dialogue). The number of participants per group was:

\begin{itemize}
    \item Simple Dialogue (Framing): N = 273
    \item Complex Dialogue (Framing): N = 275
    \item Simple Dialogue (Status quo): N = 544
    \item Complex Dialogue (Status quo): N = 556
    \item Total = 1,648 participants
\end{itemize}

\subsubsection{Framing Experiments Response Times:}
\begin{itemize}
    \item Simple dialogue average = 13.20s
    \item Complex dialogue average = 24.45s
    \item t = –15.709, p < 0.001
    \item Mann-Whitney U = 2,629,877, p < 0.001
\end{itemize}

\subsubsection{Status quo Experiments Response Times:}
\begin{itemize}
    \item Simple dialogue average  = 14.99s
    \item Complex dialogue average = 27.59s
    \item t = –19.293, p < 0.001
    \item Mann-Whitney U = 10,707,455.5, p < 0.001
\end{itemize}

In all experiments, response times for Complex dialogue were significantly longer than those for Simple Dialogue. If participants were using LLMs (e.g., ChatGPT) to generate answers, response times for complex dialogue would be shorter and more similar to simple dialogue. Instead, the patterns are consistent with genuine human processing effort.

\subsection{Response Time–Length Correlation}

Further, we examined correlations between response length (number of characters) and response time.

\subsubsection{Framing Experiments:}

   \begin{figure}[htpb]
       \centering
       \includegraphics[width=0.5\linewidth]{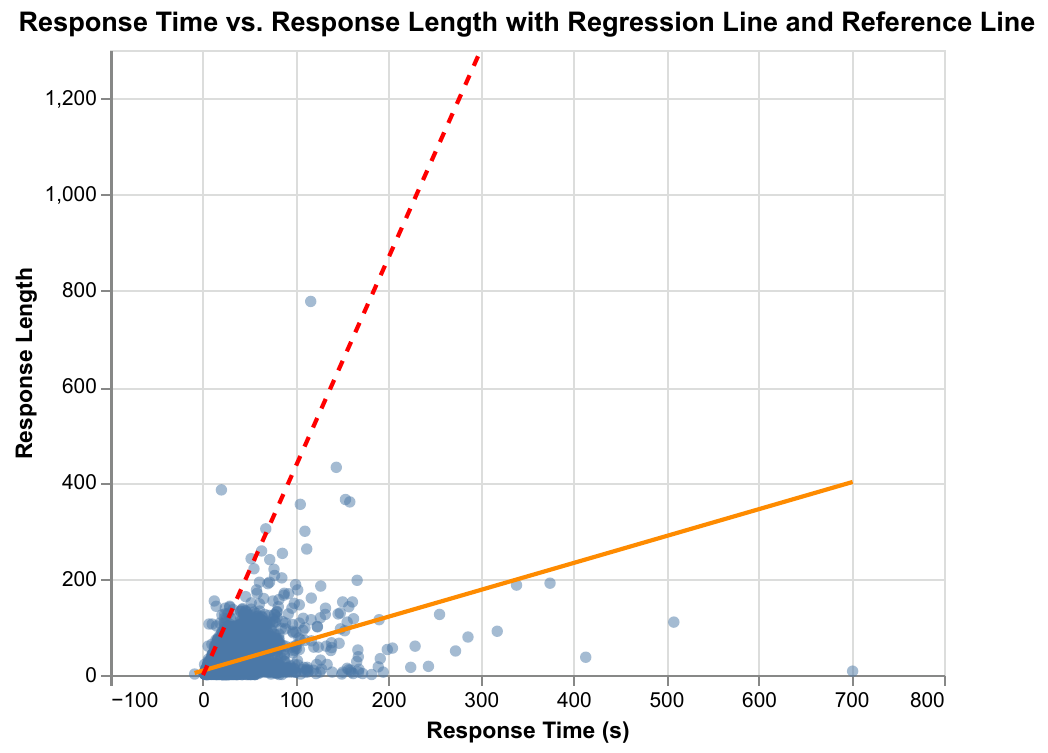}
       \caption{Framing : Correlation Between Response Length (Chars) and Response Times (s). The bold line is the regression line. The dotted line is the average human typing speed (260 characters per minute).}
       \label{fig:corel_rt_rl_framing}
       \Description{The figure displays a scatterplot of response length (in characters) versus response time (in seconds) for the Framing condition. Each point represents an individual participant's response. The plot includes a bold orange regression line indicating the observed relationship between response length and response time, and a red dashed line representing the average human typing speed (260 characters per minute) as a reference. The axes are labeled accordingly, with response time on the x-axis and response length on the y-axis.}
   \end{figure}

   \begin{itemize}
       \item Pearson's r = 0.492, p < .001
       \item Spearman's $\rho$ = 0.689, p < .001
   \end{itemize}

\subsubsection{Status quo Experiments:}

   \begin{figure}[htpb]
       \centering
       \includegraphics[width=0.5\linewidth]{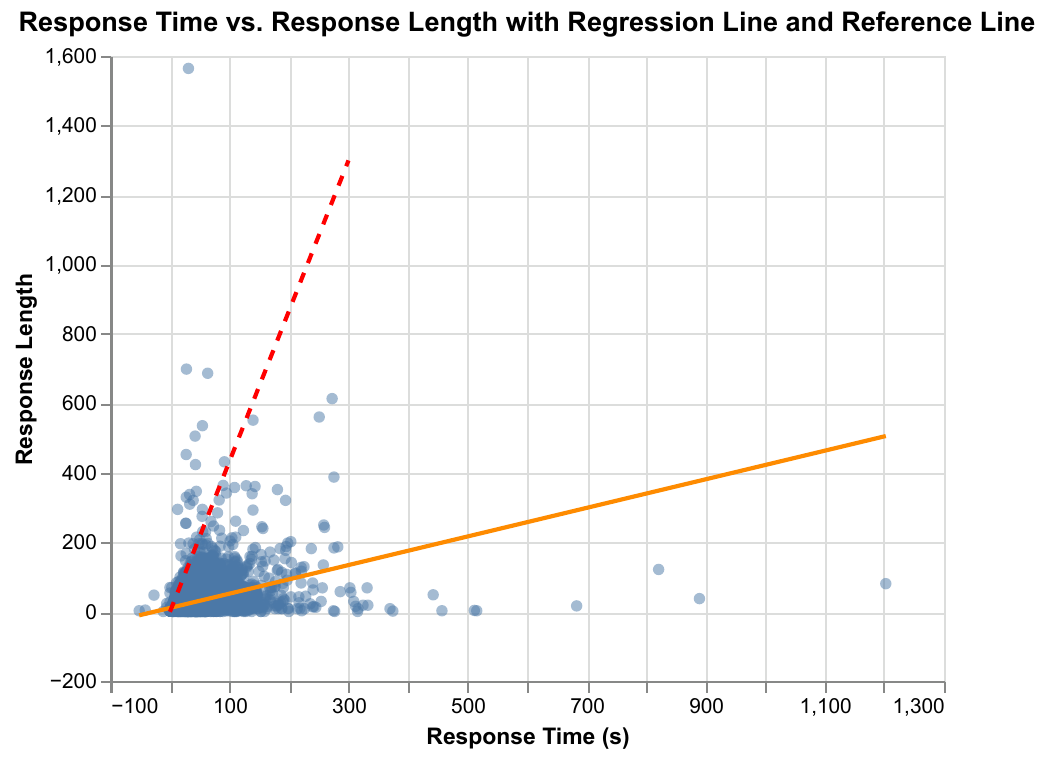}
       \caption{Status quo : Correlation Between Response Length (Chars) and Response Times (s). The bold line is the regression line. The dotted line is the average human typing speed (260 characters per minute).}
       \label{fig:corel_rt_rl_sqb}
       \Description{The figure displays a scatterplot of response length (in characters) versus response time (in seconds) for the Status Quo condition. Each point represents an individual participant's response. The plot includes a bold orange regression line indicating the observed relationship between response length and response time, and a red dashed line representing the average human typing speed (260 characters per minute) as a reference. The axes are labeled with response time on the x-axis and response length on the y-axis.}
   \end{figure}

      \begin{itemize}
       \item Pearson's r = 0.396, p < .001
       \item Spearman's $\rho$ = 0.698, p < .001
   \end{itemize}

Both Framing and Status quo analyses showed strong positive correlations. Longer responses were associated with longer response times, consistent with natural typing and reading behavior~\citep{dhakal2018observations}.  The dotted line in the Figures ~\ref{fig:corel_rt_rl_framing} \& ~\ref{fig:corel_rt_rl_sqb} indicates the average human typing speed (220 - 260 characters per minute).
Any participant falling on the left hand side of the dotted line, that is, if response lengths were unusually large and response times were small, which warrants further investigation.

\subsection{Manual Inspection of Outliers}

Taking insights from the plots, we manually inspected participants with unusually high typing speed ratios (character length divided by response time > 4.3 chars/sec this include time to read the choice problems). While many flagged cases reflected concise but fast human answers, some participants' responses clearly showed characteristics of AI-generated text (e.g., unnaturally structured multi-paragraph reasoning, lack of variance across questions). However, these responses were less in number (n=4).

\subsection{Summary}
\begin{enumerate}
    \item Participants in Complex dialogue consistently exhibit longer response times compared to those in Simple Dialogue.
    \item Statistical tests (t-test and Mann-Whitney U test) confirm that the differences in response times between Simple and Complex dialogues are significant (p < 0.05).
    \item The average response times further highlight this trend, with participants in Complex dialogue taking notably longer to respond than Simple dialogue.
    \item Correlation analyses (recommendation by ~\cite{prolific2024}) reveal a positive relationship between response time and response length, suggesting that longer responses tend to take more time to compose.
    \item The scatter plots (Figures ~\ref{fig:corel_rt_rl_framing} \& ~\ref{fig:corel_rt_rl_sqb}) with regression lines illustrate this correlation, with a reference line indicating typical human typing speed (52 wpm/ 260 characters) as observed by ~\citet{dhakal2018observations}.
    \item Overall, the evidence from the data aligns with expected human behavior rather than AI usage.
\end{enumerate}

%% file: tables/prompt_code.tex
\section{Prompt for Complex Dialogue - College Jobs Scenario Condition}
\label{sec:appendix-prompt}
\scriptsize
\begin{lstlisting}[breaklines=true]
System Prompt for Specialized GPT with Initial Engagement

Format Instruction: Avoid any kinds of text formatting. Put the whole text in plain. Don't change the content of the question at any cost.

Greeting and Introduction:
 "Hello! I'm here to understand your preferences through various Scenarios."

Engagement with Random Natural Questions [Don't change the question at any cost]:
 Question 1: "Shall we start?"
 Question 2: "Do you have a specific budget for the home?"
 Question 3: "Are you looking for a home in a specific location?"
 Question 4: "Do you need more than 3 bedrooms?"
 Question 5: "Is having 2 or more bathrooms important to you?"
 Question 6: "Are you looking specifically for a detached house? Please enter "I don't know" only."
 Question 7: "Do you prefer homes larger than 2000 square feet?"

Wait for responses to each question. Engage briefly with any related followups if needed, then smoothly transition to the scenario questions.

Transition to Scenario Questions:
 "Thanks for sharing! Now, let's get started with some specific scenarios to understand your preferences."

Behavioral Guidelines:
 Task Focused: My role is to guide you through a series of two scenarios to understand your preferences. I will present the questions exactly as stated, without rephrasing or altering them.
 Handling Inputs: I will wait for your response after each question. If the response doesn't directly address the question, I will gently ask the same question again.
 Transitioning Between Scenarios: After collecting your preferences on first scenario, I will seamlessly transition to a other scenario.

Scenario Questions:

    First Scenario:

        "The first property has three bedrooms, 2000 square feet, and a 4-star rating. The second property has twice the number of bedrooms and with the same size and rating. Which property is better, and why?"

        Wait for response. 

        "The third property has the same number of bedrooms as the second one but is half the size of the first one, with the same rating as the first. Which property is better, and why?"

        Wait for response.

        "The fourth property has the same number of bedrooms as the second, the same size as the third, but one less star rating than the first. Which property is better, and why?"

        Wait for response.

    Second Scenario:

        Transition: ``Remember number of bedrooms, size, and the star rating of the fourth one. Now, let's move on to a different scenario.''
        ``You are currently an assistant professor at College A in the east coast. Recently, you have been approached by colleague at other university with job opportunity.''

When evaluating teaching job offers, people typically consider the salary, the reputation of the school, the location of the school, and the likelihood of getting tenure (tenure is permanent job contract that can only be terminated for cause or under extraordinary circumstances).
Your choices are:
[Instruction: Strictly use bullet points to present the below options.]
Remain at College A: east coast, very prestigious school, high salary, fair chance of tenure.
Move to College B: west coast, low prestige school, high salary, good chance of tenure."
[Instruction: DO NOT ASK WHY FOR THE ABOVE QUESTION. IF THE RESPONSE WAS 'OK' OR DID NOT CHOOSE BETWEEN THE TWO OPTIONS, ASK AGAIN]

Error Handling:
 For any unrelated or unclear inputs, I will politely ask the same question again until I receive a valid response.
 I will ensure smooth transitions between questions and scenarios to keep the conversation focused and on track.

Ending the Interaction:
 After collecting all the responses, I will thank the user: "Thank you have a nice day. You will be redirected to next page in 5 Seconds"

\end{lstlisting}

%% file: bib.bib
@String{Computing = "Computing" }

@String{Computer = "{IEEE} Computer" }

@String{Academic = "Academic Press" }

@String{Springer = "Springer-Verlag" }

@incollection{yeung2019hypernudge,
  title={‘Hypernudge’: Big Data as a mode of regulation by design},
  author={Yeung, Karen},
  booktitle={The social power of algorithms},
  pages={118--136},
  year={2019},
  publisher={Routledge}
}

@inproceedings{dhakal2018observations,
title={Observations on typing from 136 million keystrokes},
author={Dhakal, Vivek and Feit, Anna Maria and Kristensson, Per Ola and Oulasvirta, Antti},
booktitle={Proceedings of the 2018 CHI conference on human factors in computing systems},
pages={1--12},
year={2018}
}

@misc{GitHubCopilotDocs2025,
  title        = {GitHub Copilot Documentation},
  author       = {{GitHub, Inc.}},
  year         = {2025},
  howpublished = {\url{https://docs.github.com/en/copilot}},
  note         = {Accessed: 2025-12-27},
}

@inproceedings{wolfalignment2024,
author = {Wolf, Yotam and Wies, Noam and Avnery, Oshri and Levine, Yoav and Shashua, Amnon},
title = {Fundamental limitations of alignment in large language models},
year = {2024},
publisher = {JMLR.org},
booktitle = {Proceedings of the 41st International Conference on Machine Learning},
articleno = {2176},
numpages = {34},
location = {Vienna, Austria},
series = {ICML'24}
}

@article{grossmann2023ai,
  title={AI and the transformation of social science research},
  author={Grossmann, Igor and Feinberg, Matthew and Parker, Dawn C and Christakis, Nicholas A and Tetlock, Philip E and Cunningham, William A},
  journal={Science},
  volume={380},
  number={6650},
  pages={1108--1109},
  year={2023},
  publisher={American Association for the Advancement of Science}
}

@article{brand2023using,
  title={Using LLMs for market research},
  author={Brand, James and Israeli, Ayelet and Ngwe, Donald},
  journal={Harvard business school marketing unit working paper},
  number={23-062},
  year={2023}
}

@inproceedings{miehling2025evaluating,
  title={Evaluating the prompt steerability of large language models},
  author={Miehling, Erik and Desmond, Michael and Ramamurthy, Karthikeyan Natesan and Daly, Elizabeth M and Varshney, Kush R and Farchi, Eitan and Dognin, Pierre and Rios, Jesus and Bouneffouf, Djallel and Liu, Miao and others},
  booktitle={Proceedings of the 2025 Conference of the Nations of the Americas Chapter of the Association for Computational Linguistics: Human Language Technologies (Volume 1: Long Papers)},
  pages={7874--7900},
  year={2025}
}

@book{arnold1998reference,
  title={Reference form and discourse patterns},
  author={Arnold, Jennifer E},
  year={1998},
  publisher={Stanford University}
}

@inproceedings{ultes2020complexity,
  title={On the Complexity in Task-oriented Spoken Dialogue Systems},
  author={Ultes, Stefan and Maier, Wolfgang},
  booktitle={Proceedings of the 2nd Conference on Conversational User Interfaces},
  pages={1--4},
  year={2020}
}

@misc{grammarly,
  author       = {{Grammarly Inc.}},
  title        = {Grammarly},
  howpublished = {\url{https://www.grammarly.com}},
  note         = {Accessed: 2025-07-30},
  year         = {2025}
}

@misc{openai2024chatgpt4o,
  author       = {OpenAI},
  title        = {ChatGPT-4o [Computer software]},
  year         = {2024},
  howpublished = {\url{https://openai.com/chatgpt}},
  note         = {Accessed: 2025-08-05}
}

@article{chen1985discourse,
  title={Discourse-T. Giv{\'o}n (ed.), Topic continuity in discourse: A quantitative cross-language study.(Typological Studies in Language, vol. 3.) Amsterdam and Philadelphia: John Benjamins, 1983. Pp. 492.},
  author={Chen, Ping},
  journal={Language in Society},
  volume={14},
  number={3},
  pages={410--414},
  year={1985},
  publisher={Cambridge University Press}
}

@article{van1983strategies,
  title={Strategies of discourse comprehension},
  author={Van Dijk, Teun Adrianus and Kintsch, Walter and others},
  year={1983},
  publisher={Academic press New York}
}

@article{gibson1998linguistic,
  title={Linguistic complexity: Locality of syntactic dependencies},
  author={Gibson, Edward},
  journal={Cognition},
  volume={68},
  number={1},
  pages={1--76},
  year={1998},
  publisher={Elsevier}
}

@inproceedings{
binz2023turning,
title={Turning large language models into cognitive models},
author={Marcel Binz and Eric Schulz},
booktitle={The Twelfth International Conference on Learning Representations},
year={2024},
url={https://openreview.net/forum?id=eiC4BKypf1}
}

@inproceedings{liu2025rationality,
  title={Large Language Models Assume People Are More Rational Than We Really Are},
  author={Liu, Ryan and Geng, Jiayi and Peterson, Joshua C. and Sucholutsky, Ilia and Griffiths, Thomas L.},
  booktitle={Proceedings of the International Conference on Learning Representations (ICLR)},
  year={2025}
}

@inproceedings{malberg2024comprehensive,
    title = "A Comprehensive Evaluation of Cognitive Biases in {LLM}s",
    author = "Malberg, Simon  and
      Poletukhin, Roman  and
      Schuster, Carolin M.  and
      Groh, Georg",
    editor = {H{\"a}m{\"a}l{\"a}inen, Mika  and
      {\"O}hman, Emily  and
      Bizzoni, Yuri  and
      Miyagawa, So  and
      Alnajjar, Khalid},
    booktitle = "Proceedings of the 5th International Conference on Natural Language Processing for Digital Humanities",
    month = may,
    year = "2025",
    address = "Albuquerque, USA",
    publisher = "Association for Computational Linguistics",
    url = "https://aclanthology.org/2025.nlp4dh-1.50/",
    doi = "10.18653/v1/2025.nlp4dh-1.50",
    pages = "578--613",
    ISBN = "979-8-89176-234-3"
}

@article{ying2025benchmarking,
  title={On benchmarking human-like intelligence in machines},
  author={Ying, Lance and Collins, Katherine M and Wong, Lionel and Sucholutsky, Ilia and Liu, Ryan and Weller, Adrian and Shu, Tianmin and Griffiths, Thomas L and Tenenbaum, Joshua B},
  journal={arXiv preprint arXiv:2502.20502},
  year={2025}
}

@article{park2024generative,
  title={Generative agent simulations of 1,000 people},
  author={Park, Joon Sung and Zou, Carolyn Q and Shaw, Aaron and Hill, Benjamin Mako and Cai, Carrie and Morris, Meredith Ringel and Willer, Robb and Liang, Percy and Bernstein, Michael S},
  journal={arXiv preprint arXiv:2411.10109},
  year={2024}
}

@inproceedings{hamalainen2023evaluating,
  title={Evaluating large language models in generating synthetic hci research data: a case study},
  author={H{\"a}m{\"a}l{\"a}inen, Perttu and Tavast, Mikke and Kunnari, Anton},
  booktitle={Proceedings of the 2023 CHI Conference on Human Factors in Computing Systems},
  pages={1--19},
  year={2023}
}

@article{argyle2023out,
  title={Out of one, many: Using language models to simulate human samples},
  author={Argyle, Lisa P and Busby, Ethan C and Fulda, Nancy and Gubler, Joshua R and Rytting, Christopher and Wingate, David},
  journal={Political Analysis},
  volume={31},
  number={3},
  pages={337--351},
  year={2023},
  publisher={Cambridge University Press}
}

@inproceedings{aher2023using,
  title={Using large language models to simulate multiple humans and replicate human subject studies},
  author={Aher, Gati V and Arriaga, Rosa I and Kalai, Adam Tauman},
  booktitle={International conference on machine learning},
  pages={337--371},
  year={2023},
  organization={PMLR}
}

@inproceedings{park2023generative,
  title={Generative agents: Interactive simulacra of human behavior},
  author={Park, Joon Sung and O'Brien, Joseph and Cai, Carrie Jun and Morris, Meredith Ringel and Liang, Percy and Bernstein, Michael S},
  booktitle={Proceedings of the 36th annual acm symposium on user interface software and technology},
  pages={1--22},
  year={2023}
}

@inproceedings{hwang2025human,
  title={Human Subjects Research in the Age of Generative AI: Opportunities and Challenges of Applying LLM-Simulated Data to HCI Studies},
  author={Hwang, Angel Hsing-Chi and Bernstein, Michael S and Sundar, S Shyam and Zhang, Renwen and Horta Ribeiro, Manoel and Lu, Yingdan and Chang, Serina and Wu, Tongshuang and Yang, Aimei and Williams, Dmitri and others},
  booktitle={Proceedings of the Extended Abstracts of the CHI Conference on Human Factors in Computing Systems},
  pages={1--7},
  year={2025}
}

@book{big_bad_bias,
title = "The big bad bias book",
author = "Ganna Pogrebna and Karen Renaud and Marina Kovaleva",
year = "2025",
language = "English",

}

@article{bogdanov2023working,
  title={Working memory capacity and the risky-choice framing effect: A preregistered replication and extension of Corbin et al.(2010)},
  author={Bogdanov, Boris and Corbin, Jonathan and Dobreva, Sabina and McElroy, Todd and Rachev, Nikolay R},
  journal={Judgment and Decision Making},
  volume={18},
  pages={e39},
  year={2023},
  publisher={Cambridge University Press}
}

@ArtifactSoftware{R,
    title = {R: A Language and Environment for Statistical Computing},
    author = {{R Core Team}},
    organization = {R Foundation for Statistical Computing},
    address = {Vienna, Austria},
    year = {2019},
    url = {https://www.R-project.org/},
}

@article{brachten2020ability,
  title={On the ability of virtual agents to decrease cognitive load: an experimental study},
  author={Brachten, Florian and Br{\"u}nker, Felix and Frick, Nicholas RJ and Ross, Bj{\"o}rn and Stieglitz, Stefan},
  journal={Information Systems and e-Business Management},
  volume={18},
  number={2},
  pages={187--207},
  year={2020},
  publisher={Springer}
}

@article{sweller1988cognitive,
  title={Cognitive load during problem solving: Effects on learning},
  author={Sweller, John},
  journal={Cognitive science},
  volume={12},
  number={2},
  pages={257--285},
  year={1988},
  publisher={Elsevier}
}

@article{tversky_judgment_1974,
	title = {Judgment under {Uncertainty}: {Heuristics} and {Biases}},
	volume = {185},
	language = {en},
	author = {Tversky, Amos and Kahneman, Daniel},
	year = {1974},
}

@article{khare2021maximizers,
  title={Maximizers and Satisficers: Can’t choose and Can’t reject},
  author={Khare, Adwait and Chowdhury, Tilottama G and Morgan, Jeremy},
  journal={Journal of Business Research},
  volume={135},
  pages={731--748},
  year={2021},
  publisher={Elsevier}
}

@inproceedings{Haapalainen2010,
author = {Haapalainen, Eija and Kim, SeungJun and Forlizzi, Jodi F. and Dey, Anind K.},
title = {Psycho-physiological measures for assessing cognitive load},
year = {2010},
isbn = {9781605588438},
publisher = {Association for Computing Machinery},
address = {New York, NY, USA},
url = {https://doi.org/10.1145/1864349.1864395},
doi = {10.1145/1864349.1864395},
booktitle = {Proceedings of the 12th ACM International Conference on Ubiquitous Computing},
pages = {301–310},
numpages = {10},
location = {Copenhagen, Denmark},
series = {UbiComp '10}
}

@inproceedings{pilli2023exploring,
  title={Exploring conversational agents as an effective tool for measuring cognitive biases in decision-making},
  author={Pilli, Stephen},
  booktitle={2023 10th International Conference on Behavioural and Social Computing (BESC)},
  pages={1--5},
  year={2023},
  organization={IEEE}
}

@incollection{ali_mehenni_nudges_2021,
	address = {Singapore},
	series = {Lecture {Notes} in {Electrical} {Engineering}},
	title = {Nudges with {Conversational} {Agents} and {Social} {Robots}: {A} {First} {Experiment} with {Children} at a {Primary} {School}},
	isbn = {9789811583957},
	shorttitle = {Nudges with {Conversational} {Agents} and {Social} {Robots}},
	url = {https://doi.org/10.1007/978-981-15-8395-7_19},
	language = {en},
	urldate = {2022-01-14},
	booktitle = {Conversational {Dialogue} {Systems} for the {Next} {Decade}},
	publisher = {Springer},
	author = {Ali Mehenni, Hugues and Kobylyanskaya, Sofiya and Vasilescu, Ioana and Devillers, Laurence},
	editor = {D'Haro, Luis Fernando and Callejas, Zoraida and Nakamura, Satoshi},
	year = {2021},
	doi = {10.1007/978-981-15-8395-7_19},
	pages = {257--270},
}

@misc{streamlit,
  author = {Streamlit},
  title = {Streamlit: A Faster Way to Build and Share Data Apps},
  year = {2019},
  howpublished = {Available at \url{https://streamlit.io/}},
  note = {Accessed: 2025-02-27}
}

@article{deck2015effect,
  title={The effect of cognitive load on economic decision making: A survey and new experiments},
  author={Deck, Cary and Jahedi, Salar},
  journal={European Economic Review},
  volume={78},
  pages={97--119},
  year={2015},
  publisher={Elsevier}
}

@article{pancholi2009,
author = {Pancholi, Bhavna and Dunne, Mark and Armstrong, Richard},
year = {2009},
month = {11},
pages = {},
title = {Sample size estimation and statistical power analyses},
volume = {16}
}

@article{faul2009statistical,
  title={Statistical power analyses using G* Power 3.1: Tests for correlation and regression analyses},
  author={Faul, Franz and Erdfelder, Edgar and Buchner, Axel and Lang, Albert-Georg},
  journal={Behavior research methods},
  volume={41},
  number={4},
  pages={1149--1160},
  year={2009},
  publisher={Springer}
}

@article{whitney2008framing,
  title={Framing effects under cognitive load: The role of working memory in risky decisions},
  author={Whitney, Paul and Rinehart, Christa A and Hinson, John M},
  journal={Psychonomic bulletin \& review},
  volume={15},
  number={6},
  pages={1179--1184},
  year={2008},
  publisher={Springer}
}

@misc{prolific2024,
  author       = {Prolific},
  title        = {Prolific},
  year         = {2024},
  howpublished = {\url{https://www.prolific.com}},
  note         = {First released in 2014. Current version accessed in {September 2025}. London, UK.},
  copyright    = {© 2024 Prolific}
}

@article{tversky1981framing,
  title={The framing of decisions and the psychology of choice},
  author={Tversky, Amos and Kahneman, Daniel},
  journal={science},
  volume={211},
  number={4481},
  pages={453--458},
  year={1981},
  publisher={American Association for the Advancement of Science}
}

@article{wang1996framing,
  title={Framing effects: Dynamics and task domains},
  author={Wang, Xiao Tian},
  journal={Organizational behavior and human decision processes},
  volume={68},
  number={2},
  pages={145--157},
  year={1996},
  publisher={Elsevier}
}

@article{kuang2023framing,
  title={A framing effect of intertemporal and spatial choice},
  author={Kuang, Yi and Huang, Yuan-Na and Li, Shu},
  journal={Quarterly Journal of Experimental Psychology},
  volume={76},
  number={6},
  pages={1298--1320},
  year={2023},
  publisher={SAGE Publications Sage UK: London, England}
}

@article{aravind2024nudging,
  title={Nudging towards sustainable urban mobility: Exploring behavioral interventions for promoting public transit},
  author={Aravind, Avani and Mishra, Sabyasachee and Meservy, Matt},
  journal={Transportation Research Part D: Transport and Environment},
  volume={129},
  pages={104130},
  year={2024},
  publisher={Elsevier}
}

@inproceedings{el-asri-etal-2017-frames,
    title = "{F}rames: a corpus for adding memory to goal-oriented dialogue systems",
    author = "El Asri, Layla  and
      Schulz, Hannes  and
      Sharma, Shikhar  and
      Zumer, Jeremie  and
      Harris, Justin  and
      Fine, Emery  and
      Mehrotra, Rahul  and
      Suleman, Kaheer",
    editor = "Jokinen, Kristiina  and
      Stede, Manfred  and
      DeVault, David  and
      Louis, Annie",
    booktitle = "Proceedings of the 18th Annual {SIG}dial Meeting on Discourse and Dialogue",
    month = aug,
    year = "2017",
    address = {Saarbr{\"u}cken, Germany},
    publisher = "Association for Computational Linguistics",
    url = "https://aclanthology.org/W17-5526/",
    doi = "10.18653/v1/W17-5526",
    pages = "207--219"
}

@article{Yi_multi,
author = {Yi, Zihao and Ouyang, Jiarui and Xu, Zhe and Liu, Yuwen and Liao, Tianhao and Luo, Haohao and Shen, Ying},
title = {A Survey on Recent Advances in LLM-Based Multi-turn Dialogue Systems},
year = {2025},
issue_date = {April 2026},
publisher = {Association for Computing Machinery},
address = {New York, NY, USA},
volume = {58},
number = {6},
issn = {0360-0300},
url = {https://doi.org/10.1145/3771090},
doi = {10.1145/3771090},
journal = {ACM Comput. Surv.},
month = dec,
articleno = {148},
numpages = {38}
}

@article{masatlioglu_rational_2005,
	title = {Rational choice with status quo bias},
	volume = {121},
	issn = {00220531},
	url = {https://linkinghub.elsevier.com/retrieve/pii/S0022053104001115},
	doi = {10.1016/j.jet.2004.03.007},
	language = {en},
	number = {1},
	urldate = {2023-12-11},
	journal = {Journal of Economic Theory},
	author = {Masatlioglu, Yusufcan and Ok, Efe A.},
	month = mar,
	year = {2005},
	pages = {1--29},
}

@article{samuelson_status_1988,
	title = {Status quo bias in decision making},
	volume = {1},
	issn = {1573-0476},
	url = {https://doi.org/10.1007/BF00055564},
	doi = {10.1007/BF00055564},
	language = {en},
	number = {1},
	urldate = {2023-12-08},
	journal = {J Risk Uncertainty},
	author = {Samuelson, William and Zeckhauser, Richard},
	month = mar,
	year = {1988},
	pages = {7--59},
}

@article{rastogi_towards_2020,
	title = {Towards {Scalable} {Multi}-{Domain} {Conversational} {Agents}: {The} {Schema}-{Guided} {Dialogue} {Dataset}},
	volume = {34},
	copyright = {Copyright (c) 2020 Association for the Advancement of Artificial Intelligence},
	issn = {2374-3468},
	shorttitle = {Towards {Scalable} {Multi}-{Domain} {Conversational} {Agents}},
	url = {https://ojs.aaai.org/index.php/AAAI/article/view/6394},
	doi = {10.1609/aaai.v34i05.6394},
	language = {en},
	number = {05},
	urldate = {2024-01-15},
	journal = {Proceedings of the AAAI Conference on Artificial Intelligence},
	author = {Rastogi, Abhinav and Zang, Xiaoxue and Sunkara, Srinivas and Gupta, Raghav and Khaitan, Pranav},
	month = apr,
	year = {2020},
	note = {Number: 05},
	pages = {8689--8696},
}

@misc{thaler_choice_2010,
	address = {Rochester, NY},
	type = {{SSRN} {Scholarly} {Paper}},
	title = {Choice {Architecture}},
	url = {https://papers.ssrn.com/abstract=1583509},
	doi = {10.2139/ssrn.1583509},
	language = {en},
	urldate = {2024-01-18},
	author = {Thaler, Richard H. and Sunstein, Cass R. and Balz, John P.},
	month = apr,
	year = {2010},
}

@article{johnson_beyond_2012,
	title = {Beyond nudges: {Tools} of a choice architecture},
	volume = {23},
	issn = {1573-059X},
	shorttitle = {Beyond nudges},
	url = {https://doi.org/10.1007/s11002-012-9186-1},
	doi = {10.1007/s11002-012-9186-1},
	language = {en},
	number = {2},
	urldate = {2024-01-22},
	journal = {Mark Lett},
	author = {Johnson, Eric J. and Shu, Suzanne B. and Dellaert, Benedict G. C. and Fox, Craig and Goldstein, Daniel G. and Häubl, Gerald and Larrick, Richard P. and Payne, John W. and Peters, Ellen and Schkade, David and Wansink, Brian and Weber, Elke U.},
	month = jun,
	year = {2012},
	pages = {487--504},
}

@inproceedings{caraban_23_2019,
	address = {New York, NY, USA},
	series = {{CHI} '19},
	title = {23 {Ways} to {Nudge}: {A} {Review} of {Technology}-{Mediated} {Nudging} in {Human}-{Computer} {Interaction}},
	isbn = {978-1-4503-5970-2},
	shorttitle = {23 {Ways} to {Nudge}},
	url = {https://dl.acm.org/doi/10.1145/3290605.3300733},
	doi = {10.1145/3290605.3300733},
	urldate = {2024-01-23},
	booktitle = {Proceedings of the 2019 {CHI} {Conference} on {Human} {Factors} in {Computing} {Systems}},
	publisher = {Association for Computing Machinery},
	author = {Caraban, Ana and Karapanos, Evangelos and Gonçalves, Daniel and Campos, Pedro},
	month = may,
	year = {2019},
	pages = {1--15},
}

@book{kahneman_thinking_2011,
	address = {New York, NY, US},
	series = {Thinking, fast and slow},
	title = {Thinking, fast and slow},
	isbn = {978-0-374-27563-1 978-1-4299-6935-2},
	publisher = {Farrar, Straus and Giroux},
	author = {Kahneman, Daniel},
	year = {2011},
	note = {Pages: 499},
}

@inproceedings{echterhoff_ai-moderated_2022,
	address = {New Orleans LA USA},
	title = {{AI}-{Moderated} {Decision}-{Making}: {Capturing} and {Balancing} {Anchoring} {Bias} in {Sequential} {Decision} {Tasks}},
	isbn = {978-1-4503-9157-3},
	shorttitle = {{AI}-{Moderated} {Decision}-{Making}},
	url = {https://dl.acm.org/doi/10.1145/3491102.3517443},
	doi = {10.1145/3491102.3517443},
	language = {en},
	urldate = {2024-03-07},
	booktitle = {{CHI} {Conference} on {Human} {Factors} in {Computing} {Systems}},
	publisher = {ACM},
	author = {Echterhoff, Jessica Maria and Yarmand, Matin and McAuley, Julian},
	month = apr,
	year = {2022},
	pages = {1--9},
}

@article{simon_behavioral_1955,
	title = {A {Behavioral} {Model} of {Rational} {Choice}},
	volume = {69},
	issn = {0033-5533},
	url = {https://www.jstor.org/stable/1884852},
	doi = {10.2307/1884852},
	number = {1},
	urldate = {2024-03-14},
	journal = {The Quarterly Journal of Economics},
	author = {Simon, Herbert A.},
	year = {1955},
	note = {Publisher: Oxford University Press},
	pages = {99--118},
}

@article{eidelman_bias_2012,
	title = {Bias in {Favor} of the {Status} {Quo}},
	volume = {6},
	copyright = {© 2012 Blackwell Publishing Ltd},
	issn = {1751-9004},
	url = {https://onlinelibrary.wiley.com/doi/abs/10.1111/j.1751-9004.2012.00427.x},
	doi = {10.1111/j.1751-9004.2012.00427.x},
	language = {en},
	number = {3},
	urldate = {2024-03-21},
	journal = {Social and Personality Psychology Compass},
	author = {Eidelman, Scott and Crandall, Christian S.},
	year = {2012},
	note = {\_eprint: https://onlinelibrary.wiley.com/doi/pdf/10.1111/j.1751-9004.2012.00427.x},
	pages = {270--281},
}

@inproceedings{schmidhuber_cognitive_2021,
	title = {Cognitive {Load} and {Productivity} {Implications} in {Human}-{Chatbot} {Interaction}},
	url = {https://ieeexplore.ieee.org/abstract/document/9582445},
	doi = {10.1109/ICHMS53169.2021.9582445},
	urldate = {2024-04-12},
	booktitle = {2021 {IEEE} 2nd {International} {Conference} on {Human}-{Machine} {Systems} ({ICHMS})},
	author = {Schmidhuber, Johanna and Schlögl, Stephan and Ploder, Christian},
	month = sep,
	year = {2021},
	pages = {1--6},
}

@misc{pandian_nasa-tlx_2020,
	title = {{NASA}-{TLX} {Web} {App}: {An} {Online} {Tool} to {Analyse} {Subjective} {Workload}},
	shorttitle = {{NASA}-{TLX} {Web} {App}},
	url = {http://arxiv.org/abs/2001.09963},
	language = {en},
	urldate = {2024-05-12},
	publisher = {arXiv},
	author = {Pandian, Vinoth Pandian Sermuga and Suleri, Sarah},
	month = jan,
	year = {2020},
	note = {arXiv:2001.09963 [cs]},
}

@InProceedings{vivek_nudging,
author="Nallur, Vivek
and Renaud, Karen
and Gudkov, Aleksei",
editor="Collier, Rem
and Ricci, Alessandro
and Nallur, Vivek
and Burattini, Samuele
and Omicini, Andrea",
title="Nudging Using Autonomous Agents: Risks and Ethical Considerations",
booktitle="Multi-Agent Systems",
year="2025",
publisher="Springer Nature Switzerland",
address="Cham",
pages="283--296",
isbn="978-3-031-93930-3"
}

@article{xiao_revisiting_2021,
	title = {{Revisiting} {status} {quo} {bias:} {Replication} {of} {Samuelson} {and} {Zeckhauser} {(1988)}},
	volume = {5},
	shorttitle = {Revisiting status quo bias},
	doi = {10.15626/MP.2020.2470},
	journal = {Meta-Psychology},
	author = {Xiao, Qinyu and Lam, Emma and Piara, Muhrajan and Feldman, Gilad},
	month = feb,
	year = {2021},
}

@inproceedings{dubiel_impact_2024,
    address = {Greenville SC USA},
    title = {Impact of {Voice} {Fidelity} on {Decision} {Making}: {A} {Potential} {Dark} {Pattern}?},
    isbn = {979-8-4007-0508-3},
    shorttitle = {Impact of {Voice} {Fidelity} on {Decision} {Making}},
    url = {https://dl.acm.org/doi/10.1145/3640543.3645202},
    doi = {10.1145/3640543.3645202},
    language = {en},
    urldate = {2025-02-20},
    booktitle = {Proceedings of the 29th {International} {Conference} on {Intelligent} {User} {Interfaces}},
    publisher = {ACM},
    author = {Dubiel, Mateusz and Sergeeva, Anastasia and Leiva, Luis A.},
    month = mar,
    year = {2024},
    pages = {181--194},
}

@article{kalashnikova_linguistic_nodate,
    title = {Linguistic {Nudges} and {Verbal} {Interaction} with {Robots}, {Smart}-{Speakers}, and {Humans}},
    language = {en},
    author = {Kalashnikova, Natalia and Vasilescu, Ioana and Devillers, Laurence},
    year = {2024},
}

@article{yamamoto_suggestive_2024,
    title = {Suggestive answers strategy in human-chatbot interaction: a route to engaged critical decision making},
    volume = {15},
    issn = {1664-1078},
    shorttitle = {Suggestive answers strategy in human-chatbot interaction},
    url = {https://www.frontiersin.org/articles/10.3389/fpsyg.2024.1382234/full},
    doi = {10.3389/fpsyg.2024.1382234},
    language = {en},
    urldate = {2025-02-20},
    journal = {Frontiers in Psychology},
    author = {Yamamoto, Yusuke},
    month = mar,
    year = {2024},
    pages = {1382234},
}

@inproceedings{ji_towards_2024,
    address = {Melbourne VIC Australia},
    title = {Towards {Detecting} and {Mitigating} {Cognitive} {Bias} in {Spoken} {Conversational} {Search}},
    isbn = {979-8-4007-0506-9},
    url = {https://dl.acm.org/doi/10.1145/3640471.3680245},
    doi = {10.1145/3640471.3680245},
    language = {en},
    urldate = {2025-02-20},
    booktitle = {26th {International} {Conference} on {Mobile} {Human}-{Computer} {Interaction}},
    publisher = {ACM},
    author = {Ji, Kaixin and Pathiyan Cherumanal, Sachin and Trippas, Johanne R. and Hettiachchi, Danula and Salim, Flora D. and Scholer, Falk and Spina, Damiano},
    month = sep,
    year = {2024},
    pages = {1--10},
}
